\documentclass[11pt,prd,groupedaddress,showpacs,doublecolumn,nofootinbib]{revtex4-1}

\pdfoutput=1 

\usepackage{graphicx}
\usepackage{dcolumn}
\usepackage{bm}
\usepackage{amssymb}
\usepackage{amsmath}
\usepackage{epsfig}
\usepackage[colorlinks,citecolor=blue]{hyperref}
\usepackage{hhline}
\usepackage{color}
\usepackage{multirow}
\usepackage[normalem]{ulem}
\usepackage{slashed}
\usepackage{slashed}
\usepackage{tikz}
\usetikzlibrary{snakes}
\usepackage{blindtext}
\usepackage[section]{placeins}
\usepackage{tikz}
\usepackage{graphicx}
\usetikzlibrary{snakes}

\definecolor{Zsug}{RGB}{0, 145, 33} 
\definecolor{Zcor}{RGB}{210, 0, 210}
\definecolor{Zque}{RGB}{0, 180, 190} 
 
\definecolor{jd}{rgb}{0.858, 0.188, 0.478}

\def\lapp{\mathrel{\rlap{\raise.5ex\hbox{$<$}}
                    {\lower.5ex\hbox{$\sim$}}}}
\def\gapp{\mathrel{\rlap{\raise.5ex\hbox{$>$}}
                    {\lower.5ex\hbox{$\sim$}}}}

{\newcommand{\lsim}{\mbox{\raisebox{-.6ex}{~$\stackrel{<}{\sim}$~}}}
{\newcommand{\gsim}{\mbox{\raisebox{-.6ex}{~$\stackrel{>}{\sim}$~}}}

\newcommand{\bmt}{\begin{pmatrix}}
\newcommand{\emt}{\end{pmatrix}}
\newcommand{\ba}{\begin{array}{c}}
\newcommand{\ea}{\end{array}}
\newcommand{\be}{\begin{equation}}
\newcommand{\ee}{\end{equation}}
\newcommand{\bea}{\begin{eqnarray}}
\newcommand{\eea}{\end{eqnarray}}

\newcommand{\bi}{\begin{itemize}}
\newcommand{\ei}{\end{itemize}}

\newcommand{\baz}{\begin{array}{cc}}

\newcommand{\mathsym}[1]{{}}

\newcommand{\bt}{\begin{tabular}}
\newcommand{\et}{\end{tabular}}

\newcommand{\benu}{\begin{enumerate}}
\newcommand{\eenu}{\end{enumerate}}

\newcommand{\gBL}{g_{B-3L_{\tau}}
}

\newcommand{\bav}{\begin{array}{cccc}}

\begin{document}

\renewcommand*{\thefootnote}{\fnsymbol{footnote}}

\begin{center}
 {\Large\bf Flavoured gauge extension of singlet-doublet fermionic dark matter: neutrino mass, high scale validity and collider signatures}
 \\
 \vskip .5cm
 {
 Basabendu Barman$^{a,}$\footnote{bb1988@iitg.ac.in},
 Debasish Borah$^{a,}$\footnote{dborah@iitg.ac.in },
 Purusottam Ghosh$^{a,}$\footnote{pghoshiitg@gmail.com},
 Abhijit Kumar Saha$^{b,}$\footnote{aks@prl.res.in},}\\[3mm]
 {\it{
 $^a$ Department of Physics, Indian Institute of Technology Guwahati, Assam 781039, India \\
 $^b$Theoretical Physics Division, Physical Research Laboratory, Ahmedabad 380009, India}\\
 }
 \end{center}
\vspace{1cm}

\begin{center}
 {\bf{Abstract}}
\end{center}

\begin{abstract}
We propose an Abelian gauged version of the singlet-doublet fermionic dark matter (DM) model where the DM, combination of a vector like fermion doublet and a fermion singlet, is naturally stabilised by the gauge symmetry without requiring any ad-hoc discrete symmetries. In order to have an enlarged parameter space for the DM, accsessible at collider experiments like the Large Hadron Collider (LHC), we consider the additional gauge symmetry to be based on the quantum $B-3L_{\tau}$. The restriction to third generation of leptons is chosen in order to have weaker bounds from the LHC on the corresponding gauge boson. The triangle anomalies arising in this model can be cancelled by the inclusion of a right handed neutrino which  also takes part in generating light neutrino masses through type I seesaw mechanism. The model thus offers a potential thermal DM candidate, interesting collider signatures and correct neutrino mass along with a stable electroweak vacuum and perturbative couplings all the way up to the Planck scale. We constrain our model parameters from these requirements as well as existing relevant constraints related to DM and colliders.
\end{abstract}


\maketitle
\flushbottom


\setcounter{footnote}{0}
\renewcommand*{\thefootnote}{\arabic{footnote}}

\section{Introduction}
\label{sec:intro}
The fact that our present universe has a large proportion of its energy density contained in a mysterious, non-luminous and non-baryonic form of matter, popularly known as dark matter (DM), has been very well established by now. Amongst notable evidences suggesting this, the galaxy cluster observations made by Fritz Zwicky~ \cite{Zwicky:1933gu} in 1933, observations of galaxy rotation curves in 1970's by Rubin and collaborators~\cite{Rubin:1970zza}, the observation of the bullet cluster 
by Chandra observatory~\cite{Clowe:2006eq} and the measurements of cosmic microwave background (CMB) by several cosmology experiments, the latest of which is the Planck experiment~\cite{Aghanim:2018eyx}. In terms of density 
parameter $\Omega_{\rm DM}$ and $h = \text{Hubble Parameter}/(100 \;\text{km} ~\text{s}^{-1} 
\text{Mpc}^{-1})$, the present DM abundance is conventionally reported as~\cite{Aghanim:2018eyx}:
$\Omega_{\text{DM}} h^2 = 0.120\pm 0.001$
ghosh1992
at 68\% CL. This corresponds to around $26\%$ of the present universe's energy density, filled up by DM. While all such evidences are purely based on gravitational interactions of DM, there exists motivations to expect that DM could have some other forms of weak interactions as well. Interestingly, if DM interactions with the standard model (SM) particles are similar to those of electroweak interactions, and particle DM's mass also remain around the electroweak scale, such a DM can be thermally produced in the early universe, followed by its freeze-out, leaving a thermal relic very close to the observed DM abundance. This remarkable coincidence is often referred to as the weakly interacting massive particle (WIMP) miracle \cite{Kolb:1990vq}.

Similarly, the origin of light neutrino masses and mixing have also been a mystery in the last few decades. While experimental evidence confirms the neutrino mass squared differences to be several order of magnitudes below the electroweak scale, their large mixing, in sharp contrast with the well known quark sector, leads to another puzzle ~\cite{Tanabashi:2018oca}. Due to the absence of the right handed neutrino, the SM can not accommodate light neutrino masses due to absence of neutrino-Higgs coupling at the renormalisable level. One can however, introduce non-renormalisable Weinberg operator~\cite{Weinberg:1979sa} $(L L H H)/\Lambda, L \equiv$ lepton doublet, $\Lambda \equiv$ unknown cut-off scale, at dimension five level, to account for tiny neutrino masses. Dynamical realisation of this operator leads to beyond standard model (BSM) scenarios \cite{Minkowski:1977sc,GellMann:1980vs,Schechter:1980gr} where introduction of heavy singlet neutrinos take part in generating light neutrino masses through type I seesaw mechanism.

Motivated by these two problems in the SM, we consider a popular DM scenario based on vector-like fermions along with an extended gauge symmetry. This extra gauge symmetry, on top of stabilizing the DM, also plays a role in generating light neutrino masses and provides additional incentive of enhanced detection aspects. The DM is an admixture of a vector like singlet fermion and neutral component of a vector like $SU(2)_L$ doublet fermion, popularly known as singlet-doublet fermion DM~\cite{Mahbubani:2005pt,DEramo:2007anh,Enberg:2007rp,Cohen:2011ec,Cheung:2013dua, Restrepo:2015ura,Calibbi:2015nha,Cynolter:2015sua, Bhattacharya:2015qpa,Bhattacharya:2017sml, Bhattacharya:2018fus,Bhattacharya:2018cgx,DuttaBanik:2018emv}. While such singlet-doublet fermion extension of the SM can also have some other motivations like, for example, electroweak baryogenesis, leading to the observed baryon asymmetry of the universe \cite{Egana-Ugrinovic:2017jib}, we confine ourselves to the discussion of DM and its relevant phenomenology. {Typically, vector like fermion (VLF) DM in such scenarios are stabilised by an additional $Z_2$ symmetry under which the DM is odd while all the SM particles are even. A purely singlet VLF DM does not have any renormalizable portal interaction with the SM to generate correct thermal relic abundance. The purely doublet VLF, on the other hand, by virtue of its electroweak gauge interactions, annihilates a lot to SM particles, thus reducing its thermal relic abundance unless its mass is beyond a TeV. A purely doublet VLF also faces stringent constraints from DM direct detection experiments like LUX~\cite{Akerib:2016vxi}, PandaX-II~ \cite{Tan:2016zwf,Cui:2017nnn} and XENON1T~ \cite{Aprile:2017iyp,Aprile:2018dbl} because of large DM-nucleon scattering mediated by electroweak gauge bosons. Even an admixture of a vector-like singlet and a doublet fermion faces tight constraints from direct detection experiments. In order to overcome that, a small Majorana mass term is introduced to split the vector-like mass eigenstates into two pseudo-Dirac ones. This splitting results in inelastic $Z$ boson coupling to DM which can be prevented kinematically.  The admixture of a singlet and a doublet fermion, therefore, remains as an interesting scenario as it can circumvent both of these problems (namely, under/over-abundant relic and too large direct detection cross section of pure singlet/doublet DM), thus allowing the possibility of sub-TeV DM. We embed the singlet-doublet fermion DM within a gauge symmetry based on the $U(1)_{B-3L_{\tau}}$ gauge charge. While gauged $B-L$ symmetric extension of the SM~ \cite{Mohapatra:1980qe,Marshak:1979fm,Masiero:1982fi,Mohapatra:1982xz} has been one of the most widely studied and well motivated BSM frameworks, we restrict it to the third lepton family only in order to evade strong bounds on the $B-L$ gauge boson from the Large Hadron Collider (LHC) as well as flavour physics~ \cite{Khachatryan:2016qkc,Aaboud:2017sjh,Aaboud:2017buh,Sirunyan:2018exx,Chun:2018ibr,Ma:1997nq,Ma:1998dp,Okada:2012sp,Lee:2010hf,Pal:2003ip,Chang:2000xy,Ma:1998dr}. In addition, such family non-universal neutral gauge boson is also motivating from flavour anomalies point of view. Several proposals have appeared in the literature in order to accommodate the reported anomalies in $B$-meson systems by incorporating additional flavoured gauge bosons (see, for example \cite{Chang:2009tx, Crivellin:2015lwa, Altmannshofer:2015mqa, Allanach:2015gkd, Ko:2017lzd, Ko:2017yrd, Allanach:2018lvl}). The possibility of a TeV scale neutral gauge boson enhances the production cross section of different components of the vector-like fermions in comparison to the usual singlet-doublet fermion DM models. Since such neutral gauge bosons also mediate DM annihilations, henc they also affect DM parameter space compared to the scenarios with universal $B-L$ gauge bosons \footnote{A recent summary of such $Z^{\prime}$ mediated DM scenarios can be found in \cite{Blanco:2019hah}.}. Also, the requirement of anomaly cancellation in this model 
introduces a heavy singlet right handed neutrino (RHN) which, along with one or two more right handed neutrinos having vanishing $B-3L_{\tau}$ charges, can take part in generating light neutrino masses through usual type I seesaw mechanism. Such right handed neutrino, charged under the additional gauge symmetry, can be produced resonantly in colliders and can lead to exotic signatures like displaced vertex, if sufficiently long lived \cite{Das:2019fee}. Apart from offering a potential DM candidate, generating light neutrino mass, and providing tantalizing collider signatures, the model also attempts to give a solution to the metastable nature of electroweak vacuum~\cite{Isidori:2001bm,Ellis:2009tp,EliasMiro:2011aa,Alekhin:2012py,Buttazzo:2013uya, Anchordoqui:2012fq,Tang:2013bz,Bhattacharya:2019fgs}. The negative fermionic contribution (primarily due to top quark and VLFs) to the renormalisation group (RG) running of the Higgs quartic coupling is compensated by respective contributions from additional scalars in the model \footnote{With VLF alone, it is possible to make the electroweak vacuum stable, if they have coloured charges, as shown in \cite{Gopalakrishna:2018uxn}.}. In particular, the constraints from the requirement of vacuum stability restricts the gauge coupling of TeV scale $B-3L_{\tau}$ gauge symmetry to $g_{B-3L_{\tau}} \lsim 0.25$ and SM Higgs coupling with VLF to $Y \lsim 0.3$. For such couplings the model also remains perturbative all the way upto the Planck scale. A recent attempt to constrain a similar Abelian gauge model with inverse seesaw origin of light neutrino mass from DM and vacuum stability criteria has appeared in \cite{Das:2019pua}.


This paper is organised as follows: In section~\ref{sec:model} we have introduced the new particles in our model and their interaction Lagrangian. In section~\ref{sec:constraint} we have discussed the constraints on the model parameters arising from stability, unitarity and perturbativity of the scalar potential, electroweak precision observables, LHC searches and generation of light neutrino mass requirements. The details of the parameter space scan for the DM phenomenology is elaborated in section~\ref{sec:dm} where in subsection~\ref{sec:relicdm} we have illustrated the relic density allowed parameter space and in subsection~\ref{sec:DD} direct search is discussed. The high scale validity and perturbativity of the model is elaborated in section~\ref{sec:highscale}, where we have also chosen some of the benchmark points satisfying all relevant constraints in order to perform the collider analysis. The collider signatures of the model, along with discovery potential in the LHC is thoroughly explained in section~\ref{sec:collider}. Finally in section~\ref{sec:conclusion} we have concluded and summarised our findings.

\section{The Model: fields and interactions}
\label{sec:model}

We first tabulate all the particles of the model in table \ref{tab:particle} with their corresponding gauge charges. On top of the familiar SM particles we have a few additional particles. In the fermion sector, there are three right handed neutrinos (RHN): $N_R$ of which $N_{R_{1,2}}$ have zero $B-3L_{\tau}$ charges and $N_{R_{3}}$ has non-zero charge under $U(1)_{B-3L_{\tau}}$. Apart from cancelling the triangle anomalies arising due to the $U(1)_{B-3L_{\tau}}$ symmetry (shown below) these right handed neutrinos help us to generate light neutrino masses via type I seesaw mechanism as we shall discuss in detail.  We also have one VLF doublet $\psi^T:\left(\psi^0~\psi^{-}\right)$ and one VLF singlet $\chi$. The lightest physical state that arises from the mixing of these two will serve as the DM, while the charged components can be produced at the collider to give interesting signatures. In the scalar sector, apart from the SM Higgs doublet $H$ we have two more singlet scalars: $\mathcal{S}$ and $\Phi$ both charged under $U(1)_{B-3L_{\tau}}$.  The additional SM singlet scalar fields take part in spontaneous symmetry breaking (SSB) of $U(1)_{B-3L_{\tau}}$, thus giving mass to the new heavy charge neutral gauge boson. While one such scalar is sufficient for spontaneous gauge symmetry breaking, we need both of them in the set up for phenomenological reasons. One of the scalars helps in providing the required texture
of RH neutrino mass matrix through its non zero vev after spontaneous breaking of $U(1)_{B-3L_\tau}$. The other scalar splits the Dirac VLFs into pseudo-Dirac states, required to avoid stringent direct detection bounds, as we shall explicitly show while discussing the DM phenomenology.

%


\begin{table}[htb]
\begin{center}
\begin{tabular}{|c| c| c| c|c|} 
\hline
Particles & $SU(3)_c$ & $SU(2)$ & $U(1)_Y$ & $U(1)_{B-3L_{\tau}}$ \\ [0.5ex] 
\hline\hline
$q^i_L$  & 3 & 2 & 1/6 & 1/3 \\
\hline
$u^i_R$  & 3 & 1 & 2/3 & 1/3 \\
\hline
$d^i_R$  & 3 & 1 & -1/3 & 1/3 \\
\hline\hline
$l_{1,2L}$  & 1 & 2 & -1/2 & 0 \\
\hline
$l_{3L}$  & 1 & 2 & -1/2 & -3 \\
\hline
$e_{1,2R}$  & 1 & 1 & -1 & 0 \\
\hline
$e_{3R}$  & 1 & 1 & -1 & -3 \\
\hline
$H$ & 1 & 2 & -1/2 & 0 \\
\hline
$\mathcal{S}$ & 1 & 1 & 0 & 3 \\
\hline
$\Phi$ & 1 & 1 & 0 & -3/2 \\
\hline\hline
$N_{R_{1,2}}$  & 1 & 1 & 0 & 0 \\
\hline
$N_{R_{3}}$  & 1 & 1 & 0 & -3 \\
\hline
$\chi$ & 1 & 1 & 0 & 3/4 \\
\hline
$\psi^T:\left(\psi^0,\psi^{-}\right)$ & 1 & 2 & -1/2 & 3/4 \\ 
\hline
\hline
\end{tabular}
\end{center}
\caption{Relevant particle content of the model and their charges under $SM\times U(1)_{B-3L_{\tau}}$.}
\label{tab:particle}
\end{table}
The condition for anomaly cancellation comes from the triangle diagrams with gauge gauge bosons at the vertices:

\bea
\sum_{LH}{\rm Tr}\left[T^a\{T^b,T^c\}\right]-\sum_{RH}{\rm Tr}\left[T^a\{T^b,T^c\}\right],
\eea
with ``Tr'' standing for trace and $T^{a,b,c}$ denoting the corresponding generators. This vanishes exactly as the VLFs contribute identically to the left and right-handed representation. However, the contributions from other fermions do not vanish in general due to their chiral nature. Demanding cancellation of gauge and gravitational anomalies we end up with the following non-trivial relations:

\bea
\begin{split}
{\text{gravity gauge}\implies}9 (-x_d+2 x_q-x_u)+(-x_{e_R}+2 x_{\ell}-x_{N_{R_3}}) =~ & 0 &\\{\text{$U(1)_{B-3L_{\tau}}^3$}\implies}9 \left(-x_d^3+2 x_q^3-x_u^3\right)+\left(-x_{e_R}^3+2 x_{\ell}^3-x_{N_{R_3}}^3\right) =~ & 0 &\\ {\text{$U(1)_{B-3L_{\tau}}^2 U(1)_Y\implies$}} 9 \left(-\frac{1}{3} (-1) x_d^2+\frac{2 x_q^2}{6}-\frac{2 x_u^2}{3}\right)+\left((-1) x_{\ell}^2-(-1) x_{e_R}^2\right) = ~& 0 &\\{\text{$U(1)_{Y}^2 U(1)_{B-3L_{\tau}}\implies$}} 9 \left(\left(-\frac{1}{3}\right)^2 (-x_d)+2 \left(\frac{1}{6}\right)^2 x_q-\left(\frac{2}{3}\right)^2 x_u\right)+\left(2 \left(-\frac{1}{2}\right)^2 x_{\ell}-(-1)^2 x_{e_R}\right) & = 0,
\label{eq:anomaly}
\end{split}
\eea
where $x_i$ stands for the $U(1)_{B-3L_{\tau}}$ charges for all the particles appearing in table~\ref{tab:particle}. The other anomalies, namely $SU(2)^2 U(1)_{B-3L_{\tau}}, SU(3)_c^2 U(1)_{B-3L_{\tau}}$ are trivially cancelled. The $U(1)_{B-3L_{\tau}}$ charges of the relevant fermions are assigned in table \ref{tab:particle} from which it is easy to check that all the anomalies are cancelled by taking all three fermion generations into account. This is a typical feature of non-universal gauge symmetry where anomalies are not cancelled generation wise, but they vanish only for all three generations combined. The $U(1)_{B-3L_{\tau}}$ charges of the scalar fields and vector fermions will be dictated through the interaction terms in the Lagrangian as we explain in the following paragraphs. It is worth mentioning that the minimal $B-L$ gauge symmetric model \cite{Mohapatra:1980qe,Marshak:1979fm,Masiero:1982fi,Mohapatra:1982xz} is anomaly free if three right handed neutrinos having $B-L$ charge $-1$ each are taken into account. However, this is not the only solution to anomaly cancellation conditions. There exist exotic charges of additional chiral fermions that can give rise to vanishing triangle anomalies \cite{Montero:2007cd, Wang:2015saa, Patra:2016ofq, Nanda:2017bmi, Bernal:2018aon}. Similarly, the anomaly cancellation solution mentioned here for our model is not the only possible one, there exists non-minimal solutions for the same which we do not discuss in our work.

%

With this particle content at our disposal, now we can proceed to write the Lagrangian for this model. The Lagrangian contains four more parts on top of the SM Lagrangian:  
\bea
\mathcal{L} =\mathcal{L}_{SM}+\mathcal{L}_{gauge}+\mathcal{L}_{f}+\mathcal{L}_{scalar}+\mathcal{L}_{yuk}.
\label{eq:lagrng}
\eea
The gauge part of the Lagrangian is written as:
\bea
\mathcal{L}_{gauge}=-\frac{1}{4}B^{'}_{\mu\nu}B^{'\mu\nu}+\epsilon B_{\mu\nu}B^{'\mu\nu},
\label{eq:gauge}
\eea
with 
\bea
B^{'}_{\mu\nu} = \partial_{\mu}\left(Z^{'}_{B-3L_{\tau}}\right)_{\nu}
-\partial_{\nu}\left(Z^{'}_{B-3L_{\tau}}\right)_{\mu}.
\eea
The first term in Eq.~\eqref{eq:gauge} is the kinetic term for the new $U(1)_{B-3L_{\tau}}$ gauge boson and the second term arises due to kinetic mixing between $U(1)_Y$ and $U(1)_{B-3L_{\tau}}$ gauge bosons. For simplicity we choose $\epsilon=0$ at the scale of $B-3L_{\tau}$ symmetry breaking as in~\cite{Okada:2018ktp}. The Lagrangian for the new fermion sector (excluding their couplings to the SM Higgs) reads:
\bea
\begin{split}
\mathcal{L}_f &=  \overline{\psi}\slashed{D}\psi +\overline{\chi}\slashed{D}\chi+ \sum_{j=1}^3\overline{N_{R_j}}\slashed{D} N_{R_j}- M_{\psi}\overline{\psi}\psi-M_\chi\overline{\chi}\chi-\frac{1}{2}\sum_{i,j=1,2}M_{ij}\overline{\left(N_{R_i}\right)^c}N_{R_j} \\&-\sum_{i=1,2}\frac{1}{2}y_{i3}\overline{\left(N_{R_3}\right)^c}N_{R_i}\mathcal{S}-y_{\chi}\overline{\left(\chi\right)^c}\chi\Phi + {\rm h.c.},
\label{eq:fermion}
\end{split}
\eea
where under $SM\otimes U(1)_{B-3L_{\tau}}$, the covariant derivative is defined as: 
\bea
D_{\mu}\equiv\left(\partial_{\mu}-i g_2\frac{\tau^a}{2}W_{\mu}^a-i g_1 Y B_{\mu}-i g_{B-3L_{\tau}} Y_{B-3L_{\tau}} Z_{(B-3L_{\tau})\mu}\right).
\label{eq:covdev}
\eea
where the second and third terms on the right hand side will be present only for the VLF doublet while for VLF singlet and RHNs, only the first and the last terms are present. In the last term on right hand side of Eq.~\eqref{eq:covdev}, $Y_{B-3L_{\tau}} $ corresponds to $B-3L_{\tau}$ charge of the respective fermion. The first two terms in Eq.~\eqref{eq:fermion} are the kinetic terms for the VLF doublet and VLF singlet respectively. The third term corresponds to the kinetic term of the RHNs. The mass term for the VLFs is given by the fourth and fifth term. The Majorana mass terms for the RHNs is given by the  sixth and seventh term. The seventh term decides the $B-3L_{\tau}$ charge of the scalar singlet $\mathcal{S}$ (+3) as $N_3$ must have $B-3L_{\tau}$ charge of -3 for the sake of anomaly cancellation. The last term is the Majorana term for the VLF singlet, which is responsible for the pseudo-Dirac splitting of the VLFs.  
For the scalar sector the Lagrangian can be expressed as: 
\bea
\mathcal{L}_{scalar}= \left(D_\mu H\right)^{\dagger}\left(D^{\mu}H\right)+\left(D_\mu\Phi\right)^{\dagger}\left(D^{\mu}\Phi\right)+\left(D_\mu\mathcal{S}\right)^{\dagger}\left(D^{\mu}\mathcal{S}\right)-V(H,\Phi,\mathcal{S}),
\eea
where
\bea
\begin{split}
D_{\mu}H &= \left(\partial_{\mu}-i g_2\frac{\tau^a}{2}W_{\mu}^a-i g_1 Y B_{\mu}\right) H,\\ \nonumber\ 
D_{\mu}\Phi &= \left(\partial_{\mu}-i g_{B-3L_{\tau}} Y_{B-3L_{\tau}} Z_{(B-3L_{\tau})\mu}\right)\Phi, \\ \nonumber\
D_{\mu}\mathcal{S} &= \left(\partial_{\mu}-i g_{B-3L_{\tau}} Y_{B-3L_{\tau}} Z_{(B-3L_{\tau})\mu}\right)\mathcal{S}.
\end{split}
\label{eq:scalar}
\eea
With one SM Higgs doublet and two non-standard singlets the renormalisable scalar potential takes the form: 
\bea
\begin{split}
V\left(H,\Phi,\mathcal{S}\right)&=\mu_H^2 \left(H^{\dagger}H\right)+\lambda_H\left(H^{\dagger}H\right)^2+\mu_{\Phi}^2|\Phi|^2+\lambda_\Phi|\Phi|^4+\mu_{\mathcal{S}}^2|\mathcal{S}|^2 \\&+\lambda_{\mathcal{S}}|\mathcal{S}|^4+\lambda_1\left(H^{\dagger}H\right)|\Phi|^2+\lambda_{\mathcal{S}\Phi}|\Phi|^2|\mathcal{S}|^2+\lambda_2\left(H^{\dagger}H\right)|\mathcal{S}|^2 \\&+\mu\left(\mathcal{S}\Phi\Phi+H.c\right).
\label{eq:scalar-pot}
\end{split}
\eea
The last term is important as it provides mass for the pseudoscalar by explicitly breaking the global symmetry of the potential, in absence of which we would have ended up with one massless Goldstone boson.  Note that, this term also determines the $B-3L_{\tau}$ charge for the second scalar singlet $\Phi$ (-3/2) as the charge for $\mathcal{S}$ is already determined from RHN Majorana mass term in Eq.~\ref{eq:fermion} as discussed earlier. At the same time, this term also restricts the choice for the $B-3L_{\tau}$ charge of the VLFs (+3/4),  as evident from the last term in Eq.~\eqref{eq:fermion}. Thus, the charge assignment of the new scalars and the VLFs is completely determined by the desired interaction Lagrangian as well as mass spectrum.

Now, the SM Higgs field and the $B-3L_{\tau}$ sector scalars are expanded around their respective vacuum expectation value (VEV):

\begin{equation}
H = 
\begin{pmatrix} 
G^{+}  \\
\frac{h+v_d+i z_1}{\sqrt{2}}
\end{pmatrix}\nonumber\\,~ \Phi=\frac{1}{\sqrt{2}}\left(\phi+v_{\Phi}+i z_2\right),~\\
\mathcal{S}= \frac{1}{\sqrt{2}}\left(s+v_{\mathcal{S}}+i z_3\right).
\end{equation}

The gauged $U(1)_{B-3L_{\tau}}$ is spontaneously broken as the two scalar singlets acquire non-zero VEV. Then the weak eigenstates of the scalars mix with each other. Thus, in order to obtain the physical mass eigenstates, we diagonalise the mass matrix as:
\bea
\begin{pmatrix} 
h_1 \\
h_2 \\
h_3
\end{pmatrix}
= U\left(\theta_{12},\theta_{13},\theta_{23} \right)\begin{pmatrix} 
h \\
\phi\\
s
\end{pmatrix},
\label{eq:scalarmass}
\eea
where $U\left(\theta_{12},\theta_{13},\theta_{23} \right)$ is  usual unitary matrix involving the mixing angles $\theta_{12},\theta_{13},\theta_{23}$ and complex phase $\delta=0$. We assume $h_1$ to be SM-like Higgs with a mass of 125 GeV. The minimisation conditions are given by:
\bea
\begin{split}
\mu_H^2=\frac{1}{2} \left(-2 \lambda_Hv_d^2-\lambda_2 v_{\mathcal{S}}^2-\lambda_1 v_{\Phi}^2\right),&\\ \mu_{\Phi}^2= \frac{1}{2} \left(-\lambda_1v_d^2-\lambda_{\Phi\mathcal{S}} v_{\mathcal{S}}^2-3 \lambda  v_{\mathcal{S}} v_{\Phi}-2 \lambda \phi  v_{\Phi}^2\right),&\\ \mu_{\mathcal{S}}^2 = \frac{-\lambda_2v_d^2 v_{\mathcal{S}}-2 \lambda_{\mathcal{S}} v_{\mathcal{S}}^3-\lambda_{\Phi\mathcal{S}} v_{\mathcal{S}} v_{\Phi}^2-\lambda  v_{\Phi}^3}{2 v_{\mathcal{S}}}.
\end{split}
\label{eq:minimcond}
\eea

The weak eigenstates $(h,~\phi,~s)$ in terms of the physical eigenstates $(h_1,~h_2,~h_3)$ and the mixing angles are given by:
\bea
\begin{split}
h &= h_1c_{12} c_{13}+h_2s_{12} c_{13}+h_3 s_{13} &\\  \phi&=c_{12} (c_{23} (h_2-h_1t_{12})+s_{23} (h_3\frac{c_{13}}{c_{12}}-s_{13} (h_1+h_2t_{12})))\nonumber\\ s &=c_{12} c_{23} (h_1t_{12} t_{23}-s_{13} (h_1+h_2t_{12})-h_2t_{23}+h_3 c_{12} c_{13}+h_3s_{12} t_{12} c_{13}).
\end{split},
\label{eq:scalarphystate}
\eea
where $c_{ij}=\cos\theta_{ij}$, $s_{ij}=\sin\theta_{ij}$ and $t_{ij}=\tan\theta_{ij}$ with $\{i,j\}=1,2, 3$ and $i\neq j$. It is easy to understand from here, for $\theta_{12},\theta_{13}=0,\theta_{23}\neq 0$ we revive the SM Higgs purely from $h_1$. Therefore, $\theta_{12}$ and $\theta_{13}$ are constrained from Higgs data, while $\theta_{23}$ is a free parameter. After diagonalising, we are left with three CP-even scalars denoted by $h_{1,2,3}$. We also have three CP-odd pseudoscalars which, after diagonalising to their physical mass eigenstates, are referred to as $A$ and $G_{1,2}$, out of which $G_{1,2}$ turn out to be the Goldstone modes of $B-3L_{\tau}, Z$ gauge bosons giving $m_{G_1}=m_{G_2}=0$. Now, $h_1$ is  the lightest CP-even Higgs that has been seen at the LHC, hence $m_{h_1}=125~\rm GeV$. Also, the mixing angles $\theta_{12}$ and $\theta_{13}$ are constrained from Higgs data which we shall elaborate in section~\ref{sec:constraint}. Essentially the scalar sector has the following free parameters:
\bea
\{m_{h_{2,3}}, m_A, \theta_{23}\}.
\eea

All the quartic couplings appearing in the scalar potential can be expressed in terms of the physical masses and mixings as follows:

\bea
\begin{split}
&2 v_d^2\lambda_H = m_{h_1}^2 c_{12}^2 c_{13}^2+m_{h_2}^2 s_{12}^2 c_{13}^2+m_{h_3}^2 s_{13}^2,\\ & 2v_{\phi}^2\lambda_\Phi = m_{h_1}^2 (c_{12} s_{13} s_{23}+s_{12} c_{23})^2+m_{h_2}^2 (c_{12} c_{23}-s_{12} s_{13} s_{23})^2+m_{h_3}^2 c_{13}^2 s_{23}^2,\\&  2 v_S^2\lambda_S = m_{h_1}^2 s_{12}^2 s_{23}^2+c_{12}^2 \left(m_{h_1}^2 s_{13}^2 c_{23}^2+m_{h_2}^2 s_{23}^2\right)-s_{12} c_{12} s_{13} (1-2 s_{23}^2)\left(m_{h_1}^2-m_{h_2}^2\right)\\&~~~~~~~~~~~+m_{h_2}^2 s_{12}^2 s_{13}^2 c_{23}^2+m_{h_3}^2 c_{13}^2 c_{23}^2-m_A^2, \\ & v_d v_S\lambda_2 = c_{13} \left(s_{12} c_{12} s_{23} \left(m_{h_1}^2-m_{h_2}^2\right)+s_{13} c_{23} \left(-m_{h_1}^2 c_{12}^2-m_{h_2}^2 s_{12}^2+m_{h_3}^2\right)\right), \\ & v_Sv_{\phi}^2 \lambda_{\phi S}=v_{\phi} c_{12}^2 s_{23} c_{23} \left(m_{h_1}^2 s_{13}^2-m_{h_2}^2\right)+v_{\phi} s_{12} c_{12} s_{13} (2 c_{23}^2-1) \left(m_{h_1}^2-m_{h_2}^2\right)\\ &~~~~~~~~~~~~-m_{h_1}^2v_{\phi} s_{12}^2 s_{23} c_{23}+m_{h_2}^2v_{\phi} s_{12}^2 s_{13}^2 s_{23} c_{23}+m_{h_3}^2v_{\phi} c_{13}^2 s_{23} c_{23}+2 m_A^2 v_S, \\ & \mu =- \frac{\sqrt{2} m_A^2 v_S}{v_{\phi}^2}.
\end{split}
\label{eq:lambdas}
\eea

After SSB we are also left with a new massive charge neutral gauge bosons corresponding to broken $U(1)_{B-3L_{\tau}}$. The mass of the new gauge boson is given by:
\bea
m_{Z_{B-3L_{\tau}}}^2=\frac{9}{4} v_{\Phi}^2 g_{B-3L_{\tau}}^2\left(1+4\tilde{v}^2\right),
\label{eq:zmass}
\eea
where $\tilde{v}=\frac{v_{\mathcal{S}}}{v_{\Phi}}$. Eq.~\eqref{eq:zmass} is important in our analysis as depending on $m_{Z_{B-3L_{\tau}}}\gsim 2.4~\rm GeV$ \footnote{We choose this conservative lower bound in order to be in agreement with relevant experimental constraints on flavoured gauge bosons \cite{Khachatryan:2016qkc,Aaboud:2017sjh,Aaboud:2017buh,Sirunyan:2018exx, Chun:2018ibr}.} we can constrain our parameter space. 
Finally, the Lagrangian for the Yukawa sector involving the SM Higgs reads as: 
\bea
\begin{split}
-\mathcal{L}_{yuk} &= \sum_{i=1}^2\sum_{\alpha=e,\mu} y_{\alpha i}~\overline{L_{\alpha}}\widetilde{H} N_{R_{i}}+y_{\tau_3}\overline{L_{3}}\widetilde{H} N_{R_3}+Y\overline{\psi}\tilde{H}\chi +{\rm h.c.},
\label{eq:yuk}
\end{split}
\eea
where the first two terms are the interactions of the SM leptons with the RHNs. Note that, $N_{R_3}$ can only have interaction with the third generation SM leptons because of the $U(1)_{B-3L_{\tau}}$ charge assignment. The last term is the mixing of the two VLFs mediated by SM Higgs. This gives rise to the two physical eigenstates for the VLFs as mentioned below.

\subsection{VLF mass eigenstates}
\label{sec:vlsmasseig}

The Dirac mass matrix in the basis $\{\chi,\psi^0\}$, containing the singlet and doublet VLF can be written as:
\bea
\mathcal{M}_{VLF}=\begin{pmatrix} 
M_\chi & m_D \\
m_D & M_\psi \\
\end{pmatrix}, 
\eea
where $m_D=\frac{Y v_d}{\sqrt{2}}<<M_{\chi}\lsim M_{\psi}$. Thus, the physical eigenstates arise as:
\bea
\begin{pmatrix}
\psi_1\\
\psi_2
\end{pmatrix}=\begin{pmatrix} 
\cos\theta & \sin\theta\\
-\sin\theta & \cos\theta \\
\end{pmatrix}\begin{pmatrix}
\chi\\ \psi^0
\end{pmatrix}, 
\label{eq:vlfphys}
\eea
where $\theta$ is the mixing angle given by:
\bea
\tan\left(2\theta\right) = \frac{2 m_D}{M_\chi-M_\psi}.
\label{eq:mixangle}
\eea
The mass for the physical eigenstates are:
\bea
M_{\psi_1} \simeq M_{\chi} + m_D\sin2\theta\\
M_{\psi_2}\simeq M_{\psi} - m_D\sin2\theta.
\eea

The lightest physical charge neutral fermion from above is a viable DM candidate in this model and we choose it to be $\psi_1$ $(M_{\psi_1} < M_{\psi_2})$. The DM is naturally stable due to our particular choice of the $U(1)_{B-3L_{\tau}}$ charge. In the small mixing limit the charged component of the VLF doublet $\psi^{\pm}$ acquires a mass:
\bea
M_{\psi^{\pm}}=M_{\psi}=M_{\psi_1}\sin^2\theta+M_{\psi_2}\cos^2\theta.
\label{eq:ch-vlf}
\eea
For small $\sin\theta$ ($\approx 0$) limit, $M_{\psi^{\pm}} \simeq M_{\psi_2} $. From Eq.~\eqref{eq:mixangle}, we see that the VLF Yukawa $Y$ is related to the mass difference between the two physical eigenstates, and is no more a free parameter:
\bea
Y = -\frac{(M_{\psi_2}-M_{\psi_1})\sin 2\theta}{\sqrt{2} v_d} = -\frac{\Delta M\sin 2\theta}{\sqrt{2} v_d}.
\label{eq:vlfyuk}
\eea
\subsection{Pseudo-Dirac mass splitting}
\label{sec:pseudsplit}

Due to presence of the Majorana term : $y_{\chi}\overline{\chi^c}\chi\Phi$, the pseudo-Dirac mass matrix for the VLF singlet $\chi$ can be expressed in the basis $(\chi^c~~\chi)^T$:
\bea
\mathcal{M}_{p-Dirac} = \begin{pmatrix} 
m_{\chi} & M_{\chi}\\
M_{\chi} & m_\chi \\
\end{pmatrix},
\eea
where $m_{\chi}=y_{\chi}\frac{v_\Phi}{\sqrt{2}}~\ll M_\chi$. The mass matrix can be expressed in terms of physical states as:
\bea
\begin{split}
\mathcal{L}_{p-Dirac}=&
\overline{\left(\chi~~\chi^c \right)} \begin{pmatrix}  
m_\chi & M_\chi \\
M_\chi & m_\chi \\
\end{pmatrix} \begin{pmatrix} 
\chi^c \\
\chi
\end{pmatrix}\nonumber \\&
\equiv\overline{\left(\chi^a~~\chi^b \right)} \begin{pmatrix}  
M_\chi - m_\chi & 0 \\
0 & M_\chi + m_\chi  \\
\end{pmatrix} \begin{pmatrix} 
\chi^a \\
\chi^b
\end{pmatrix} \equiv \overline{\chi}M_{\chi}\chi,
\end{split}
\label{eq:p-dirac-split}
\eea
where $(\chi^a~~\chi^b)^T$ is the physical pseudo-Dirac eigenstate. Since $m_\chi<<M_\chi$, which we can always assume the last equality in above equation by taking either $y_\chi$ or $v_{\Phi}$ or both to be small. In the presence of the Majorana term, the singlet $\chi$ is split into two pseudo-Dirac states, $\{\chi^a,~\chi^b\}$ which are propagated into the physical states $\{\psi_1^{a,b},~\psi_2^{a,b}\}$ via VLF mixing. As these two pseudo-Dirac states ($\psi^{a,b}$) are nearly degenerate i.e, $\delta m\sim\mathcal{O}\left(100\rm keV\right)$, we consider them to be a single state $\psi=\left(\psi^a~~\psi^b\right)^T$ with mass $M_\psi$, identical to the Dirac mass of $\psi$. This will not make difference in DM relic abundance calculations, while for direct detection, such splitting will play a crucial role in preventing spin independent elastic scattering mediated by neutral gauge bosons.

\section{Constraints on the model parameters}
\label{sec:constraint}
The phenomenology of the model is mainly dictated by following free parameters :
\bea
\{g_{B-3L_{\tau}},~\tilde{v},~v_{\Phi},~\theta_{23},~\sin\theta,~M_{\psi_1},~\Delta M\},
\label{eq:freeparam}
\eea
where $\Delta M=M_{\psi_2}-M_{\psi_1}$ is the difference between heavy and light VLF mass eigenstates. All these parameters are important for both DM as well as collider phenomenology as we shall see later. But before going into the details of parameter space scan, here we would like to explain how different parameters arising in the model are already constrained from theoretical as well as existing experimental bounds. Especially existing collider bound on the mass of the neutral gauge boson $Z_{B-3L_{\tau}}$ puts stringent constraint on the model parameters. Apart from that, there are bounds from stability of the scalar potential and perturbativity of dimensionless couplings, collider bounds on non-standard scalar masses and mixings, and bounds from light neutrino mass. 
\subsection{Stability, perturbativity and tree-level unitarity}
\label{sec:stabpot}
\subsubsection{Stability}
Stability of the scalar potential is mainly dictated by the quartic terms of the scalar potential, $V(H,\Phi,\mathcal{S})$ which is defined as:
\bea\label{co-positivity}
\begin{split}
V^{(4)}\left(H,\mathcal{S},\Phi\right)&=\lambda_H |H|^4 +\lambda_\Phi |\Phi|^4+\lambda_{\mathcal{S}}|\mathcal{S}|^4 +\lambda_{1}|H|^2 |\Phi|^2+\lambda_{2}|H|^2 |\mathcal{S}|^2 + \lambda_{\mathcal{S}\Phi}|\Phi|^2|\mathcal{S}|^2.
\end{split}
\eea
In order to ensure the bounded-from-below condition in any field direction, the quartic couplings of the potential (Eq.\eqref{co-positivity}) must obey the following co-positivity conditions \cite{Kannike:2012pe,Chakrabortty:2013mha}: 
\bea
\lambda_H \geq 0 ,~~~ \lambda_{\Phi} \geq 0, ~~~\lambda_{\mathcal{S}}\geq  0 \nonumber \\
\lambda_{1}+2\sqrt{\lambda_H \lambda_{\Phi}} ~\geq 0 ,~\lambda_{2}+2\sqrt{\lambda_H \lambda_{\mathcal{S}}} ~\geq 0,~\lambda_{\mathcal{S}\Phi}+2\sqrt{\lambda_\Phi \lambda_{\mathcal{S}}}~ \geq 0, \nonumber \\
\sqrt{\lambda_H \lambda_{\mathcal{S}}\lambda_{\Phi}} + \frac{\lambda_{1}}{2}\sqrt{\lambda_{\mathcal{S}}}+\frac{\lambda_{2}}{2}\sqrt{\lambda_{\Phi}}+\frac{\lambda_{\mathcal{S}\Phi}}{2}\sqrt{\lambda_{H}}  \nonumber \\ 
+\sqrt{2\left(\frac{\lambda_{1}}{2}+\sqrt{\lambda_H \lambda_{\Phi}}\right)\left(\frac{\lambda_{2}}{2}+\sqrt{\lambda_H \lambda_{\mathcal{S}}}\right)\left(\frac{\lambda_{\mathcal{S}\Phi}}{2}+\sqrt{\lambda_{\mathcal{S}} \lambda_{\Phi}}\right)}~\geq 0 ~.
\label{eq:StabC}
\eea 
\subsubsection{Perturbativity}
To prevent perturbative breakdown of the model, all quartic, Yukawa and gauge couplings should obey the following limits at any energy scale:

\bea
|\lambda_H| < 4 \pi,~|\lambda_\Phi| < 4 \pi,~|\lambda_{\mathcal{S}}| < 4 \pi,~\nonumber\\
|\lambda_1| < 4 \pi,~|\lambda_2| < 4 \pi,~|\lambda_{\mathcal{S} \Phi}| < 4 \pi,~\nonumber\\
|y_{\alpha j}| < \sqrt{4 \pi},~ |y_{\tau_3 }| < \sqrt{4 \pi},~ |y_{j3}| < \sqrt{4 \pi},~\nonumber \\
|y_{\chi}| < \sqrt{4 \pi},~|Y|< \sqrt{4 \pi}, \nonumber \\
|g_{i=1,2,3}| < \sqrt{4\pi},~|g_{B-3L_{\tau}}| < \sqrt{4\pi} ,
\label{eq:PerC}
\eea

where $j=1,2$ and $\alpha=e,\mu$. 
\subsubsection{Tree-level unitarity}
The quartic couplings of the scalar potential which are shown in Eq.~\eqref{co-positivity} are also constrained from the following tree level perturbative unitarity conditions \cite{Horejsi:2005da,Bhattacharyya:2015nca,Kang:2013zba}:
\bea
|\lambda_{H}| \leq 4 \pi , ~~|\lambda_{\mathcal{S}}| \leq 4\pi , \nonumber \\
|\lambda_1| \leq 8\pi,~~|\lambda_2| \leq 8 \pi,~~|\lambda_{\mathcal{S}\Phi}|\leq 8\pi, \nonumber \\
|x_{1,2,3}| \leq 16 \pi ,
\label{eq:UniC}
\eea
where, $x_{1,2,3}$ are the cubic roots of the polynomial equation detailed in Appendix~\ref{sec:apunit}.

\subsection{Constraint from electroweak precision observables (EWPO)}
\label{sec:ewpo}

Since our model has two BSM scalars and two vector like fermions, hence there should be corrections to the SM electroweak precision observables (EWPO) i.e, $S$,$T$,$U$ parameters~\cite{Peskin:1991sw,delAguila:2008pw,Erler:2010sk,Cynolter:2008ea}. Here we would like to estimate the effect of the BSM particles on those parameters. We have four parameters, namely $\hat S$, $\hat T$, $W$ and $Y$~\cite{Barbieri:2004qk} where $\hat S$ is related to the Peskin-Takeuchi parameter $S$: $\hat S=\frac{\alpha S}{4 s_w^2}$ and $\hat T=\alpha T$ where $\alpha$ is the fine structure constant and $s_w\equiv\sin\theta_w$ corresponds to sine of the Weinberg angle $\theta_w$. The parameters, $W$ and $Y$ on the other hand, are new set of parameters. $T$ parameter is more significant for small mixing in the scalar sector and the constraint on $T$-parameter is parametrised by the following data~\cite{Tanabashi:2018oca}: $\Delta T = T^{\rm xSM}-T^{\rm SM}=0.07\pm 0.12$, where we consider contributions from both the VLFs and the non-standard scalars to $T^{\rm xSM}$. In this situation $\Delta T$ is given by~\cite{Barger:2007im,Ghosh:2015apa}:

\bea
\begin{split}
\Delta T = T^{\rm scalar}(h_1,h_2,h_3)-T^{\rm SM~ Higgs}(h_1) + T^{\rm VLF},
\end{split}
\label{eq:tparam}
\eea
which indicates how much the oblique parameter is shifted from the SM value. Now,



\bea
\begin{split}
T^{\rm scalar}&= -\frac{3}{16\pi s_w^2}\Bigg\{ c_{12}^2\left(\frac{m_{h_1}^2 \log \left(\frac{m_{h_1}^2}{m_Z^2}\right)}{c_w^2 \left(m_{h_1}^2-m_Z^2\right)}-\frac{m_{h_1}^2 \log \left(\frac{m_{h_1}^2}{m_W^2}\right)}{m_{h_1}^2-m_W^2}\right)\\&+s_{12}^2\left( \frac{\left(m_{h_2}^2 \log \left(\frac{m_{h_2}^2}{m_Z^2}\right)\right)}{c_w^2 \left(m_{h_2}^2-m_Z^2\right)}-\frac{m_{h_2}^2 \log \left(\frac{m_{h_2}^2}{m_W^2}\right)}{m_{h_2}^2-m_W^2}\right)\Bigg\}\\&-\frac{3}{16\pi s_w^2}\Bigg\{ c_{13}^2 \left(\frac{m_{h_1}^2 \log \left(\frac{m_{h_1}^2}{m_Z^2}\right)}{c_w^2 \left(m_{h_1}^2-m_Z^2\right)}-\frac{m_{h_1}^2 \log \left(\frac{m_{h_1}^2}{m_W^2}\right)}{m_{h_1}^2-m_W^2}\right)\\& + s_{13}^2\left( \frac{\left(m_{h_3}^2 \log \left(\frac{m_{h_3}^2}{m_Z^2}\right)\right)}{c_w^2 \left(m_{h_3}^2-m_Z^2\right)}-\frac{m_{h_3}^2 \log \left(\frac{m_{h_3}^2}{m_W^2}\right)}{m_{h_2}^2-m_W^2}\right)\Bigg\},
\end{split}
\label{eq:tscalar}
\eea

and $T^{\rm SM~ Higgs}(h_1)$ can be obtained by using the decoupling limits $s_{12}\to 0$ and $s_{13}\to 0$ in Eq.\eqref{eq:tscalar}. The contribution from VLF DM is followed as,

\bea
\begin{split}
T^{\rm VLF}&=\frac{g_2^2}{16\pi m_W^2}\left(-2 \sin^2\theta~\Pi (M_{\psi},M_{\psi_1})\right)\\&-\frac{g_2^2}{16\pi m_W^2}\left(2 \cos^2\theta~\Pi (M_{\psi},M_{\psi_2})\right)\\&+\frac{g_2^2}{16\pi m_W^2}\left(2 \cos^2\theta \sin^2\theta~\Pi (M_{\psi_1},M_{\psi_2})\right),
\end{split}
\label{eq:tvlf}
\eea

where 

\bea
\begin{split}
\Pi(m_i,m_j)&=-\frac{1}{2} \left(m_i^2+m_j^2\right) \left(\text{div}+\log \left(\frac{\mu_{EW}^2}{m_i m_j}\right)\right)\\&+m_i m_j \left(\text{div}+\frac{\left(m_i^2+m_j^2\right) \log \left(\frac{m_j^2}{m_i^2}\right)}{2 \left(m_i^2-m_j^2\right)}+\log \left(\frac{\mu_{EW}^2}{m_i m_j}\right)+1\right)\\&-\frac{1}{4} \left(m_i^2+m_j^2\right)-\frac{\left(m_i^4+m_j^4\right) \log \left(\frac{m_j^2}{m_i^2}\right)}{4 \left(m_i^2-m_j^2\right)},
\end{split}
\label{eq:pi}
\eea

In Eq.~\eqref{eq:tscalar} $m_{h_1}$ refers to the SM Higgs boson with mass 125 GeV, while $m_{h_{2,3}}$ are the two non-standard Higgs bosons appearing in our model. $m_W$ and $m_Z$ are the masses of SM $W$ and $Z$ bosons respectively. The expression for $T$-parameter corresponding to the contribution from the VLFs is given by Eq.~\eqref{eq:tvlf}, where $g_2$ is the $SU(2)_L$ SM gauge coupling. The $\Pi$'s appearing in the expression are given as in Eq.~\eqref{eq:pi} and these correspond to the gauge boson propagator correction due to the VLFs. Here `div' is the usual expression that appears in dimensional regularisation: ${\rm div}=\frac{1}{\epsilon}+\ln 4\pi-\gamma_{\epsilon}$, with $\gamma_{\epsilon}=0.577$ is the Euler-Mascheroni constant ($\epsilon =4-d, d \equiv$ spacetime dimension in dimensional regularisation). Note that the divergence appearing in the last term in Eq.~\eqref{eq:tvlf} (due to the divergences in Eq.~\eqref{eq:pi}) is cancelled by the first two terms . The physical mass eigenstates appearing in this case are $M_{\psi}$, $M_{\psi_1}$ and $M_{\psi_2}$ ($M_{\psi}\equiv M_{\psi^\pm}$ according to Eq.~\eqref{eq:ch-vlf}). Once more we would like to remind that $M_{\psi_2}\approx M_{\psi^{\pm}}$ under small mixing limit as apparent from Eq.~\eqref{eq:ch-vlf}.

The bound on $\hat S$ comes from a global fit: $10^3\hat S=0.0\pm 1.3$~\cite{Barbieri:2004qk}. For $S$-parameter, we consider contribution only due to the VLFs{\footnote{As only $T$-paramter is important for scalar extension of the SM in small mixing case~\cite{Ghosh:2015apa,Barger:2007im}.}} as given by~\cite{Bhattacharya:2018fus,Cynolter:2008ea}: 

\bea
\begin{split}
\hat S &=\frac{g_2^2}{16\pi^2}\left(\tilde\Pi^{'}\left(M_{\psi^{\pm}},M_{\psi^{\pm}},0\right)-\cos^4\theta\tilde\Pi^{'}\left(M_{\psi_1},M_{\psi_1},0\right)-\sin^4\theta\tilde\Pi^{'}\left(M_{\psi_2},M_{\psi_2},0\right)\right)\\&-\frac{g_2^2}{16\pi^2}\left(2\sin^2\theta\cos^2\theta\tilde\Pi^{'}\left(M_{\psi_2},M_{\psi_1},0\right)\right), 
\end{split}
\label{eq:s-param}
\eea

where $g_2$ is the $SU(2)_L$ gauge coupling. The expression for vacuum polarization for identical masses (at $q^2=0$)~\cite{Cynolter:2008ea}:

\bea
\tilde\Pi^{'}\left(m_i,m_i,0\right)=\frac{1}{3}{\text{div}}+\frac{1}{3}\ln\left(\frac{\mu_{EW}^2}{m_i^2}\right).
\label{eq:vp1}
\eea

For two different masses ($m_i\neq m_j$) the expression for vacuum polarization reads~\cite{Cynolter:2008ea}:

\bea
\begin{split}
\tilde\Pi^{'}\left(m_i,m_j,0\right)&=\left(\frac{1}{3}{\text{div}}+\frac{1}{3}\ln\left(\frac{\mu_{EW}^2}{m_i m_j}\right)\right)+\frac{m_i^4-8 m_i^2 m_j^2+m_j^4}{9\left(m_i^2-m_j^2\right)^2}\\&    +\frac{\left(m_i^2+m_j^2\right)\left(m_i^4-4 m_i^2 m_j^2+m_j^4\right)}{6\left(m_i^2-m_j^2\right)^3}\ln\left(\frac{m_j^2}{m_i^2}\right)\\&+m_i m_j \left(\frac{1}{2}\frac{m_i^2+m_j^2}{\left(m_i^2-m_j^2\right)^2}+\frac{m_i^2 m_j^2}{\left(m_i^2-m_j^2\right)^3}\ln\left(\frac{m_j^2}{m_i^2}\right)\right). 
\end{split}
\label{eq:vp2}
\eea

Note that all the divergences appearing in Eq.~\eqref{eq:vp1} and \eqref{eq:vp2} along with the renormalization scale $\mu_{EW}$, are cancelled on substitution in Eq.~\eqref{eq:s-param}.


\begin{figure}[htb!]
$$
\includegraphics[scale=0.36]{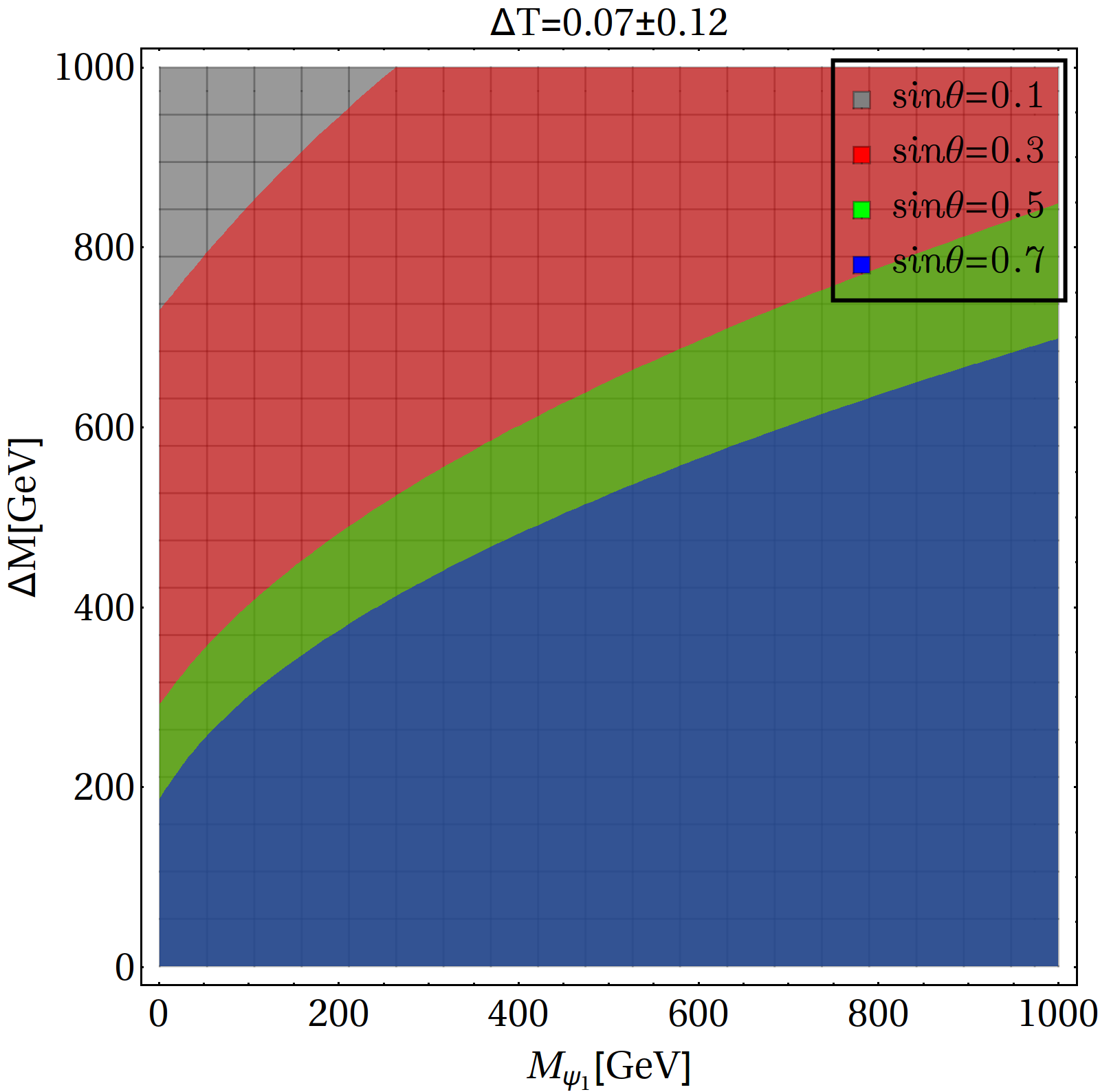} 
\includegraphics[scale=0.36]{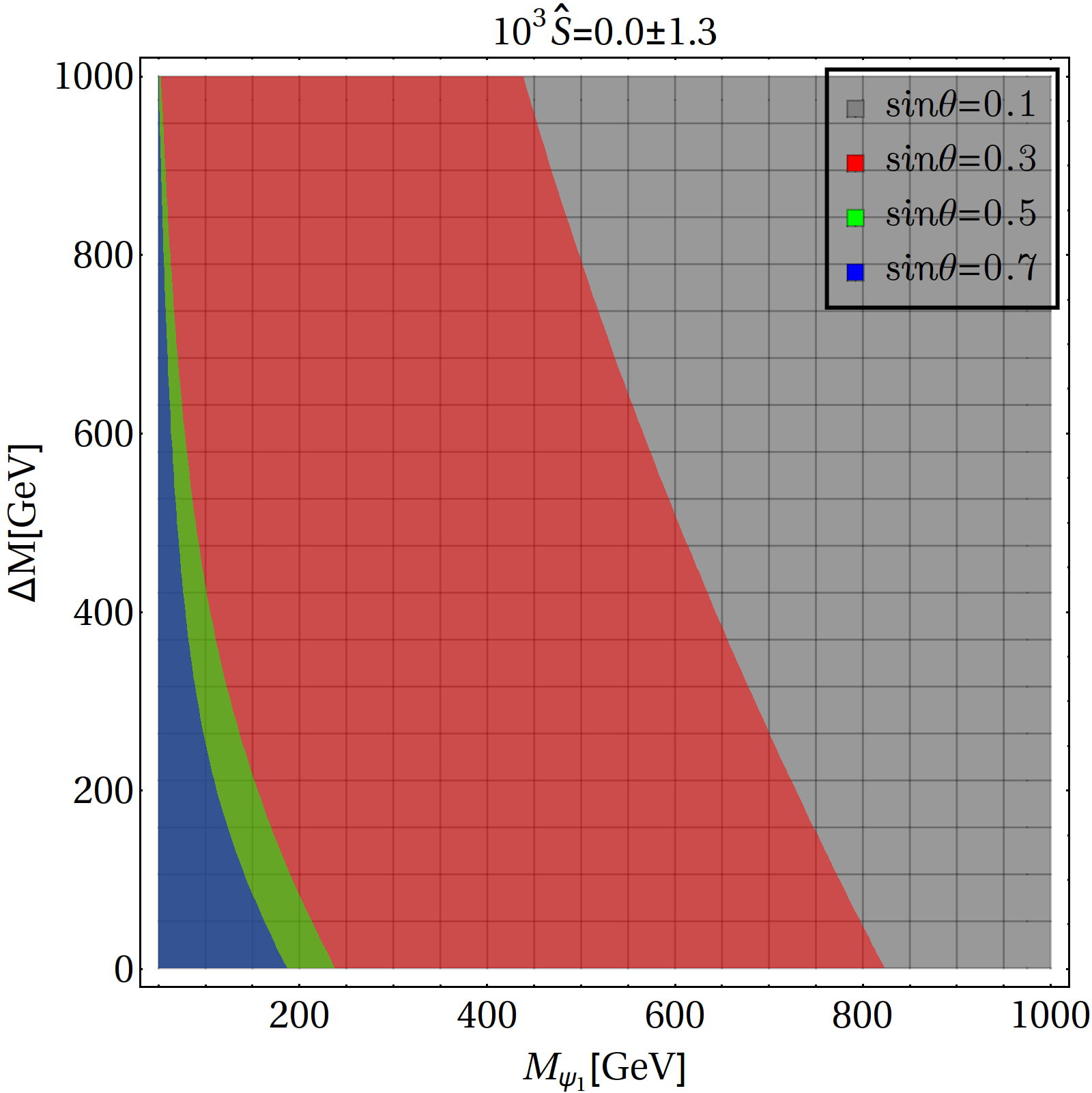}
$$
\caption{Left: Constraints on DM mass $M_{\psi_1}$ and $\Delta M$ from $T$ parameter measurement for $\sin\theta:\{0.1,0.3,0.5,0.7\}$ shown in red, green, blue and black respectively. Right: Limit from $\hat S$ on DM mass $M_{\psi_1}$ and $\Delta M$ for different choices of $\sin\theta:\{0.1,0.3,0.5,0.7\}$ shown respectively in orange, green, blue and red}
\label{fig:stmass}
\end{figure}

We have constrained the two most important free parameters of our model, namely DM mass $M_{\psi_1}$ and $\Delta M$ using these constraints. These are depicted in Fig.~\ref{fig:stmass}. On the left hand side (LHS) of Fig.~\ref{fig:stmass} we have shown the allowed values of $M_{\psi_1}$ and $\Delta M$ that obey the constraint from $T$-parameter given by Eq.~\eqref{eq:tparam}. What we see from the plot in the LHS of Fig.~\ref{fig:stmass}, for small VLF mixing ($\sin\theta\leq 0.5$) it is always possible to get large $\Delta M\gsim 500~\rm GeV$ for DM mass upto 1 TeV within the permissible range of the $T$-parameter. For $\sin\theta\gsim 0.7$, however, the parameter space a is  constrained as $\Delta M$ as large as 1 TeV can not be achieved for low DM mass. In the RHS of Fig.~\ref{fig:stmass} we have shown the allowed range of $S$-parameter in $M_{\psi_1}$-$\Delta M$ plane. Here we see, for small $\sin\theta\sim 0.1$ the whole parameter space is allowed (gray region), however for larger $\sin\theta\gsim 0.3$ one has to stick to lower DM mass. Thus, $S$ parameter constraints the DM mass for large $\sin\theta$. But in any case it is always possible to have a large $\Delta M$ within the observed range of $S$ and $T$ parameter.
\subsection{Constraint on $Z_{B-3L_{\tau}}$ mass from LHC}
\label{sec:lhczbl}

\begin{figure}[htb!]
$$
\includegraphics[scale=0.4]{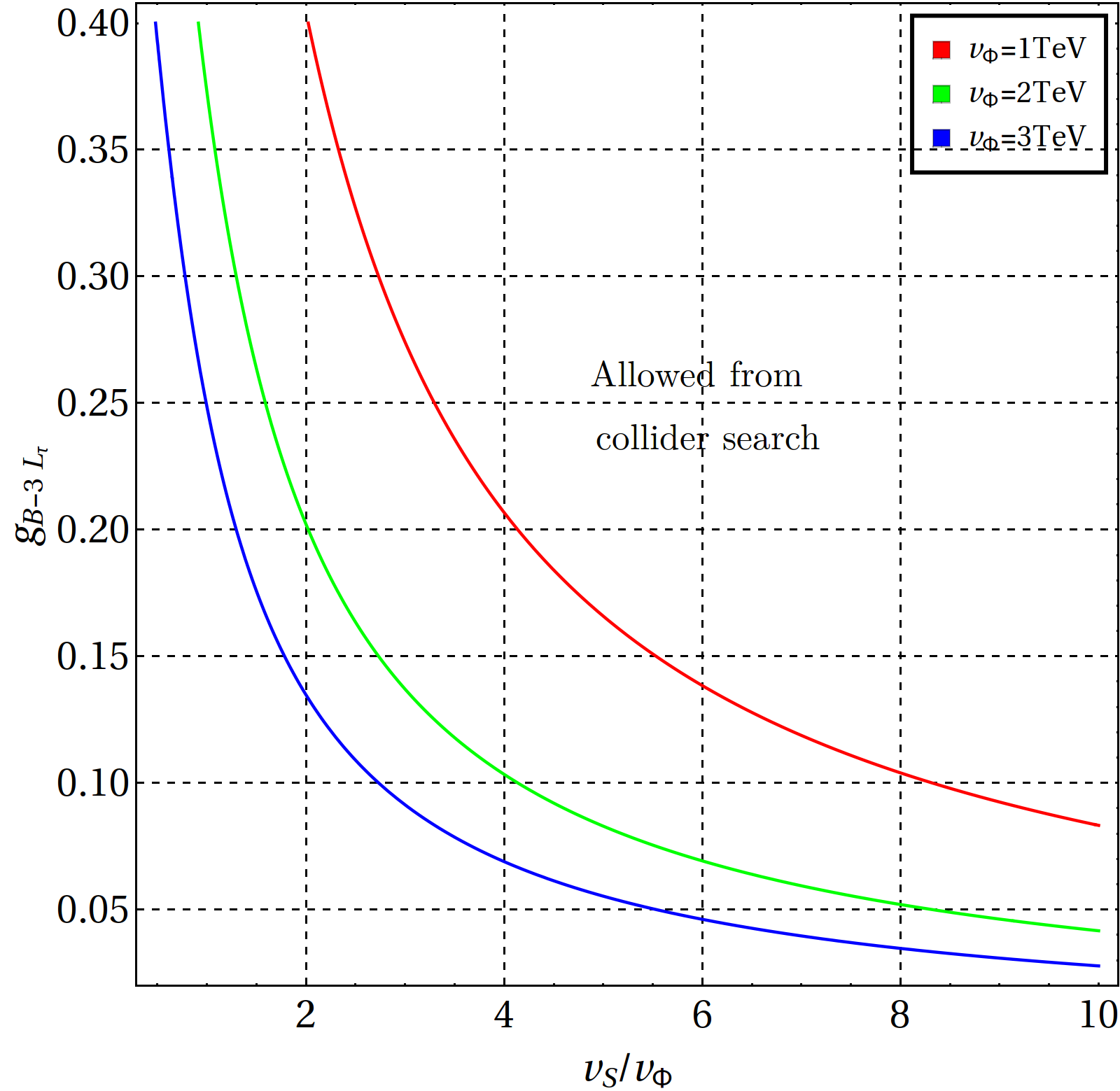}
$$
\caption{Contours satisfying $m_{Z_{B-3L_{\tau}}}\simeq 2.5~\rm TeV$ are shown following Eq.~\eqref{eq:zmass} for three different choices of $v_\Phi:\{1.0,2.0,3.0\}~\rm TeV$ in red, green and blue respectively.}
\label{fig:zprmass}
\end{figure}

Experimental limits from LEP II constrains such new gauge sector by putting a lower bound on the ratio of new gauge boson mass to the new gauge coupling $M_{Z'}/g' \geq 7$ TeV \cite{Carena:2004xs, Cacciapaglia:2006pk}. The corresponding bounds from the LHC experiment have become stronger than this by now. As the main motivation to choose family non-universal gauge boson is to have weaker collider bounds on its mass, hence it is of utmost importance to realise what choice of the free parameters can give rise to right $Z_{B-3L_{\tau}}$ mass satisfying LHC bounds.

Search for heavy neutral Higgs and $Z_{B-3L_{\tau}}$ resonances have been performed at the LHC~\cite{Aaboud:2017sjh}, with the assumption that the heavy resonances decay to $\tau^+\tau^-$ final states. These searches rule out $m_{Z_{B-3L_{\tau}}}<2.42~\rm TeV$ at 95\% CL for sequential SM and $m_{Z_{B-3L_{\tau}}}<2.25~\rm TeV$ at 95\% CL for non-universal $G\left(221\right)$ model. We choose $m_{Z_{B-3L_{\tau}}}\gsim 2.5~\rm TeV$ for a conservative limit. This puts a bound on three parameters in our model, namely: $\{g_{B-3L_{\tau}},\tilde{v},v_{\Phi}\}$. This is shown in Fig.~\ref{fig:zprmass}, where each contour corresponds to $m_{Z_{B-3L_{\tau}}}=2.5~\rm TeV$ and hence the region right to each of the contours is allowed from collider constraint. We have chosen three different VEVs $v_{\Phi}:\{1.0,2.0,3.0\}~\rm TeV$ corresponding to red, green and blue contours respectively. As it is seen, larger $v_{\Phi}$ allows larger gauge boson mass, which is in accordance with Eq.~\eqref{eq:zmass}. One should note here, Eq.~\eqref{eq:zmass} has a consequence. It does not allow to fix either the gauge coupling or $\tilde v$ for a fixed $v_{\Phi}$. As a result, for the parameter space scan we have varied both $g_{B-3L_{\tau}}$ and $\tilde v$ for a fixed $v_{\Phi}$ to keep the $Z_{B-3L_{\tau}}$ mass in the right ballpark ($\gsim 2.5~\rm TeV$). Here we would like to mention that a combination of $B_s$-$\overline{B_s}$ mixing from flavour physics, together with ATLAS' $Z^{'}$ search puts a bound on third family hypercharge models by requiring $m_{Z^{'}}>1.9~\rm TeV$~\cite{Allanach:2019mfl}. However, as we are considering even more conservative bound, our models is safe from such constraints arising from flavour physics measurements.

\subsection{Bounds on singlet scalar from collider}
\label{sec:singltlhc}

The bounds on singlet scalars typically arise due to their mixing with the SM Higgs boson. The bound on such scalar mixing angles would come from both theoretical and experimental constraints
\cite{Robens:2015gla,Chalons:2016jeu}. In case of scalar singlet extension of SM, the strongest bound on scalar-SM Higgs mixing angle ($\theta_m$) comes form $W$ boson mass correction \cite{Lopez-Val:2014jva} at NLO for $250 {\rm ~ GeV} \lesssim m_{h_2} \lesssim 850$ GeV as ($0.2\lsim \sin\theta_m \lsim 0.3$) where $m_{h_2}$ is the mass of  other physical Higgs. Whereas, for $m_{h_2}>850$ GeV, the bounds from the requirement of perturbativity and unitarity of the theory turn dominant  which gives $\sin\theta_m\lesssim 0.2$. For lower values {\it i.e.} $m_{h_2}<250$ GeV, the LHC and LEP direct search \cite{Khachatryan:2015cwa,Strassler:2006ri} and measured Higgs signal strength \cite{Strassler:2006ri} restrict the mixing angle $\sin\theta_m$ dominantly ($\lesssim 0.25$). The bounds from the measured value of EW precision parameter are mild for $m_{h_2}< 1$ TeV. In our analysis we have two singlet scalars which we intend to keep below TeV range. Now considering all the possible bounds, we make conservative choices of the mixing angles (with SM Higgs) as $\sin\theta_{12},\sin\theta_{13}\sim 0.1$. We also fix $v_\Phi\gsim 1~\rm TeV$ which helps in keeping the perturbativity of the theory intact. The other mixing angle $\sin\theta_{23}$ is a free parameter. We keep it below 0.2.

\subsection{Neutrino mass}
\label{sec:neutrino}
The light neutrino mass matrix can be generated via type I seesaw mechanism
\bea
M_{\nu} = -M_D M_R^{-1} \left(M_D\right)^T 
\label{eq:ltnumass},
\eea
where \begin{center}
        $M_D=\begin{pmatrix} 
y_{e_{1}}v_d & y_{e_{2}}v_d & 0 \\
y_{\mu_{1}}v_d & y_{\mu_{2}}v_d & 0 \\
0 & 0 & y_{\tau_{3}}v_d
\end{pmatrix}$ and $M_R=\begin{pmatrix} 
M_{11} & M_{12} & y_{13} v_{\mathcal{S}} \\
M_{12} & M_{22} & y_{23} v_{\mathcal{S}} \\
y_{13} v_{\mathcal{S}}  & y_{23} v_{\mathcal{S}}  & 0
\end{pmatrix}$
       \end{center}
as obtained from the Yukawa interaction of SM leptons in Eq.~\eqref{eq:yuk} and singlet neutral fermions in Eq. \eqref{eq:fermion}. Now, the neutrino mixing angles and mass squared differences are precisely measured from neutrino oscillation experiments~\cite{PhysRevD.98.030001}. This, in turn, puts bound on model parameters including the VEV $v_{\mathcal{S}}$ and relevent Yukawa couplings. This can be understood from the light neutrino mass matrix itself. Diagonalising the mass matrix in Eq.~\eqref{eq:ltnumass} with the usual $3\times 3$ PMNS matrix (choosing the charged lepton mass matrix diagonal) gives the light neutrino masses. We can choose: $y_{e_{1}}=y_{e_{2}}=y_{\mu_{1}}=y_{\mu_{1}}\equiv y_l$, $y_{13}=y_{23}\equiv y$, $M_{11}=M_{22}=M$. With this, if we assume $v_{\mathcal{S}}\sim\mathcal{O}\left(\rm TeV\right)$ and $y_l\simeq y_{\tau_3}\sim\mathcal{O}(10^{-7})$, then we can produce correct order of light neutrino mass for $y\sim 0.1$ and $M\simeq 1~\rm TeV$. Even if we take $M\simeq 10~\rm TeV$ or a different order of magnitude for  $v_S$, correct light neutrino mass can still be obtained with Yukawa couplings of the similar order. However, in that case, the RHNs are beyond the present collider reach. We have kept $v_{\mathcal{S}}\sim\mathcal{O}(\rm TeV)$ such that $Z_{B-3L_{\tau}}$ can be produced at the coillders, which determines the collider signature for this model. And as we have shown above, this choice is not in contradiction with the neutrino mass generation. For simplicity, in our analysis we  shall assume that the annihilation of the DM to RHN final state is kinematically forbidden. In that case the Yukawa couplings $y$, $y_l$ and $y_{\tau_3}$ do not play important role in the DM or collider analysis of this model, hence we can fix them to produce the neutrino mass (as well as mixing) in the right ballpark without disturbing the outcome of the DM or collider phenomenology. Note that, the requirement of generating correct neutrino mass does not put a very tight constraint on the choice of the VEV $v_{\mathcal{S}}$. Thus, in this model, the DM sector and neutrino sector are closely connected even though the bounds on dark sector from right neutrino mass requirement is not very stringent. It should be noted from the structure of $M_R$ that if we had considered a singlet scalar having $B-3L_{\tau}$ charge $6$ instead of $3$, $M_R$ will have $3-3$ element non-zero but $1-3, 2-3$ elements zero. This, as can be checked by using the light neutrino mass formula in Eq. \eqref{eq:ltnumass}, will give rise to a phenomenologically unacceptable light neutrino mass matrix. This once again justifies the choice of singlet scalars and their $B-3L_{\tau}$ charges made in our model. 

Here we would also like to mention that the TeV scale RHNs have decay lifetime $\tau_N\sim 10^{-13}~\rm sec$ considering SM neutrino with scalar final state: $N\to\nu, h_1$ for our choice of Yukawa couplings. This shows that the RHNs do not contribute to the DM relic abundance as $\tau_N<<\tau_{universe}(\sim 10^{17}~\rm sec)$. Also, since they decay very fast ($<< 1~\rm sec$) to SM final states, they do not perturb the standard Big Bang Nucleosynthesis (BBN) picture and hence unconstrained from BBN data. However, for certain choices of lightest neutrino mass, RHN having such gauge interactions can be long lived enough to give interesting collider signatures like displaced vertices, as studied recently by the authors of \cite{Das:2019fee}. Before ending this subsection, we note that, although neutrino mass and mixing do not constrain the mass of $B-3L_{\tau}$ gauge boson directly, constraints on neutrino non-standard interactions (NSI) can be used to set a lower bound on such gauge boson mass as $m_{Z_{B-3L_{\tau}}}/g_{B-3L_{\tau}} > 4.8$ TeV \cite{Heeck:2018nzc}. Since we are considering the stronger bounds from the LHC and LEP II in our analysis, such weaker bounds are trivially satisfied.
\section{Dark matter phenomenology}
\label{sec:dm}

The lightest charge neutral state, $\psi_1$ in VLF sector is the stable DM candidate in our model. It is naturally stable in this set-up precisely due to the $B-3L_{\tau}$ charge assignment (discussed in the subsection~\ref{sec:vlsmasseig}). In this section we have explored in detail the parameter space appearing in Eq.~\eqref{eq:freeparam} allowed by observed relic abundance ($\Omega h^2=0.120\pm0.001$\cite{Aghanim:2018eyx}) and direct search limits (particularly from the XENON 1T experiment \cite{Aprile:2018dbl}).

\subsection{Relic abundance of the DM}
\label{sec:relicdm}

Relic density of DM, $\psi_1$ is governed by SM Higgs ($h_1$) and heavy Higgs ($h_2,~h_3$) mediated annihilation and co-annihilation types number changing processes along with SM gauge boson ($Z,~W^\pm,~\gamma$) and additional heavy gauge boson ($Z_{B-3L_{\tau}}$) mediated annihilation and co-annihilation type processes. All the relevant Feynman graphs contributing to the DM relic abundance are listed in Appendix~\ref{sec:feyn-diag}. The number density of  DM  can be computed by solving the Boltzmann equation~\cite{Kolb:1990vq} of the form: 
 \bea
 \frac{dn}{dt} + 3 H n = -{\langle \sigma v\rangle}_{eff} \Big(n^2-n_{eq}^2\Big),
 \eea
 with $n=n_{\psi_1}$ and $H$ being the Hubble parameter in radiation dominated universe. All types of DM number changing processes are taken into account inside ${\langle \sigma v\rangle}_{eff}$ \cite{Griest:1990kh, Edsjo:1997bg} which is given by
 \bea
{\langle \sigma v\rangle}_{eff}&&= \frac{\bar g_1^2}{g_{eff}^2} {\langle \sigma v \rangle}_{\overline{\psi_1}\psi_1}+\frac{2 \bar g_1 \bar g_2}{g_{eff}^2} {\langle \sigma v \rangle}_{\overline{\psi_1}\psi_2}\Big(1+\frac{\Delta M}{M_{\psi_1}}\Big)^\frac{3}{2} e^{-x \frac{\Delta M}{M_{\psi_1}}} \nonumber \\
&&+\frac{2 \bar g_1 \bar g_3}{g_{eff}^2} {\langle \sigma v \rangle}_{\overline{\psi_1}\psi^-}\Big(1+\frac{\Delta M}{M_{\psi_1}}\Big)^\frac{3}{2} e^{-x \frac{\Delta M}{M_{\psi_1}}} \nonumber \\
&& +\frac{2 \bar g_2 \bar g_3}{g_{eff}^2} {\langle \sigma v \rangle}_{\overline{\psi_2}\psi^-}\Big(1+\frac{\Delta M}{M_{\psi_1}}\Big)^3 e^{- 2 x \frac{\Delta M}{M_{\psi_1}}} \nonumber \\
&& +\frac{\bar g_2^2}{g_{eff}^2} {\langle \sigma v \rangle}_{\overline{\psi_2}\psi_2}\Big(1+\frac{\Delta M}{M_{\psi_1}}\Big)^3 e^{- 2 x \frac{\Delta M}{M_{\psi_1}}} \nonumber \\
&& +\frac{\bar g_3^2}{g_{eff}^2} {\langle \sigma v \rangle}_{{\psi^+}\psi^-}\Big(1+\frac{\Delta M}{M_{\psi_1}}\Big)^3 e^{- 2 x \frac{\Delta M}{M_{\psi_1}}}, 
\label{eq:vf-ann}
\eea
In above equation, $g_{eff}$ is defined as the effective degrees of freedom, given by
\bea
g_{eff}=\bar g_1 + \bar g_2 \Big(1+\frac{\Delta M}{M_{\psi_1}}\Big)^\frac{3}{2} e^{-x \frac{\Delta M}{M_{\psi_1}}} + \bar g_3\Big(1+\frac{\Delta M}{M_{\psi_1}}\Big)^\frac{3}{2} e^{-x \frac{\Delta M}{M_{\psi_1}}} ,
\label{eq:eff-cs}
\eea
where $\bar g_1 ,~ \bar g_2$ and $\bar g_3$ are the internal degrees of freedom of $\psi_1, ~\psi_2 \rm ~and~ \psi^\pm$ respectively, and $x=x_f=\frac{M_{\psi_1}}{T_f}$, where $T_f$ is the freeze out temperature.

Relic density of $\psi_1$ one can approximately expressed as~\cite{Bhattacharya:2016ysw}:

\bea
\Omega_{\psi_1} h^2 \simeq \frac{x_f}{\sqrt{g_*}} \frac{854.45\times 10^{-13} {~~\rm GeV^{-2}}}{ ~{\langle \sigma v\rangle}_{eff}}
\eea

where $x_f \approx 20$ and $g_* = 106.7$, the degrees of freedom for all SM particles . Note here that we have not used the above approximate formula for computing DM ($\psi_1$) relic density. In order to calculate relic density, we have used the package {\tt MicrOmegas}~\cite{Belanger:2001fz} for which the model files are generated from {\tt LanHEP}~\cite{Semenov:2014rea}.

To see the behaviour of DM ($\psi_1$) relic density, we have fixed the VEV $v_\Phi=3.0~\rm TeV$ such that the mass of $Z_{B-3L_{\tau}}$ is always above the collider bound ($m_{Z_{B-3L_{\tau}}} > 2.5$ TeV) for suitable choices of $\tilde v$ and $\gBL$ as shown in Fig.~\ref{fig:zprmass}. We have also fixed the masses of all the non-standard scalars as $\{m_{h_2},~m_{h_3},~m_A\}=\{200,~300,~250\}~\rm GeV$ obeying existing collider bounds as described in subsection~\ref{sec:singltlhc} . All scalar mixing angles are also kept fixed: $\{\sin\theta_{12},~\sin\theta_{13},~\sin\theta_{23}\}=\{0.1,0.1,0.2\}$. We have kept fixed the above parameters throughout our analysis. For the above choice of free parameters, other dependent quartic couplings are determined by Eq.~\eqref{eq:lambdas}. With this choice of parameters, we first illustrate how the relic abundance of the DM varies with DM mass for different choices of VLF mixing $\sin\theta$, $\Delta M$, the new gauge coupling $g_{B-3L_{\tau}}$ and the ratio of the VEVs $\tilde v$, keeping all other parameters fixed at their values mentioned before. 

\begin{figure}[htb!]
$$
\includegraphics[scale=0.4]{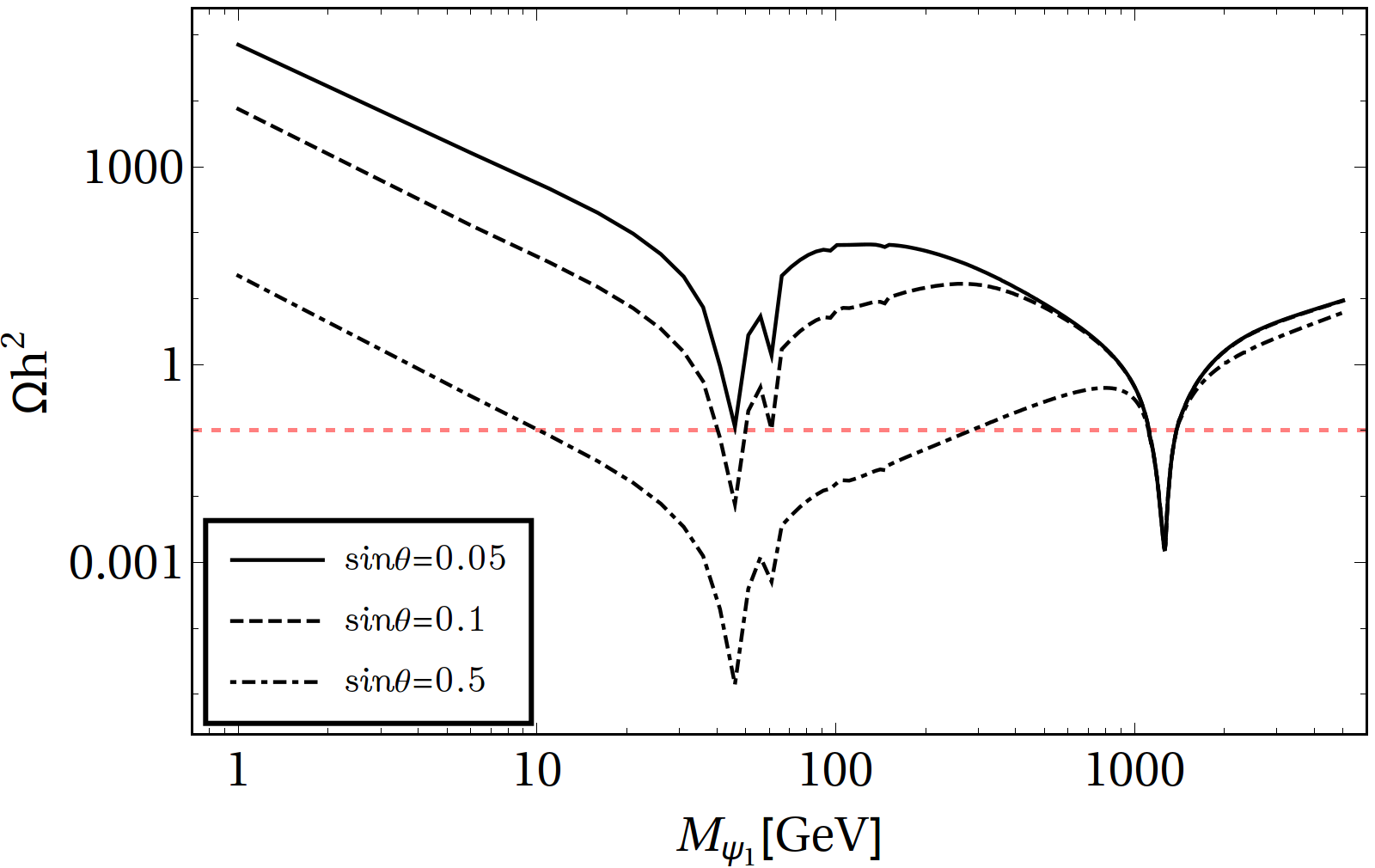} 
\includegraphics[scale=0.4]{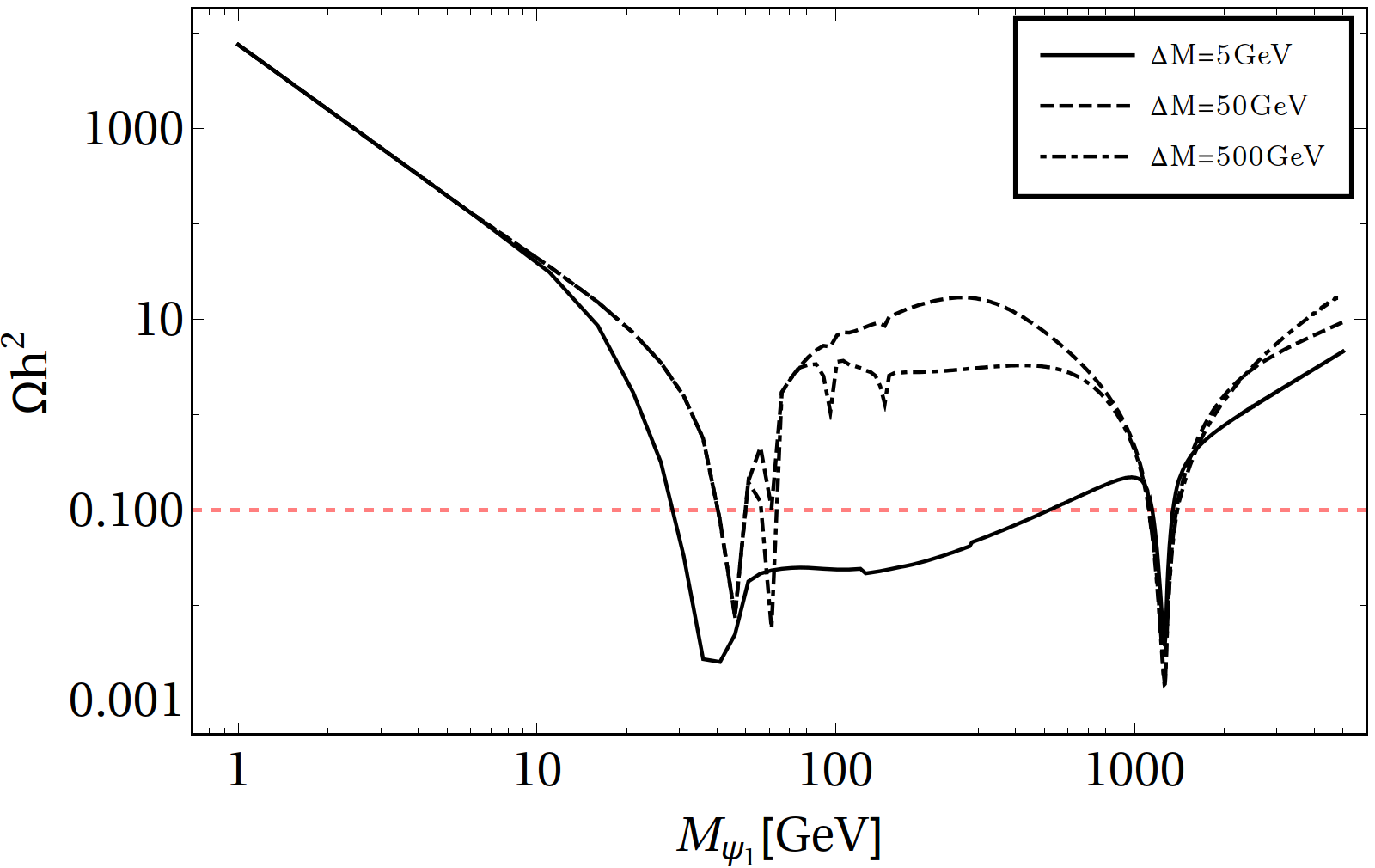}
$$
$$
\includegraphics[scale=0.4]{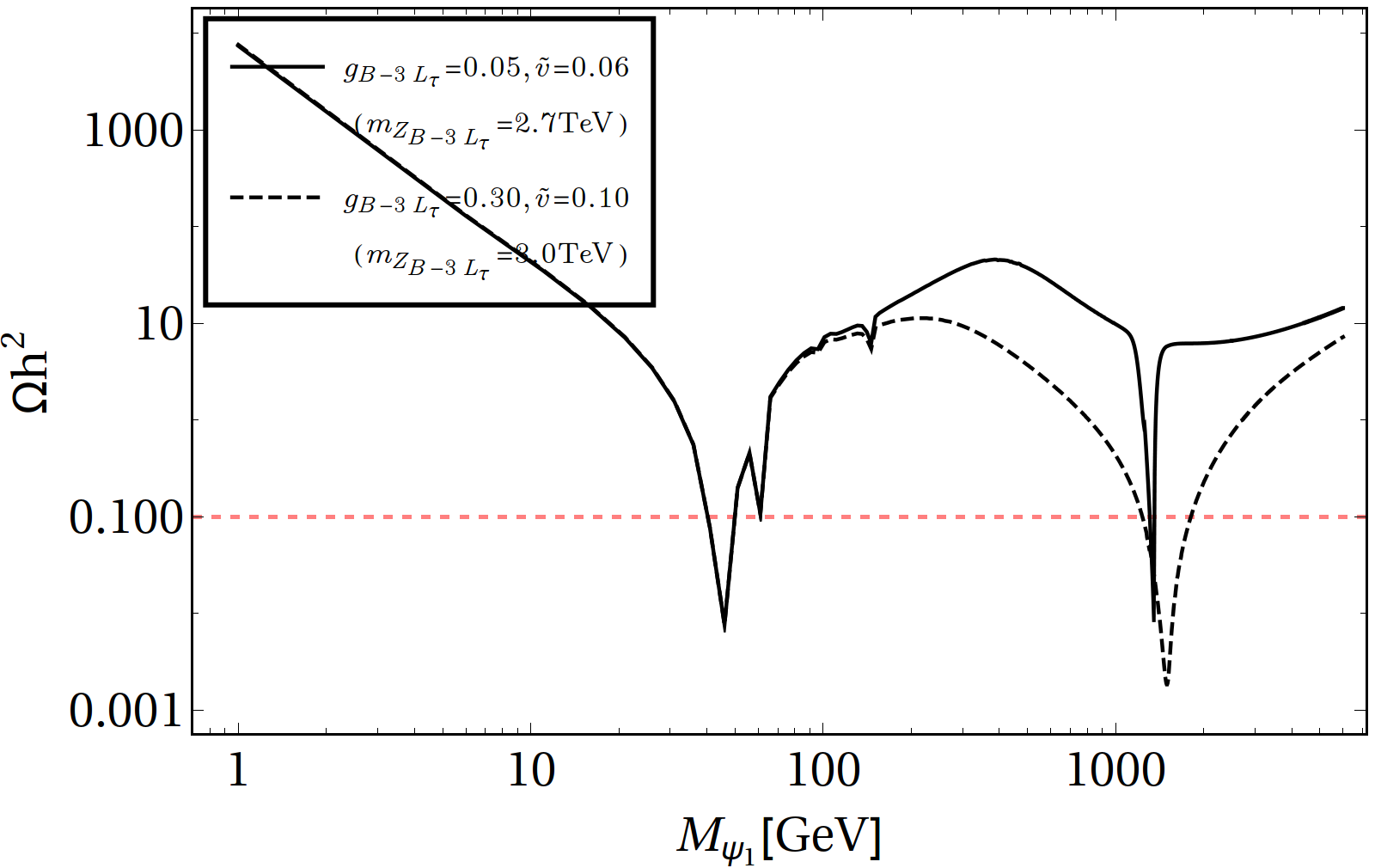}
$$
\caption{Top: Variation of relic abundance of $\psi_1$ with $M_{\psi_1}$ for different choices of the VLF mixing $\sin\theta: \{0.05,0.1,0.5\}$  keeping $\Delta M=50~\rm GeV~$ fixed (top left) and for different choices of $\Delta M:\{5,100,500\}~\rm GeV$ keeping $\sin\theta=0.1$ fixed (top right). $g_{B-3L_{\tau}}=0.2$ and $\tilde v=1.3$ are kept fixed for both of these plots.  Bottom: Variation of relic density with $M_{\psi_1}$ plotted for two different choices of $\gBL:\{0.2,0.3\}$ for fixed value of $\tilde v=1.3$, $\Delta M=50~\rm GeV$ and $\sin\theta=0.1$. In all three plots, the horizontal dashed line (red coloured) corresponds to the central value of Planck limit on DM relic \cite{Aghanim:2018eyx}.}
\label{fig:relicST}
\end{figure}

In the top left panel of Fig.~\ref{fig:relicST} we have shown how the relic abundance of the DM changes with its mass for different choices of the VLF mixing $\sin\theta:\{0.05,0.1,0.5\}$ in solid black, black dashed and black dot-dashed lines respectively. We have kept $\Delta M=50~\rm GeV$ fixed and chose $g_{B-3L_{\tau}}=0.2$ with $\tilde v=1.3$ such that $Z_{B-3L_{\tau}}$ mass satisfies the collider bound ($ > 2.5 $ TeV). The very first feature that one should note is the presence of three major resonant drops due to SM Higgs ($M_{\psi_1}\sim m_{h_1}/2$), SM $Z~(M_{\psi_1}\sim m_Z/2)$ and new gauge boson $Z_{B-3L_{\tau}}~(M_{\psi_1}\sim m_{Z_{B-3L_{\tau}}}/2)$. As one can notice, with the increase in $\sin\theta$, the DM becomes more and more under-abundant as the annihilation via $Z$ and Higgs bosons $h_{1,2,3}$ become more dominant, increasing the total annihilation cross-section. $\Delta M$ in this case is large enough and we can safely ignore the effects of co-annihilation.  In the top right panel of Fig.~\ref{fig:relicST} we have illustrated how relic abundance varies with the DM mass for three choices of $\Delta M:\{5,100,500\}~\rm GeV$ shown in solid black, dashed black and dot-dashed black curves respectively while keeping $\sin\theta$ fixed at 0.1 along with $\gBL=0.2$ and $\tilde v=1.3$. For small $\Delta M$ the co-annihilation plays dominant role, making the DM under-abundant. On the other hand, for large $\Delta M$, co-annihilation becomes sub-dominant, and as a result the Higgs ($h_{1,2,3}$) mediated resonance peaks become more prominent. Here one can notice that $h_{1,2,3}$ mediated resonances are more visible for large value of $\Delta M$, as the corresponding annihilation processes dominate. It is interesting to note from this plot that as we increase $\Delta M$ from 100 GeV to 500 GeV, the relic abundance decreases for $M_{\psi_1} \gtrsim 50$ GeV. This is due to the fact that, although increased $\Delta M$ decreases the efficiency of coannihilation processes, it increases the VLF coupling with SM Higgs, thereby increasing the scalar mediated annihilation processes.

Finally, in the bottom panel of Fig.~\ref{fig:relicST} we have shown two different sets of $\{\tilde v,g_{B-3L_{\tau}}\}:\{6,~0.05\};\{1,0.3\}$ in solid black and dashed black curves respectively. This gives two different resonances at DM mass $M_{\psi_1}$ $\sim \frac{2.7}{2}~\rm TeV$ and $M_{\psi_1} \sim \frac{3.0}{2}~\rm TeV$ due to two different masses of $Z_{B-3L_{\tau}}$. The nature of the two curves is almost identical except for two different resonances at $m_{Z_{B-3L_{\tau}}}/2$. This clearly tells the fact that the dependence of DM relic abundance on the new gauge coupling $\gBL$ is mild compared to the dependence on the VLF mixing $\sin\theta$ and mass difference $\Delta M$. In each of the plots the dashed red straight line corresponds to the central value of Planck limits on DM relic abundance \cite{Aghanim:2018eyx}.

We now scan all the free parameters of our analysis in the following range:
\bea
\begin{split}
M_{\psi_1}:\{1-4000~\rm GeV\};~\Delta M:\{1-1000~\rm GeV\}; ~\sin\theta:\{0.01-0.5\};&\\~g_{B-3L_{\tau}}:\{0.01-0.3\};~\tilde v:\{0.1-5.0\}.
\label{eq:param-val}
\end{split}
\eea
Here we remind the readers once again that the choices for other parameters are kept fixed in the analysis as: 
$\sin\theta_{12}=0.1,~\sin\theta_{13}=0.1,~\sin\theta_{23}=0.2,~ m_{h_2}=200\; {\rm GeV},~ m_{h_2}=300 \;{\rm GeV}~, m_{A}=250 \;{\rm GeV} {\rm and} ~ v_{\Phi}=3.0\; {\rm TeV} $ . The gauge coupling $g_{B-3L_{\tau}}$ is varied upto $ 0.3$, such that the model remains valid at high scale which we shall discuss in detail later. Throughout the scan we have ensured that $m_{Z_{B-3L_{\tau}}}\gsim 2.5~\rm TeV$ by properly adjusting $\gBL$ and $\tilde v$, keeping $v_{\Phi}$ fixed at 3 TeV as mentioned earlier.

The allowed parameter space from relic density requirement set by Planck experiment is shown in Fig.~\ref{fig:relicparam} in $M_{\psi_1}$-$\Delta M$ plane. In the top left corner of Fig.~\ref{fig:relicparam} we have shown this parameter space for different ranges of the VLF mixing $\sin\theta$ shown in red $(0.01\leq\sin\theta<0.05)$, green $(0.05\leq\sin\theta<0.1)$, blue $(0.1\leq\sin\theta<0.3)$ and black $(0.3\leq\sin\theta<0.5)$ where $0.05\leq g_{B-3L_{\tau}}\leq 0.3$. The relic abundance criteria is satisfied by moderate to large $\sin\theta$, while small $\sin\theta$'s are confined near SM $Z$ and SM Higgs resonance and near $\Delta M\sim 10~\rm GeV$ for $M_{\psi_1}\gsim 100~\rm GeV$. In order to understand the pattern more clearly we have chosen a fixed $\sin\theta=0.2$ and plotted the same parameter space in the top right corner of Fig.~\ref{fig:relicparam}. For DM mass around $M_{\psi_1}\sim 20~\rm GeV$ there are only a few annihilation channels open for the DM. Now, for small $\Delta M$ co-annihilation comes into picture, increasing the effective annihilation cross-section (in Eq.~\eqref{eq:vf-ann}). This causes the initial under abundance for small $\Delta M$. On further increasing $\Delta M$, co-annihilation becomes sub-dominant. As a result the DM becomes over abundant since the effective cross-section (Eq.~\eqref{eq:vf-ann}) diminishes. For $M_{\psi_1}\simeq 40-70~\rm GeV$ there is a huge under abundant region (green points) due to Z and SM Higgs resonances. Upon further increasing DM mass, we again get under-abundant regions in low $\Delta M$ region due to enhanced coannihilation and high $\Delta M$ region due to increased scalar portal annihilations as well as the resonance of the heavy scalars ($h_{2,3}$), which we noticed while discussing the behaviour of Fig.~\ref{fig:relicST} as well. As mentioned earlier, the Higgs portal Yukawa $Y$ becomes large enough in such a case (being proportional to $\Delta M$ for a fixed $\sin\theta$) resulting a net increase in the annihilation cross-section. For DM mass $\sim 1~\rm TeV$ there is a huge overabundant region in the parameter space. This is due to the $1/M_{\psi_1}^2$ suppression in the annihilation cross-section due to heavy DM mass. Correct relic abundance is still possible to reach at a very large $\Delta M$ as then the Yukawa $Y$ becomes large enough to compensate the decrease in cross-section due to mass suppression. Large $Y$ can also be achieved by increased the VLF mixing $\sin\theta$ and for $\sin\theta\lsim\mathcal{O}(1)$ we can satisfy correct abundance in this region even with moderate $\Delta M$. As we go beyond DM mass of 1 TeV, $Z_{B-3L_\tau}$ resonance shows up (as the minimum value of $m_{Z_{B-3L_\tau}}$ is 2.5 TeV). Now, since both $\tilde v$ and $\gBL$ are being varied, $m_{Z_{B-3L_\tau}}$ is not fixed (according to Eq.~\eqref{eq:zmass}). As a result, the resonance region is not sharp but broad due to different $m_{Z_{B-3L_\tau}}$. For all possible choices of $\tilde v$ and $\gBL$ according to Eq.~\eqref{eq:param-val}, $m_{Z_{B-3L_\tau}}$ is being varied between $\sim 2.5~\rm TeV$ to $\sim 12~\rm TeV$. Because of this, the resonance band lies between $M_{\psi_1}\simeq\{\frac{2.5}{2}-\frac{12}{2}\}~\rm TeV$ for all possible $\Delta M$. It is seen that regions corresponding to over-abundance, under-abundance and right relic overlap on each other in the $Z_{B-3L_\tau}$ resonance region due to different values of $m_{Z_{B-3L_\tau}}$ and $\Delta M$. Note that if we go to even higher DM mass we will still find relic abundance allowed parameter space due to resonances from different $m_{Z_{B-3L_\tau}}$.

\begin{figure}[htb!]
$$
\includegraphics[scale=0.41]{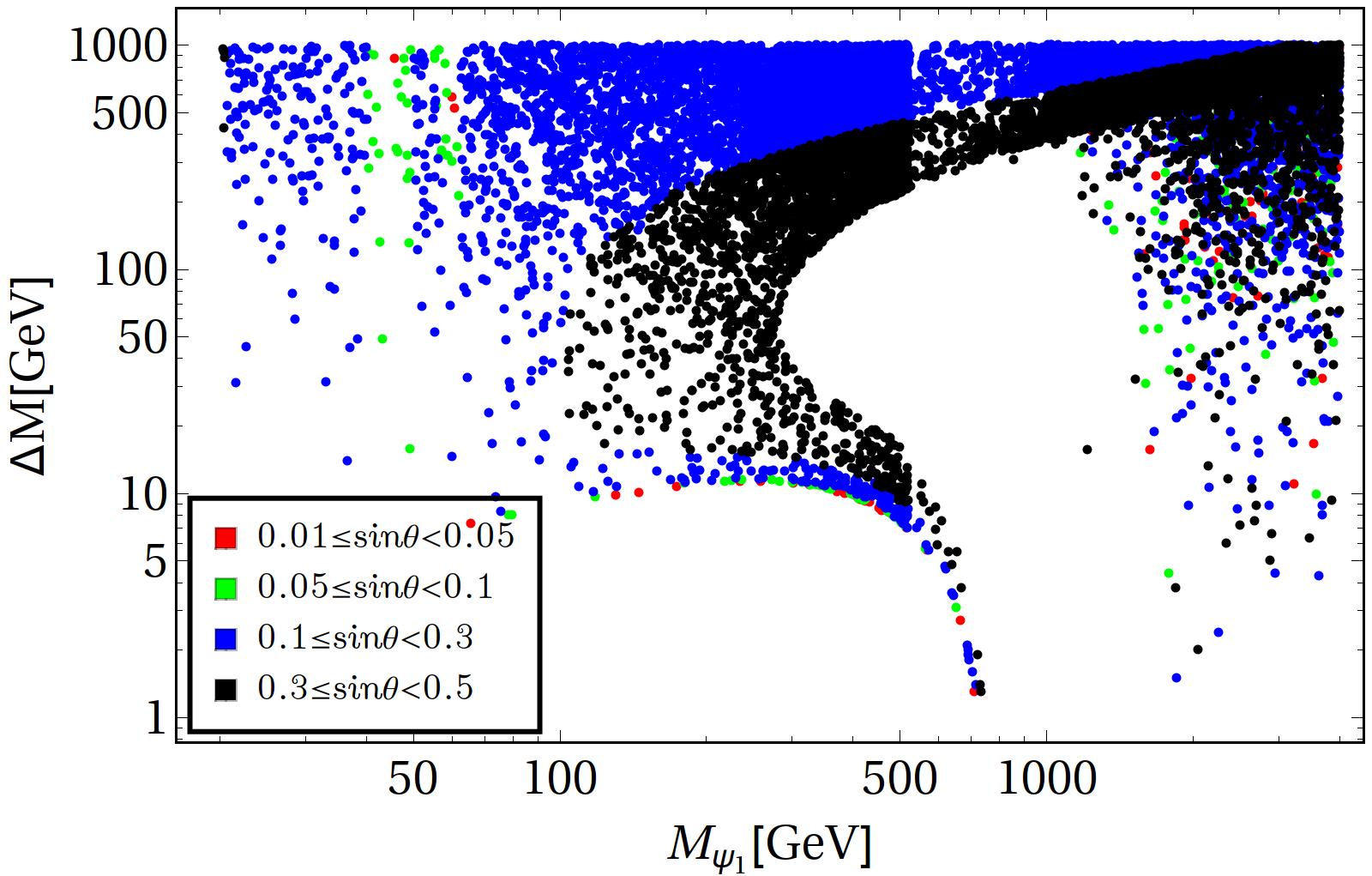}
\includegraphics[scale=0.41]{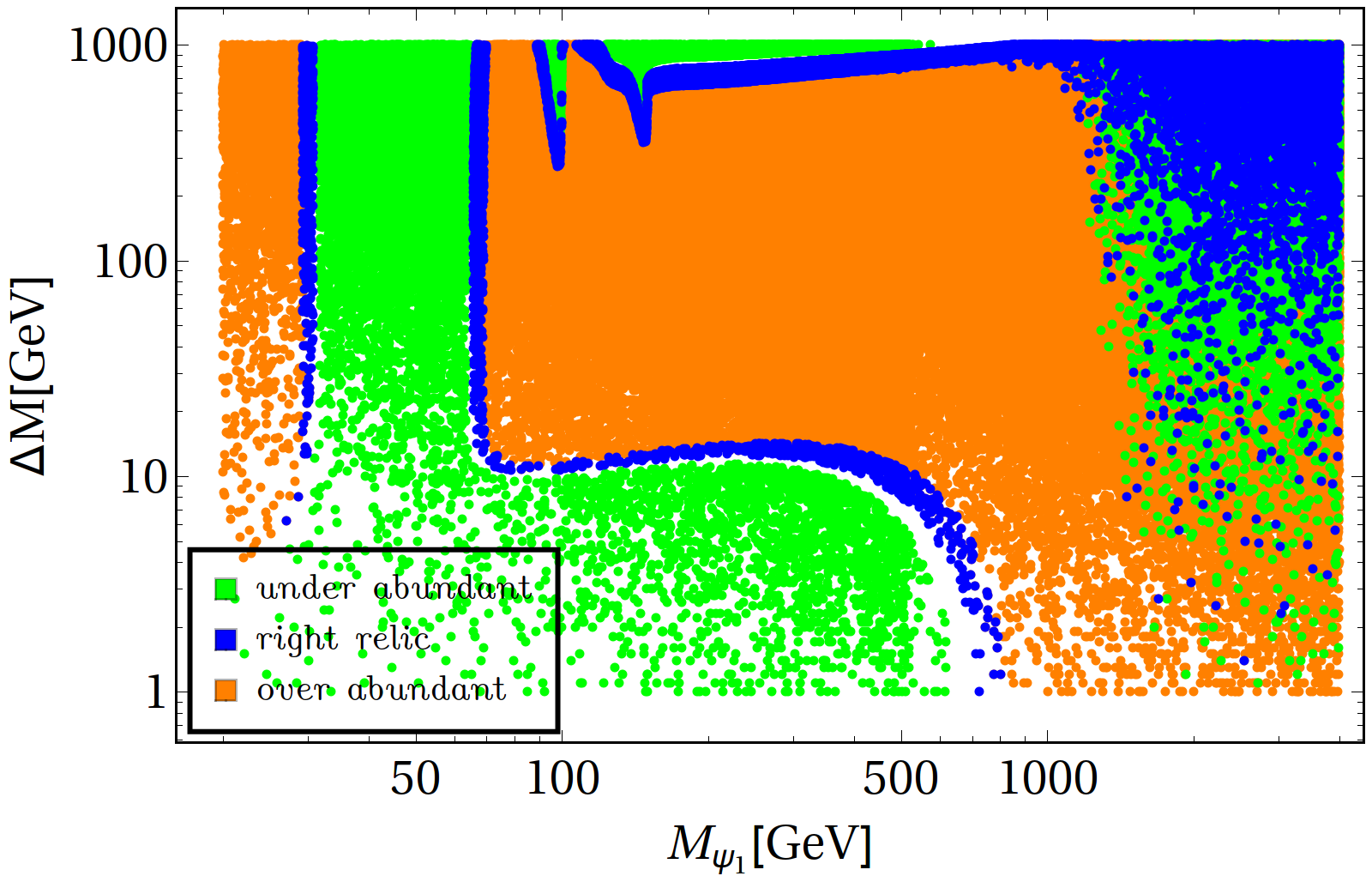} 
$$
$$
\includegraphics[scale=0.42]{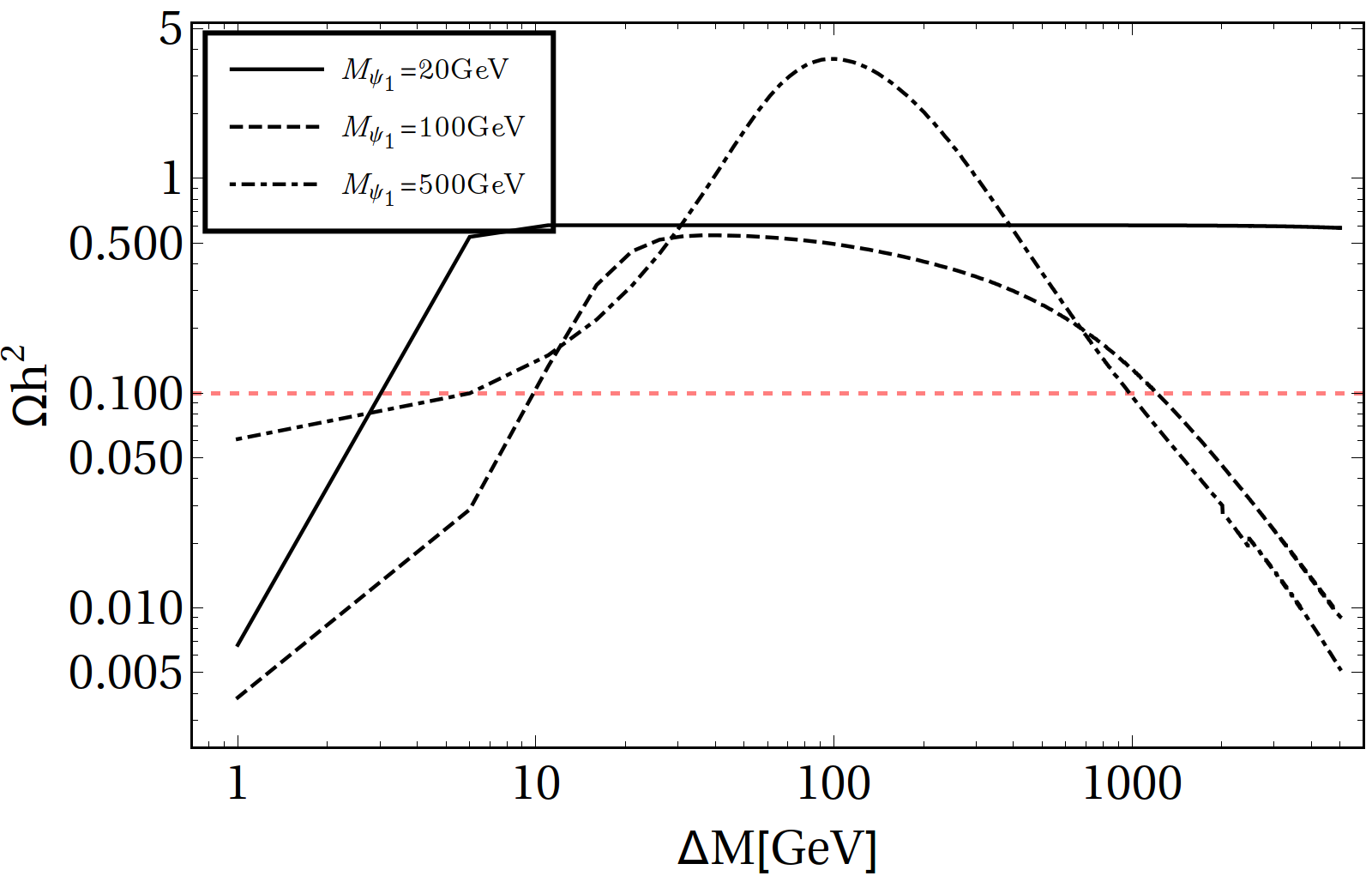}
\includegraphics[scale=0.41]{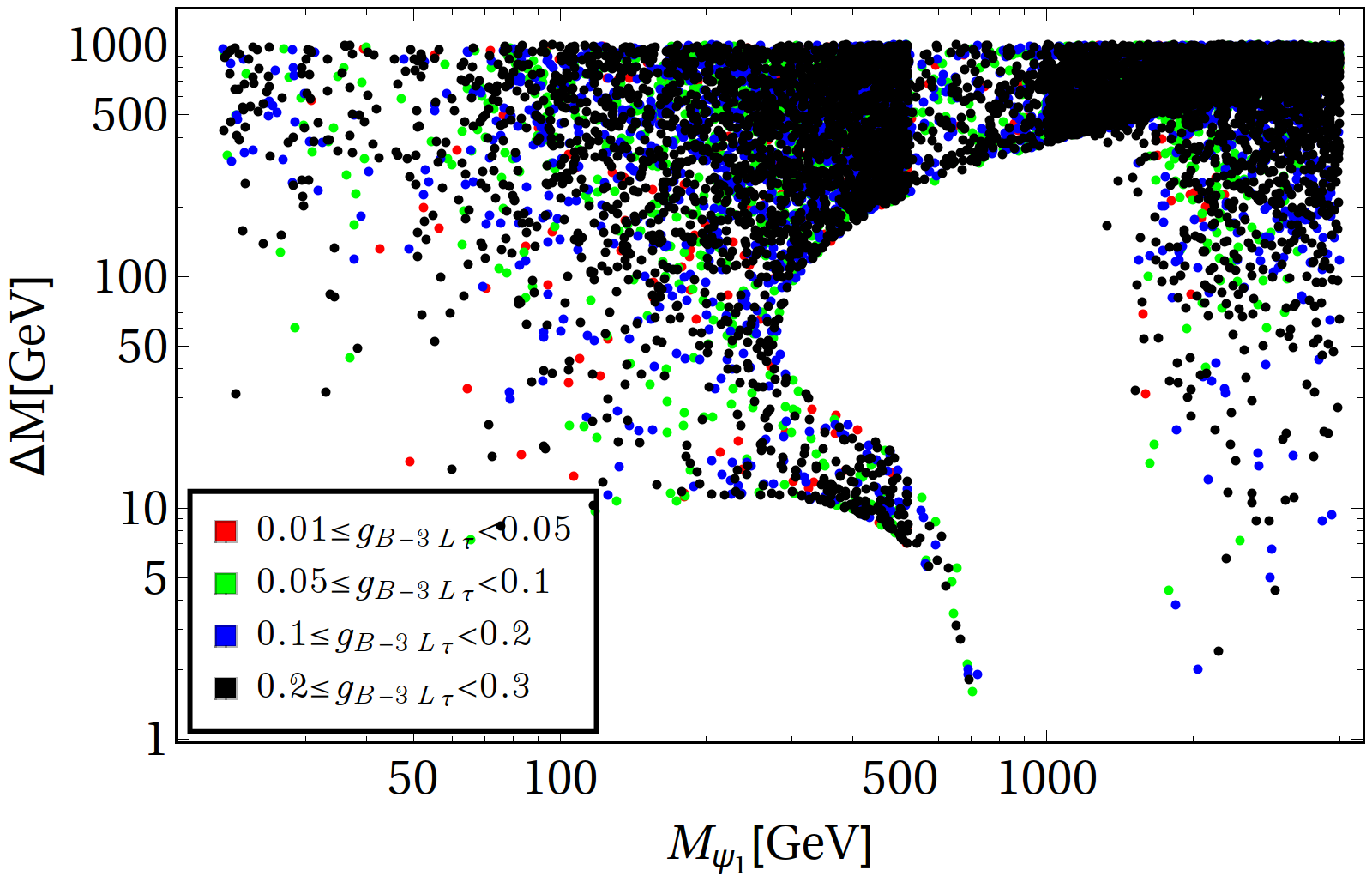} 
$$
\caption{Top Left:  Relic density allowed parameter space for DM $\psi_1$ in $M_{\psi_1}$-$\Delta M$ plane for different ranges of $\sin\theta$ depicted in the inset of figure where $0.05\leq g_{B-3L_{\tau}}\leq 0.3$. Top Right: Under abundance (green region), over abundance (orange region) and right relic density (blue) regions are shown in same $M_{\psi_1}$-$\Delta M$ plane for a fixed $\sin\theta = 0.2$ . Bottom left: Variation of relic abundance with $\Delta M$ for some choices of DM mass: 20 GeV (solid black curve), 100 GeV (dashed black curve) and 500 GeV (dot dashed black curve). For this plot $\gBL=0.2$, $\tilde v=0.32$ and $\sin\theta=0.2$. The horizontal dashed line (red coloured) corresponds to the central value of Planck limit on DM relic \cite{Aghanim:2018eyx} Bottom right: Relic density allowed parameter space in $M_{\psi_1}$-$\Delta M$ plane for different ranges of $\gBL$ depicted in the inset of figure and $\sin\theta:\{0.01-0.5\}$.}
\label{fig:relicparam}
\end{figure}

As seen from the top left plot of Fig.~\ref{fig:relicparam}, for DM mass $M_{\psi_1} \gtrsim 100$ GeV (with $ \sin \theta >0.1$, there exists two different $\Delta M$ for same $M_{\psi_1}$ which satisfies correct relic density requirement. To understand this, we plot relic density versus $\Delta M$ for different values of $M_{\psi_1}$ shown in the bottom left panel of Fig.~\ref{fig:relicparam}. Here also, the two different values of $\Delta M$, giving correct relic for same DM mass are visible for $M_{\psi_1} \gtrsim 100$ GeV.  As it is observed, for DM mass say, 100 GeV (black dashed curve) the relic abundance first rises with increase in $\Delta M$. This happens due to the fact that the co-annihilation becomes sub-dominant due to increase in $\Delta M$, which, in turn, reduces the effective annihilation cross-section. At some point right relic density is reached ($\Delta M\sim 10~\rm GeV$) as the annihilation and co-annihilation are just sufficient to produce the correct abundance. After that, the relic abundance becomes more or less constant for a small $\Delta M$ range and then again the abundance starts going downhill as the Yukawa $Y$ becomes large enough making the net annihilation cross-section larger. The cumulative result top right and bottom left panels of  Fig.~\ref{fig:relicparam} is reflected in the top left corner of the same figure for $0.01\lsim\sin\theta\lsim 0.5$. For completeness, in the bottom right corner of Fig.~\ref{fig:relicparam} we have shown the relic density allowed parameter space for different choices of the new gauge coupling $\gBL$. As we noticed earlier in Fig.~\ref{fig:relicST}, there is no strong dependence of the relic abundance on $\gBL$. This is evident from this plot as different coloured points (corresponding to different $\gBL$) are scattered within the allowed region of $\Delta M-M_{\psi_1}$ parameter space. In passing we would like to comment that the annihilation to RHN final states is suppressed because of heavy mass of the RHNs and also their contribution to total annihilation cross-section is negligible compared to the SM quarks.

\subsection{Direct detection of dark matter}
\label{sec:DD}

The presence of the $Z$ and $Z_{B-3L_{\tau}}$ mediated DM-nucleon scattering diagrams highly constrain the parameter space of singlet-doublet model by pushing $\sin\theta$ to a very small value which, in turn, forces $\Delta M$ to be small~\cite{Bhattacharya:2015qpa}. This can be avoided by exploiting the pseudo-Dirac splitting of the VLFs~\cite{Bhattacharya:2017sml,Barman:2019tuo}. As mentioned in section~\ref{sec:model}, the presence of the VLF singlet Majorana term splits the Dirac states into two pseudo-Dirac states with a mass difference between the two. From Eq.~\ref{eq:p-dirac-split} we see that due to the presence of the Majorana term (generated by the singlet scalar $\Phi$) and mixing between the singlet-doublet fermions, the physical mass eigenstate $\psi_1$ splits into two pseudo-Dirac states: $\{\psi_1^i,\psi_1^j\}$. In such a scenario, interaction of the DM with $Z$ ($Z_{B-3L_{\tau}}$) can be written as~\cite{Bhattacharya:2017sml,Barman:2019tuo}:

\bea
\mathcal{L}\supset \bar{\psi_1^{i}}i\slashed{\partial}\psi_1^{i}+\bar{\psi_1^{j}}i\slashed{\partial}\psi_1^{j}+g_z \bar{\psi_1^{i}}\gamma_{\mu}\psi_1^{j}\mathcal{Z}^{\mu}
\label{eq:zinteraction},
\eea

where $\mathcal{Z}\in \{Z,Z_{B-3L_{\tau}}\}$ and $g_z=\frac{g_L\sin^2\theta}{2 c_W}$ for SM $Z$ and $g_z=\frac{3}{4}g_{B-3L_{\tau}}\sin^2\theta$ for $Z_{B-3L_{\tau}}$ mediation, $c_W\equiv\cos\theta_W$ stands for cosine of the Weinberg angle. Note that, the $\mathcal{Z}$-mediated interaction term is off-diagonal unlike the kinetic terms due to the pseudo-Dirac nature of the VLFs. This results in an inelastic scattering of the DM in which the DM is scattered to an excited state via $\mathcal{Z}$ mediation. As pointed out in~\cite{TuckerSmith:2001hy}, such an inelastic scattering can occur only if the splitting between the two pseudo-Dirac states $\psi_1^{i}$ and $\psi_1^j$ satisfies:

\bea
\delta_{max} < \frac{\beta^2}{2}\frac{M_{\psi_1} M_N}{M_{\psi_1}+M_N}.
\label{eq:deltmax}
\eea

\begin{figure}[htb!]
$$
\includegraphics[scale=0.42]{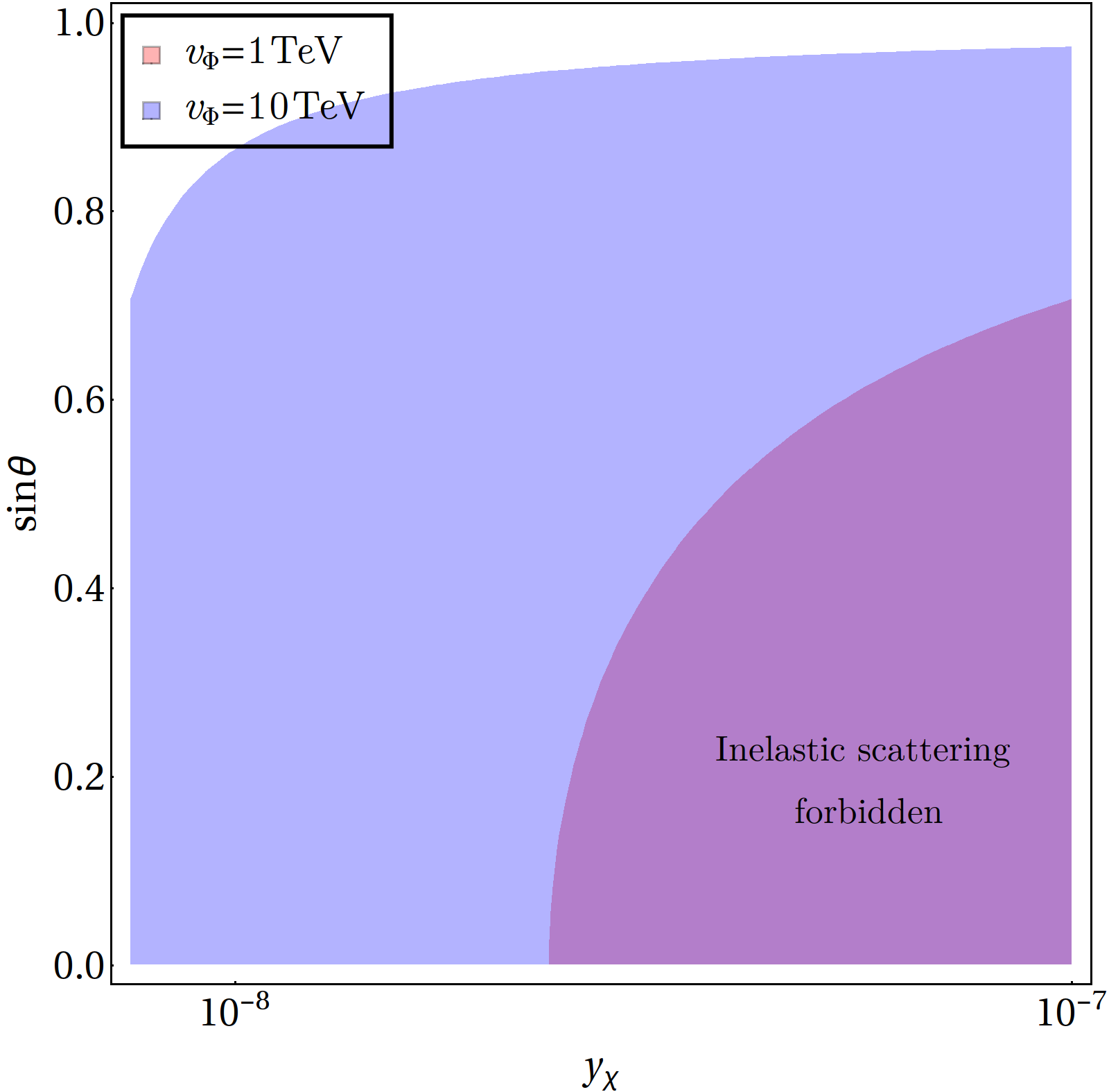}
$$
\caption{Choice of Yukawa $y_{\chi}$ and VLF mixing $\sin\theta$ (via Eq.~\eqref{eq:deltmchi}) in order to forbid the inelastic scattering via heavy neutral gauge bosons. The purple and pale blue regions correspond to $v_{\Phi}:\{1,10\}~\rm TeV$ respectively. Each colored region is where inelastic scattering gets disallowed.}
\label{fig:delmcntr}
\end{figure}

As computed in~\cite{Barman:2019tuo}, $\delta\sim 100~\rm keV$ can forbid the inelastic scattering mediated by $Z(Z_{B-3L_{\tau}})$ for a DM mass $\sim\mathcal{O}(100~\rm GeV)$ with $\beta\lsim 220~\rm km/s$. This is to be noted here, the splitting between the two pseudo-Dirac states is so small ($\sim 100~\rm keV$) that it can be ignored in determining the relic abundance of the DM, but has to be taken into account for computing the direct detection cross-section (as emphasised earlier). The mass splitting between $\psi_1^i$ and $\psi_1^j$ in terms of our model parameter is given by:

\bea
\delta m_{\chi}=y_{\chi}\cos^2\theta v_{\Phi}.
\label{eq:deltmchi}
\eea

From Eq.~\eqref{eq:deltmchi} one can put a bound on the VLF mixing and the Yukawa $y_{\chi}$ for $\delta m\sim 100~\rm keV$ such that the heavy neutral gauge boson mediated diagrams are switched off. This is depicted in Fig.~\ref{fig:delmcntr}. As we can see, in order to forbid such inelastic scattering one can choose $y_{\chi}\sim \mathcal{O}(10^{-8})$, then for all small mixing the inelastic scattering can be forbidden. Now, such a choice of scalar VEVs and small Yukawa is not in conflict with the neutrino mass generation. Again, in order to satisfy the direct detection bound, we need to confine $\sin\theta\lsim 0.5$ which is safe even if we consider the conservative bound from Fig.~\ref{fig:delmcntr} corresponding to $v_{\Phi}=1~\rm TeV$, which anyway we require to keep all the masses within the reach of the ongoing collider experiment. Therefore, for all practical purposes, we can safely ignore the scattering via heavy neutral gauge bosons in order to evade the stringent direct detection exclusion limit.

The direct detection of the VLF DM in this case, therefore, takes place dominantly via the elastic scattering mediated by the scalars ($h_{1,2,3}$). This is depicted in Fig.~\ref{fig:ddvlf}. The spin-independent (SI) direct detection cross section per nucleon is given by~\cite{Duerr:2015aka}:

\bea
\sigma^{SI} = \frac{1}{\pi A^2} \mu^2 \left|\mathcal{M}\right|^2,
\label{eq:sidd}
\eea

\begin{figure}[htb!]
$$
\includegraphics[scale=0.35]{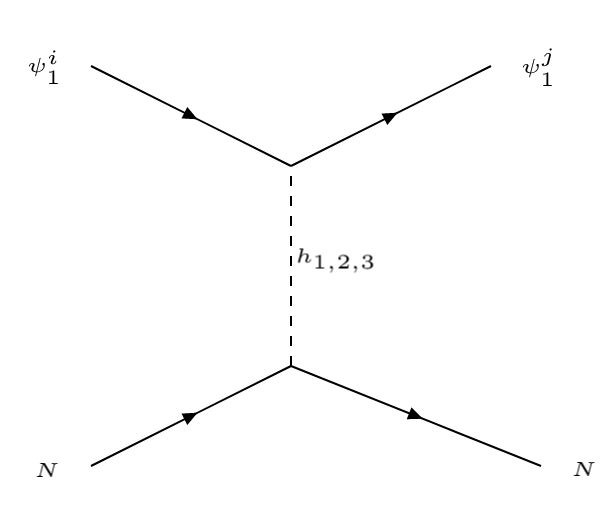}
$$
\caption{Feynman graph showing the elastic spin independent direct detection scattering for DM $\psi_1$  and nucleus via scalars $h_{1,2,3}$. Note that, in this case $i=j$, while $\mathcal{Z}:~\{Z,~Z_{B-3L_\tau}\}$ mediated diagrams are forbidden due to inelastic stattering as $i\neq j$.}
\label{fig:ddvlf}
\end{figure}

where $A$ is the mass number of the target nucleus, $\mu=\frac{M_{\psi_1}M_N}{M{\psi_1}+M_N}$ is the DM-nucleus reduced mass and $\left|\mathcal{M}\right|$ is the DM-nucleus amplitude, which reads:

\bea
\mathcal{M} = \sum_{i=1,2}\left[Z f_p^i+\left(A-Z\right)f_n^i\right].
\label{eq:amp}
\eea

The effective couplings (with form factors~\cite{Durr:2015dna}) in Eq.~\eqref{eq:amp} are:

\bea
f_{p,n}^i = \sum_{q=u,d,s} f_{T_q}^{p,n} \alpha_q^{i} \frac{m_{p,n}}{m_q}+\frac{2}{27} f_{T_G}^{p,n}\sum_{Q=c,t,b} \alpha_Q^{i}\frac{m_{p,n}}{m_Q},
\label{eq:coupling}
\eea

with

\bea
\alpha_q^{1} = -\frac{m_q}{v_d} Y \sin\theta\cos\theta\left(\frac{c_{12}^2 c_{13}^2}{m_{h_1}^2}\right)\\
\alpha_q^{2} = -\frac{m_q}{v_d} Y \sin\theta\cos\theta\left(\frac{s_{12}^2 c_{13}^2}{m_{h_2}^2}\right)\\
\alpha_q^{3} = -\frac{m_q}{v_d} Y \sin\theta\cos\theta\left(\frac{s_{13}^2}{m_{h_3}^2}\right),
\label{eq: alpha}
\eea

\begin{figure}[htb!]
$$
\includegraphics[scale=0.42]{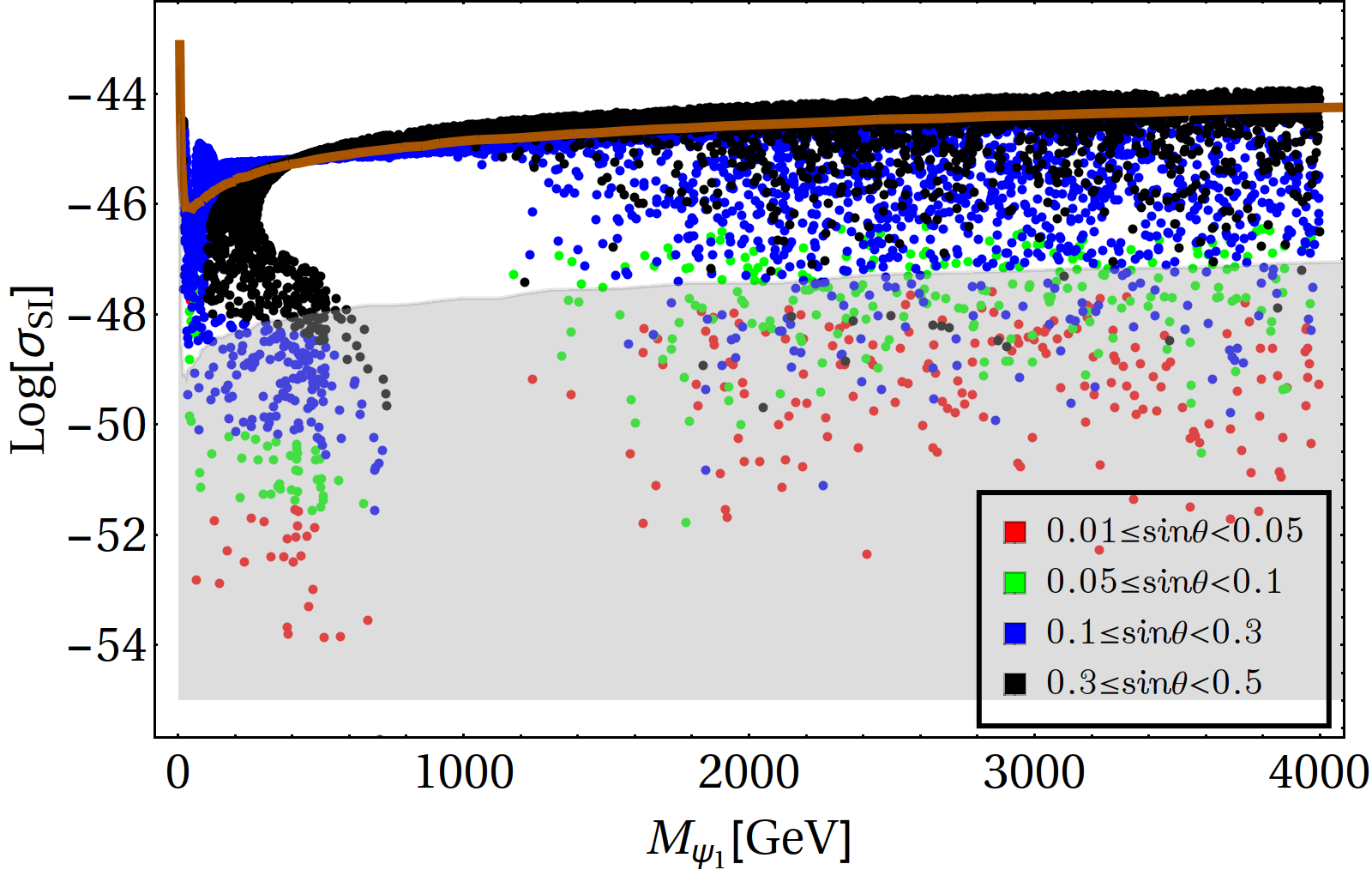} \includegraphics[scale=0.42]{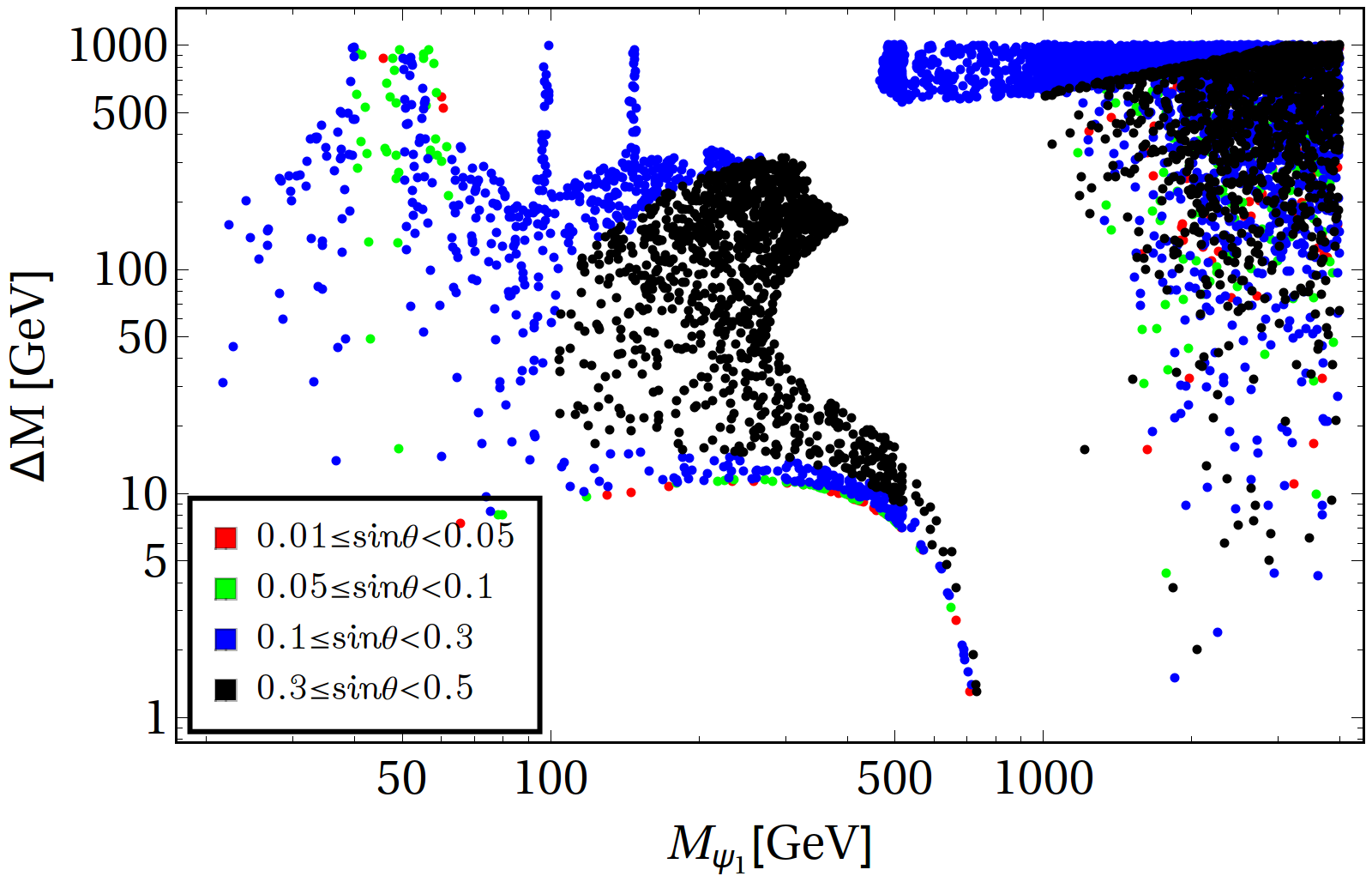}
$$
\caption{Left: Relic density allowed parameter space in $M_{\psi_1}$-$\sigma^{SI}$ plane for different choices of $\sin\theta$ (the colour codes are same as Fig.~\ref{fig:relicparam}). The present bound from XENON1T is shown by the orange thick dashed curve, while the grey region below corresponds to the neutrino floor: neutrino-nucleon coherent elastic scattering. Right: Parameter space available in $M_{\psi_1}$-$\Delta M$ plane after imposing bounds from both relic abundance and direct detection (colour codes are same as before).}
\label{fig:ddparam}
\end{figure}
where $c_{12}=\cos\theta_{12},c_{13}=\cos\theta_{12}$ and $s_{13}=\sin\theta_{13}$ are the scalar mixing angles, defined earlier. The parameter space satisfying right DM relic abundance in comparison to the present bound from direct search experiment is shown in the LHS Fig.~\ref{fig:ddparam} for different choices of VLF mixing $\sin\theta$ and gauge coupling $g_{B-3L_{\tau}}$. We have also shown how much of the parameter space is under the infamous neutrino floor~\cite{Billard:2013qya} where it is extremely difficult or even impossible to distinguish DM signal from the SM neutrino background (light grey region). In the LHS of Fig.~\ref{fig:ddparam} we see near the $Z$ and Higgs resonance, small and moderate $\sin\theta$'s are allowed by direct search ($0.01\lsim\sin\theta\lsim 0.3$). For $M_{\psi_1}\gsim 100~\rm GeV$ larger $\sin\theta$'s are also allowed as the direct search cross-section has a suppression from heavy scalars: $\sigma_{SI}\sim\frac{\mu^2\sin^2\theta}{m_{h_i}^4}$. Small $\sin\theta\lsim 0.1$ are always allowed by direct search because they produce smaller scattering cross-section, but they are mostly devoured by the neutrino floor as shown by the grey region in Fig.~\ref{fig:ddparam}. On the RHS of Fig.~\ref{fig:ddparam} we see the relic density allowed parameter space that also satisfies direct search bound. In the low DM mass region, specifically near $Z$ and Higgs resonances we can achieve large $\Delta M$ for moderate $\sin\theta$. But if $\Delta M$ becomes too large $\gsim 500~\rm GeV$ then one has to resort to small $\sin\theta$ to tame down the Yukawa $Y$ in order to satisfy both relic abundance and direct search limits. For larger DM mass large $\Delta M$ is still allowed near the non-standard scalar resonances $\sim 100~\rm GeV$ and $\sim 150~\rm GeV$. Beyond $\sim 150~\rm GeV$ large $\Delta M$ is achieved with larger $\sin\theta$, while the DM remains still allowed by direct search due to suppression from heavy scalars mentioned earlier. With DM mass $\gsim 400~\rm GeV$ the points move towards smaller $\Delta M$ in order to reach right relic exploiting co-annihilation as we have seen earlier in Fig.~\ref{fig:relicST}.  Beyond 1 TeV DM mass, the direct detection bound becomes weak as the DM mass is large, while because of $Z_{B-3L_\tau}$ resonance there is a huge parameter space that satisfy relic abundance. As a result, for $M_{\psi_1}\gsim 1~\rm TeV$, almost all of the parameter space is allowed from direct detection for all possible $\Delta M$. We would like to remind here once more that this is the novel feature of the pseudo-Dirac states that this model offers, which helps to achieve larger $\Delta M$ without constraining $\sin\theta$ to a great extent. Larger $\sin\theta$ is required to distinguish this model at the colliders as we shall elaborate in section~\ref{sec:collider}.


\section{High scale stability and perturbativity}
\label{sec:highscale}

In this section we will discuss
discuss the high scale feature of the model. To be specific, here we will constrain the relic density, direct detection  satisfied points by applying perturbativity/unitarity and vacuum stability 
bounds till some high energy scale. For this purpose we need to consider the RG running of associated couplings through $\beta$ functions. We have used PyR@Te 2.0.0~\cite{Lyonnet:2016xiz} to extract the $\beta$ functions corresponding to the gauge couplings, relevant scalar and fermionic couplings present in the model which are listed in Appendix~\ref{ap2}. For simplicity, we show only the one loop $\beta$ functions for both SM and BSM parameters in Appendix~\ref{ap2} while in our numerical calculations, we consider the three loop beta functions for SM particles due to better precision of SM parameters.

The non violation of perturbativity/unitarity conditions (Eq.~\eqref{eq:PerC} and Eq.~\eqref{eq:UniC}) for various couplings can be assured by analysing their runnings using the $\beta$ functions. In our analysis, some of the Yukawa like couplings ($y_\chi,~y_{\alpha_i},~y_{13},~ y_{23}$) are assumed to be very small. Hence they have negligible influence in the RG running of themselves and other parameters. In addition, we also fix the VEV of the singlet scalar fields within TeV range and scalar mixing angles $\lesssim 0.2$. These in turn fix the magnitude of the scalar couplings which are positive and stays below $\sim 0.1$. With these order of magnitude initial values, they are not expected to break the perturbativity conditions at high energy scale. The other two important parameters we consider are $Y$ and $g_{B-3L_{\tau}}$ which play vital role in DM phenomenology as well as in collider analysis. Largeness of these two parameters
could destroy the high scale perturbativity/unitarity of the theory. Therefore we will focus on $Y-g_{B-3L_{\tau}}$ plane and see the
the bounds coming from the requirement of satisfying perturbativity/unitarity criteria. While focusing on this particular plane of our interest, we also make sure that none of the other parameters violate the above mentioned criteria. First, in left panel of Fig. \ref{fig:Per1} we show the points (blue colored) in $Y-g_{B-3L_{\tau}}$ plane which satisfy relic, direct detection bounds. Then we constrain the same plane using the perturbativity criteria till $\mu=M_P$ (orange colored points), where $M_P$ is the Planck scale. It is clear that applying the perturbativity criteria significantly cuts the earlier parameter space with $Y\gtrsim 0.8$ and $g_{B-3L_{\tau}}\gtrsim 0.25$. Similar exercise has been done in $\Delta M-g_{B-3L_{\tau}}$ plane which is shown in right panel of Fig. \ref{fig:Per1}. 
 \begin{figure}[h]
$$
\includegraphics[scale=0.32]{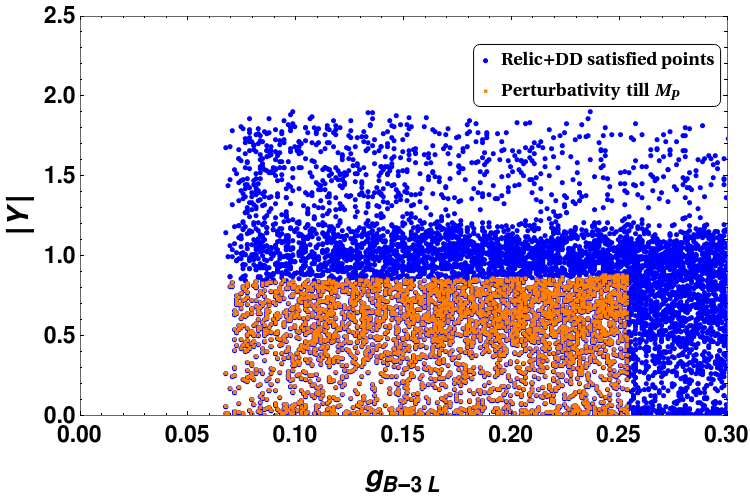} 
\includegraphics[scale=0.32]{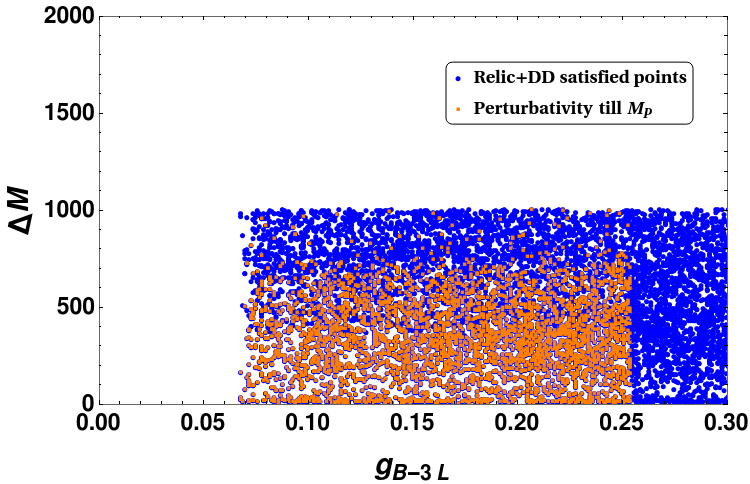} 
$$
\caption{ Parameter space satisfying DM relic and direct detection bounds (blue points) and perturbativity/unitarity till $M_P$ (orange points) are shown in (left) $Y-g_{B-3L_{\tau}}$ and in (right) $\Delta M-g_{B-3L_{\tau}}$ plane.}
\label{fig:Per1}
\end{figure}

Now the conditions of the stability or the boundedness of the scalar potential in the model till some high energy scale along various field directions are provided in Eq.~\eqref{eq:StabC}. Recall that the EW vacuum stability (stability of Higgs potential) is dictated by the condition $\lambda_H>0$. However for more accurate analysis, one should consider the radiatively improved Higgs potential where the one loop correction will be provided by SM fields and other BSM fields. The radiatively corrected one loop effective Higgs potential (at high energies $h\gg v_d$) can be written as \cite{Casas:1994qy,Casas:1996aq},
\begin{align}
 V_h^{\textrm{eff}}=\frac{\lambda_H^{\textrm{eff}}(\mu)}{4}h^4,
\end{align}
with $\lambda_H^{\rm eff}=\lambda_H^{\rm SM, eff}+\lambda_H^{(S,\Phi),~ \textrm{eff}}+\lambda_H^{(\psi,\chi), \textrm{ eff}}$ where $\lambda_H^{\rm SM, eff}$ is the SM
contribution to $\lambda_H$. The other two contributions 
$\lambda_H^{(S,\Phi),~ \textrm{eff}}$ and $\lambda_H^{(\psi,\chi), \textrm{ eff}}$ are due to the newly added fields
in the present model as provided below \cite{Khan:2012zw,Gonderinger:2012rd}. 
\begin{align}
 &\lambda_H^{(S,\Phi), \textrm{eff}}(\mu)=e^{4\Gamma(h=\mu)}\Big[\frac{\lambda_1^2}{32\pi^2}\Big(\textrm{ln}\frac{\lambda_1}{2}-\frac{3}{2}\Big)+\frac{\lambda_2^2}{32\pi^2}\Big(\textrm{ln}\frac{\lambda_2}{2}-\frac{3}{2}\Big)\Big],\\
&\lambda_H^{(\psi,\chi),\textrm{eff}}(\mu)= -e^{4\Gamma(h=\mu)}\Big[\frac{Y^2}{16\pi^2}\Big(\textrm{ln}\frac{Y^2}{2}-\frac{3}{2}\Big)\Big],
\end{align}
where $\Gamma(h)=\int_{m_t}^h\gamma(\mu)d \textrm{ln}\mu$ and $\gamma(\mu)$ is the anomalous dimension of the Higgs field \cite{Buttazzo:2013uya}. Note that we have ignored
the radiative corrections involving $y_\chi,y_{\alpha_i}, y_{13}, y_{23}$ as they are 
fixed to negligibly small values in our analysis. Now with the inclusion of radiative correction to Higgs potential, the stability condition of Higgs vacuum will be modified as $\lambda_H^{\rm eff}>0$. The remaining co-positivity conditions in Eq. (\ref{eq:StabC}) will determine the boundedness of the scalar potential in different field directions.

We numerically solve the three loop RG equations for all the SM couplings and one loop RG equations
for the other relevant BSM couplings in the model from $\mu = m_t$ to $M_P$ energy scales considering $m_t=173.1$ GeV \cite{Tanabashi:2018oca}, SM Higgs mass $m_H=125.09$ GeV \cite{Tanabashi:2018oca} and strong coupling constant $\alpha_s= 0.1184$. We also use the initial boundary values of all the SM couplings as provided in~\cite{Buttazzo:2013uya}. The boundary values have been determined at $\mu = m_t$ in~\cite{Buttazzo:2013uya} by taking various threshold corrections and mismatch between top pole mass and $\overline{\rm MS}$ renormalised couplings into account. One important point is to note that during the running of couplings, we will ignore the small mass differences between the masses of heavy BSM Higgs bosons and DM particles for the sake of simplicity. The $\beta$ function of $\lambda_H$ includes positive contributions from the scalar couplings and negative contributions from fermionic couplings. Therefore, with $y_t\sim \mathcal{O}(1)$ in SM, large value of $Y$ could destabilise the EW vacuum. The initial value of $\lambda_H$ also gets a positive shift due to the presence of additional scalars in the set up as evident from Eq.~\eqref{eq:lambdas}. The amount of shift depends on the masses of the heavier Higgs bosons and also the corresponding mixing angles. With our choices for them as specified earlier the magnitude of the shift comes out to be $\sim 0.02$. Note that we have also considered all the other scalar couplings positive and $\sim \mathcal{O}(0.1)$ in our analysis. Hence considering small order of magnitude of Yukawa like couplings ($y_{\alpha_i}, ~y_{13},~y_{23},~ y_\chi$), the BSM scalar couplings are expected to remain positive in their evolution, thus automatically guaranteeing the stability of the total scalar potential in the corresponding field directions (when $\lambda_H>0$).

 \begin{figure}[h]
$$
\includegraphics[height=6cm,width=8cm]{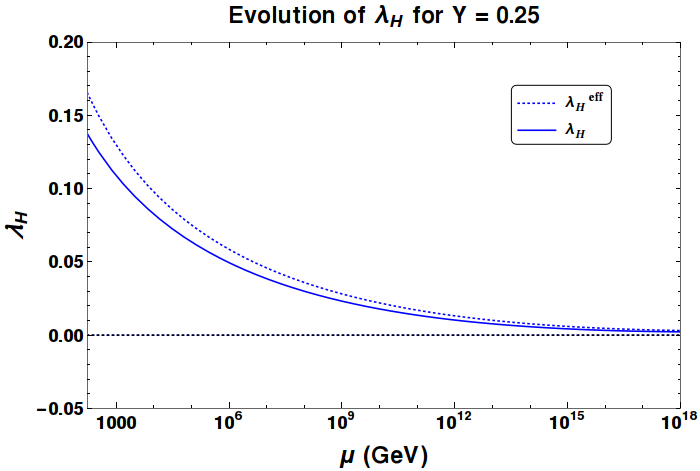} 
\includegraphics[height=6cm,width=8cm]{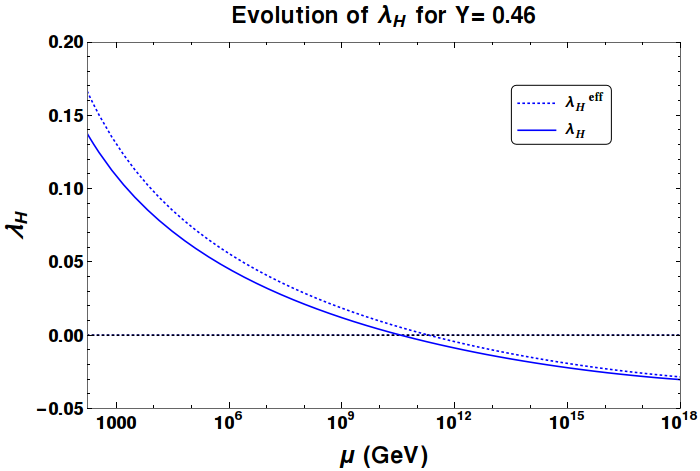} 
$$
\caption{Left: Running of $\lambda_H(\lambda_H^{\rm  eff})$ for two DM relic and direct detection bound satisfying points with (left) $Y\sim 0.25$ and (right) $Y\sim$ 0.46.}
\label{fig:RunningLH}
\end{figure}

Now we further constrain the $Y-g_{B-L}$ parameter space which is allowed from perturbativity criteria (Fig. \ref{fig:Per1}) along with correct DM related observables using vacuum stability conditions. Before that in Fig. \ref{fig:RunningLH}, we show running of 
$\lambda_H(\lambda_H^{\textrm{eff}})$ for two different DM relic + direct detection + perturbativity bounds satisfying points having $Y\sim$ 0.25 and 0.46 respectively. As it can be seen, for lower value of $Y$
$\lambda_H (\lambda_H^\textrm{eff})$ remains positive throughout its running
till $M_P$ energy scale thereby establishing the stability of EW vacuum.
On the other hand for $Y\sim 0.46$, $\lambda_H (\lambda_H^\textrm{eff})$ crosses zero around $\mu\sim 10^{15}$ GeV and ends with negative value at $\mu=M_P$. Hence it is clear that large values of $Y$ are disfavoured
in our analysis as it could destabilise the Higgs vacuum. The plots in Fig. \ref{fig:Stab1} also shows that the running of $\lambda_H$ and $\lambda_H^\textrm{eff}$ are similar and they almost merge near the energy scale $\mu=M_P$. Finally in Fig. \ref{fig:Stab1}, we constrain left panels of Fig. \ref{fig:Per1}, using both perturbativity and the vacuum stability criteria in both $Y-g_{B-L}$ and $\Delta M-g_{B-3L_{\tau}}$ planes. Now when we compare Fig. \ref{fig:Per1} with Fig. \ref{fig:Stab1}, it clearly shows that the upper limit on $Y$ is significantly reduced from 0.8 to 0.3 due to the application of vacuum stability criteria till
energy scale $M_P$. However upper limit on $g_{B-3L_{\tau}}$ remains more or less unaltered ($\lesssim 0.25$) as it does not have direct role in stability analysis. Similar conclusion can be drawn for $\Delta M$ also. As before, all the points in these plots satisfy DM related bounds.

\begin{figure}[htb!]
$$
\includegraphics[height=6cm,width=8cm]{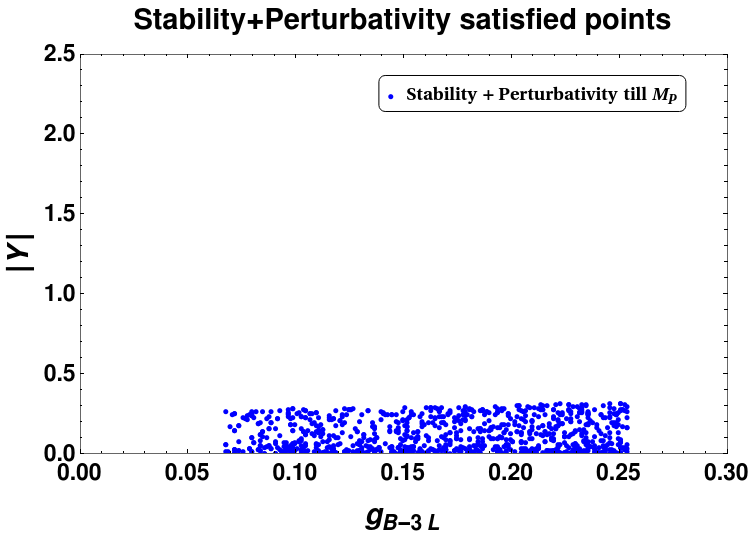}
\includegraphics[height=6cm,width=8cm]{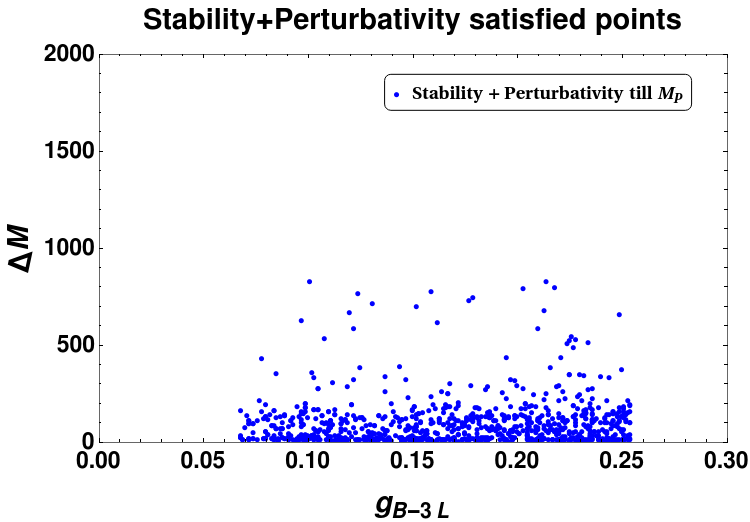}
$$
\caption{Parameter space satisfying DM relic abundance, direct detection, perturbativity and vacuum stability (till $M_P$) bounds in (left) $Y-g_{B-3L_{\tau}}$ plane and (right) $\Delta M-g_{B-3L_{\tau}}$ plane.}
\label{fig:Stab1}
\end{figure}

Before we move on further, let us first choose a few benchmark points (BPs) which we shall be using for the collider study. Note that, all these BPs need to satisfy correct relic abundance, direct detection bound, vacuum stability and perturbativity constraints and on top of that should give rise to $Z_{B-3L_{\tau}}$ mass in correct range. These are enlisted in table~\ref{tab:bp}. We also include the values of relevant EW precision parameters in table~\ref{tab:bp} for all the benchmark points which show they fall within correct experimental range. Another point is to note that these BPs are selected in the decreasing order to $\gBL$ from top to
\begin{table}[htb!]
\begin{center}
\begin{tabular}{|c|c|c|c|c|c|c|c|c|c|c|c|}
\hline
Benchmark & $\tilde v$ & $\gBL$ &$\sin\theta$ & $\Delta M$ & $M_{\psi_1}$ & $|Y|$ & $10^3\hat S$ & $\Delta T$ & $\sigma_{DD}$ & $\Omega h^2$  & $m_{Z_{B-3L_{\tau}}}$ \\ [0.5ex] 
Point     &  ($v_\mathcal{S}/v_{\Phi}$) &    &   & (GeV)   &  (GeV)     &     &  & & $(cm^2)$  &  & (TeV) \\ [0.5ex] 
\hline\hline

BP1 & 1.75 & 0.20 & 0.28 & 194.9 & 128.5 & 0.301 & 0.43 & 0.005 & $10^{-45.96}$ & 0.122 & 2.57 \\
\hline
BP2 & 2.33 & 0.16 & 0.11 & 517.4 & 55.4 & 0.311 & 0.06 & 0.004 & $10^{-46.71}$ & 0.119 & 2.73 \\
\hline
BP3 & 3.50 & 0.11 & 0.46 & 130.1 & 300.5 & 0.305 & 1.30 & 0.004 & $10^{-45.59}$ & 0.121 & 2.79 \\
\hline
BP4 & 3.63 & 0.10 & 0.14 & 274.0 & 36.5 & 0.218 & 0.03 & 0.003 & $10^{-46.78}$ & 0.120 & 2.61 \\
\hline
BP5 & 5.49 & 0.07 & 0.42 & 111.4 & 245.5 & 0.248 & 1.23 & 0.003 & $10^{-45.81}$ & 0.119 & 2.68 \\
\hline\hline
\end{tabular}
\end{center}
\caption {Choices of the benchmark points used for collider analysis. Masses, mixings, relic density and direct search cross-sections for the DM candidate are tabulated. In each case corresponding mass of $Z_{B-3L_{\tau}}$ is also quoted.} 
\label{tab:bp}
\end{table}
bottom where BP1 has highest $\gBL$ and BP5 has the smallest $\gBL$. As we shall see in section~\ref{sec:collider}, the production cross-section of $\psi^{\pm}$ will be large for small $\Delta M$ and not for large $\gBL$. This is due to the fact that larger $\gBL$ results in heavier $m_{Z_{B-3L_{\tau}}}$ (for fixed VEV), which, in turn causes propagator suppression for $\psi^{\pm}$ production (via $Z_{B-3L_{\tau}}$) leading to decrease in cross-section. However, even if $\Delta M$ is small, but $M_{\psi_1}$ is large, the production cross-section may still be small. All the BPs satisfy the invisible SM Higgs and SM $Z$ decay constraint as shown in the Appendix.~\ref{sec:invdecay}. Finally we would like to highlight that LEP has set a lower limit on pair-produced charged heavy vector-like leptons: $m_L>101.2~\rm GeV$ at 95 \%\ C.L. for $L^{\pm}=\nu W$ final states~\cite{Achard:2001qw}. Thus all our benchmark points are safe from LEP bounds. 



\section{Collider phenomenology}
\label{sec:collider}

\begin{figure}[htb!]
$$
\includegraphics[scale=0.35]{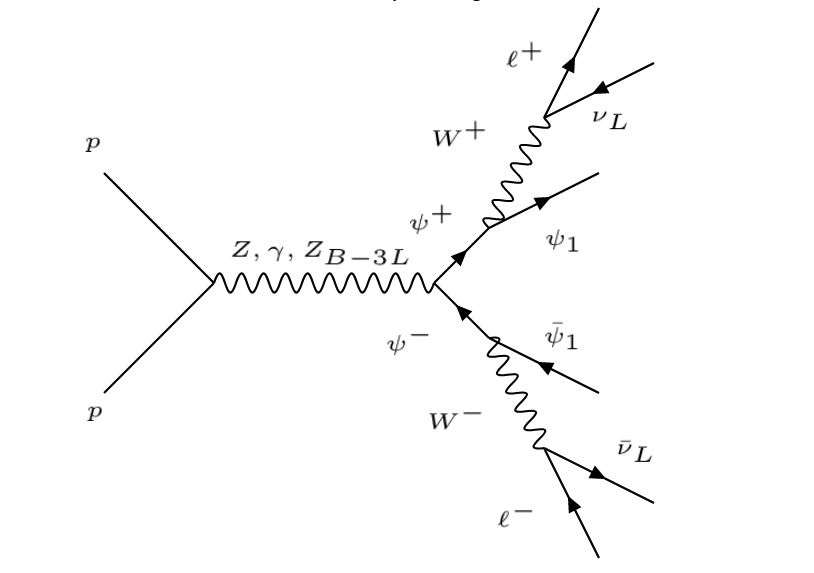} 
$$
\caption{Pair production of charged VLFs and their subsequent decay to OSD+$\slashed{E_T}$ final state.}
\label{fig:vlfdecay}
\end{figure}

The detailed study of collider signature for vector like fermions can be found in~\cite{Bahrami:2016has,Barman:2019tuo}. As we have already seen, due to the pseudo-Dirac nature of the VLFs large $\Delta M$ can be achieved satisfying both relic abundance and direct search. Such large $\Delta M$'s are actually beneficial in order to distinguish this model at the collider from the SM background~\cite{Barman:2019tuo}. It is to be noted that the charged component of $SU(2)_L$ doublet VLF can be produced at the LHC via SM $Z$, $Z_{B-3L_{\tau}}$ and photon mediation. The charged VLF can further decay via on-shell and/or off-shell $W$ (depending on whether $\Delta M\gsim80~\rm GeV$ or $\Delta M\lsim 80~\rm GeV$) to the following final states:

\begin{itemize}
 \item Hadronically quiet opposite sign dilepton (OSD) with missing energy $\left(\ell^{+}\ell^{-}+\slashed{E_T}\right)$.
 \item Single lepton, with two jets plus missing energy $\left(\ell^{\pm}+jj+\slashed{E_T}\right)$.
 \item Four jets plus missing energy $\left(jjjj+\slashed{E_T}\right)$.
\end{itemize}

\begin{figure}[htb!]
$$
\includegraphics[scale=0.42]{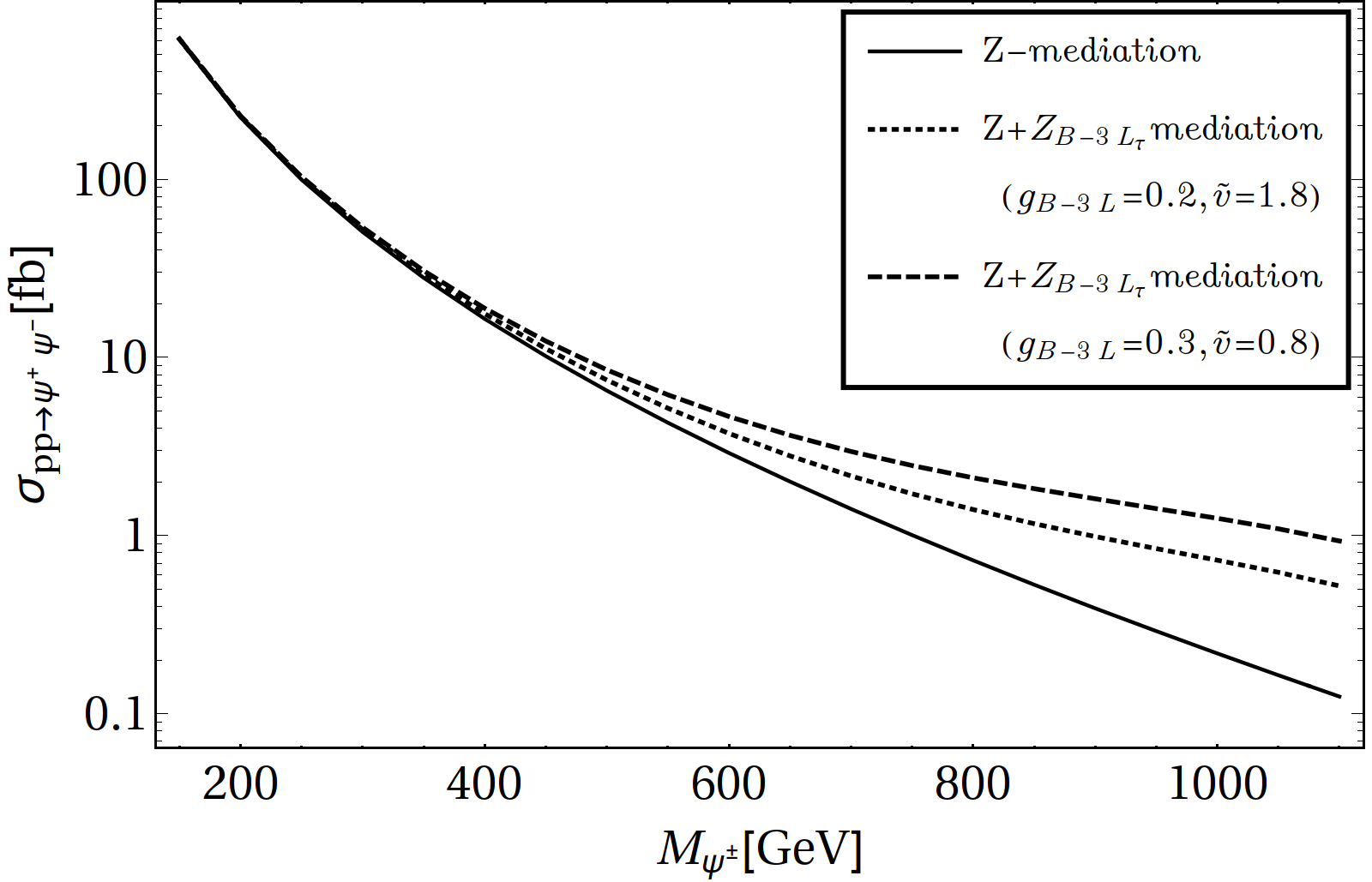} 
\includegraphics[scale=0.42]{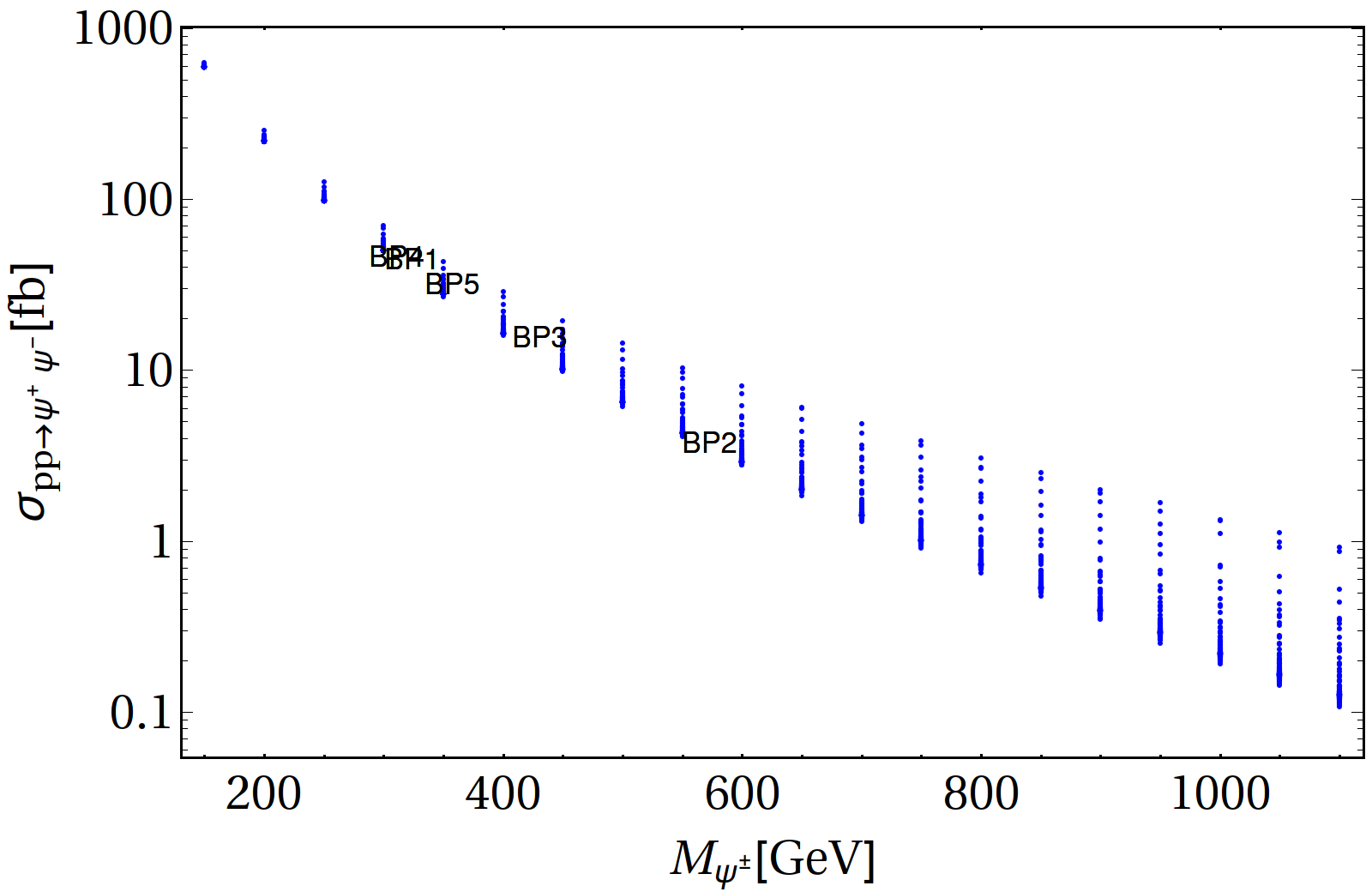} 
$$
\caption{Left: Plot showing improvement in production cross-section of the charged VLF pair due to $Z_{B-3L_{\tau}}$ mediation for two choices of $\{g_{B-3L_{\tau}},\tilde v\}:\{0.2,1.8\}\& \{0.3,0.8\}$ shown in black dashed and black dotted curves respectively. Right: Variation of $\psi^{\pm}$ pair production cross-section with $M_{\psi^{\pm}}$ for $\gBL$ $\tilde v$ varying in the range: $\{0.05,0.3\}$ and $\{0.8,5.5\}$ when both $Z$ and $Z_{B-3L_{\tau}}$ mediation are taken into account. The BPs tabulated in table~\ref{tab:bp} are also indicated in black. Note that BP1 and BP4 have almost overlapped because of similar production cross-section. All the points on the scan satisfy $m_{Z_{B-3L_{\tau}}}\gsim 2.5~\rm TeV$. For both the plots $\sqrt{s}=14~\rm TeV$ is chosen with {\tt CTEQ6l} as parton distribution function.}
\label{fig:prodctn}
\end{figure}

We shall focus only on the leptonic final states as they are much cleaner compared to others, the Feynman diagram for which is depicted in Fig.~\ref{fig:vlfdecay}. To be more specific, we shall only look into the hadronically quiet dilepton final states as we are interested to see how the presence of $Z_{B-3L_{\tau}}$ can affect the coillder signatures compared to purely $Z$ mediated scenarios studied earlier. The presence of $Z_{B-3L_{\tau}}$ significantly increases the production cross-section of the charged VLF pairs at the collider. This is due to the fact that the decay of the BSM neutral gauge boson to the charged VLFs now happens on-shell in contrast to models where this decay takes place off-shell via SM $Z$ boson and photon~\cite{Bhattacharya:2017sml,Barman:2019tuo}. Also, as there is no negative interference between the $Z$ and $Z_{B-3L_{\tau}}$ mediated charged VLF production channels, hence the addition of new channel always improves the production cross-section. In order to illustrate this improvement, we first show the variation of production cross-section $\sigma_{pp\to\psi^+ \psi^-}$ in the LHS of Fig.~\ref{fig:prodctn} for two different choices of $\{\gBL,\tilde v\}$. One noteworthy feature of this plot is that the production cross-section is lower for the choice $\{\gBL,\tilde v\}=\{0.2,1.8\}$ (black dotted curve) than for $\{\gBL,\tilde v\}=\{0.3,0.8\}$ (black dashed curve). This is simply attributed to the propagator suppression due to larger mass of $Z_{B-3L_{\tau}}$ in the former case ($m_{Z_{B-3L_{\tau}}}=3.36~\rm TeV$) over the latter ($m_{Z_{B-3L_{\tau}}}=2.54~\rm TeV$). On the RHS of Fig.~\ref{fig:prodctn} we have illustrated how the production cross-section changes for different choices of $\gBL$ and $\tilde v$ (and hence $m_{Z_{B-3L_{\tau}}}$) keeping $v_{\Phi}$ fixed at 3 TeV when both $Z$ and $Z_{B-3L_{\tau}}$ mediations are present. $\gBL$ and $\tilde v$ are chosen in such a way that $m_{Z_{B-3L_{\tau}}}$ is always above the LHC lower bound. In the same plot we have also shown our chosen BPs appearing in table~\ref{tab:bp}. Note that, the production cross-section for BP2 is the least, while it is highest for BP1 and BP4 (overlapped on each other). This tells the fact that though large $\Delta M$ is necessary in distinguishing the model at the collider (as we shall see) but at the same time we need to compromise with the production cross-section. Again, the production cross section for $M_{\psi^{\pm}}\gsim 800~\rm GeV$ is either very small (kinematically) or discarded by stability and perturbativity bound on $Y$ as larger $M_{\psi^{\pm}}$ requires larger $\Delta M$, which in turn makes $Y$ large and that is constrained from Fig.~\ref{fig:Stab1}. Although $Y$ can be tamed down by choosing a small $\sin\theta$ as per Eq.~\eqref{eq:vlfyuk} but the production cross-section will still remain small. Therefore, we have overlooked all such benchmarks. In both the plots the production cross-section decrease with the increase in charged VLF mass showing the unitarity of the cross-section remains valid.

\subsection{Object reconstruction and simulation details}
\label{sec:simul}

As already mentioned, we implemented this model in {\tt LanHEP} and the parton level events are generated in {\tt CalcHEP}~\cite{Belyaev:2012qa}. Those events are then passed through {\tt PYTHIA}~\cite{Sjostrand:2006za} for showering and hadronisation. All the SM backgrounds that can mimic our final state are generated in {\tt MADGRAPH}~\cite{Alwall:2011uj} and the corresponding production cross-sections are multiplied with appropriate $K$-factor~\cite{Alwall:2011uj} in order to match with the next to leading order (NLO) cross-sections. For all cases we have used {\tt CTEQ6l} as the parton distribution function (PDF)~\cite{Placakyte:2011az}. Now, in order to re-create the collider environment, all the leptons, jets and unclustered objects have been reconstructed using the following set of criteria:

\begin{itemize}
 \item {\it Lepton ($l=e,\mu$):} Leptons are identified with a minimum transverse momentum $p_T>20$ GeV and pseudorapidity $|\eta|<2.5$. Two leptons can be distinguished separately if their mutual distance 
 in the $\eta-\phi$ plane is $\Delta R=\sqrt{\left(\Delta\eta\right)^2+\left(\Delta\phi\right)^2}\ge 0.2$, while the separation between a lepton and a jet needs to be $\Delta R\ge 0.4$.
 
 \item {\it Jets ($j$):} All the partons within $\Delta R=0.4$ from the jet initiator cell are included to form the jets using the cone jet algorithm {\tt PYCELL} built in {\tt PYTHIA}. We demand $p_T>20$ GeV for a clustered object to be considered as jet. Jets are isolated from unclustered objects if $\Delta R>0.4$.  
 
 
 \item {\it Unclustered Objects:}  All the final state objects which are neither clustered to form jets, nor identified as leptons, belong to this category. Particles with $0.5<p_T<20$ GeV and $|\eta|<5$ are considered as unclustered. Although unclustered objects do not interfere with our signal definition but they are important in constructing the missing energy of the events.
 
 \item {\it Missing Energy ($\slashed{E}_T$):} The transverse momentum of all the missing particles (those are not registered in the detector) can be estimated from the momentum imbalance in the transverse direction associated to the visible particles. Missing energy (MET) is thus defined as:
 \bea
 \slashed{E}_T = -\sqrt{(\sum_{\ell,j} p_x)^2+(\sum_{\ell,j} p_y)^2},
 \eea
 where the sum runs over all visible objects that include the leptons, jets and the unclustered components. 
 
 \item {\it Invariant dilepton mass $\left(m_{\ell\ell}\right)$}: We can construct the invariant dilepton mass variable for two opposite sign leptons by defining:
 \bea
 m_{\ell\ell}^2 = \left(p_{\ell^{+}}+p_{\ell^{-}}\right)^2.
 \eea
 Invariant mass of OSD events, if created from a single parent, peak at the parent mass, for example, $Z$ boson. As the signal events (Fig.~\ref{fig:vlfdecay}) do not arise from a single parent particle, invariant mass cut plays key role in eliminating the $Z$ mediated SM background. 
 
 \item $H_T$: $H_T$ is defined as the scalar sum of all isolated jets and lepton $p_T$'s:
 \bea
 H_T = \sum_{\ell,j} p_T. 	
 \eea
 For our signal the sum only includes the two leptons that are present in the final state. 
 \end{itemize}

We shall use different cuts on these observables to separate the signal from the SM backgrounds and predict the significance as a function of the integrated luminosity. This is shown in the next section. 

\subsection{Event rates and signal significance}
\label{sec:event}

\begin{figure}[htb!]
$$
\includegraphics[scale=0.4]{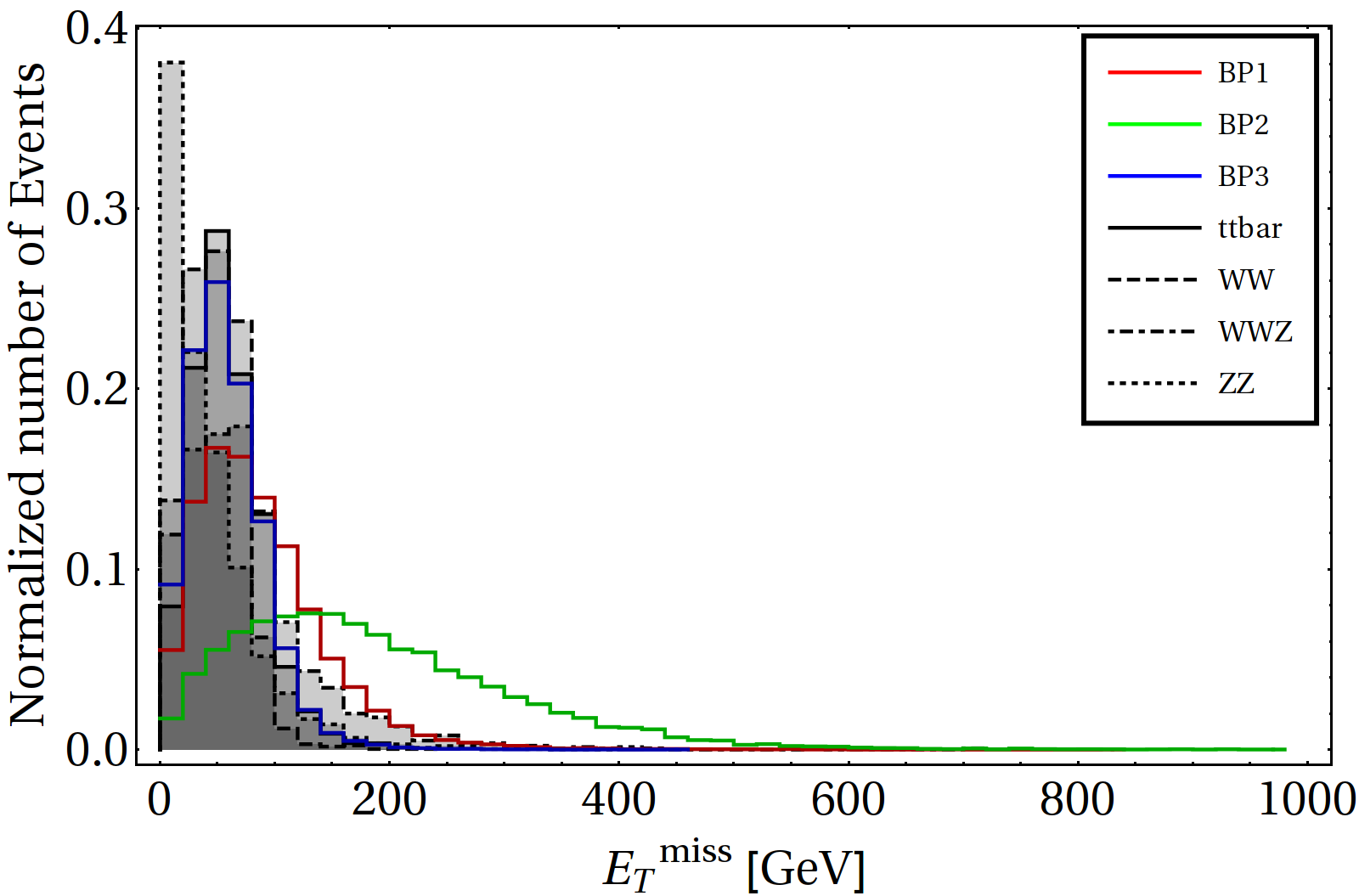} 
\includegraphics[scale=0.4]{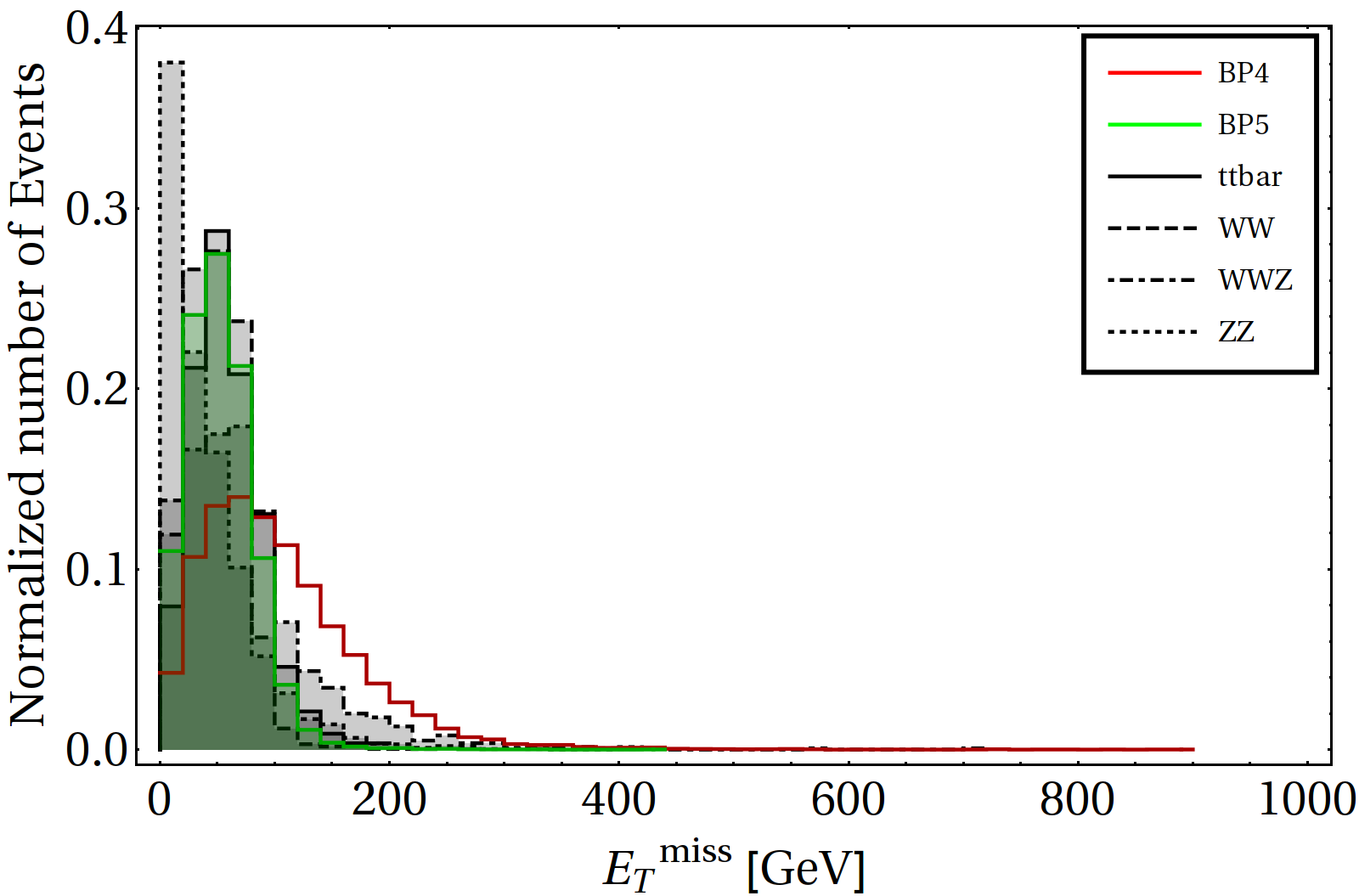} 
$$
$$
\includegraphics[scale=0.4]{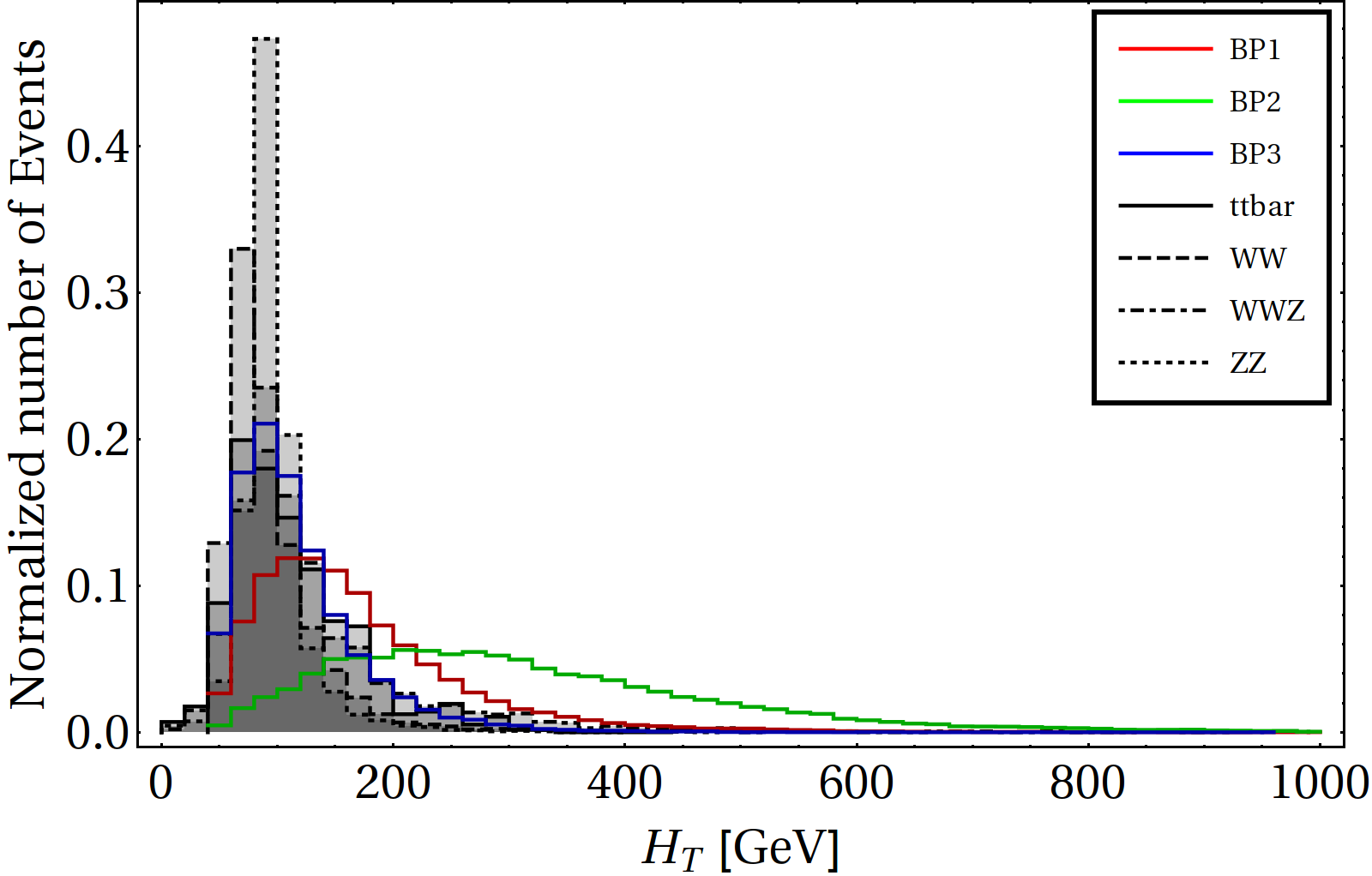}
\includegraphics[scale=0.4]{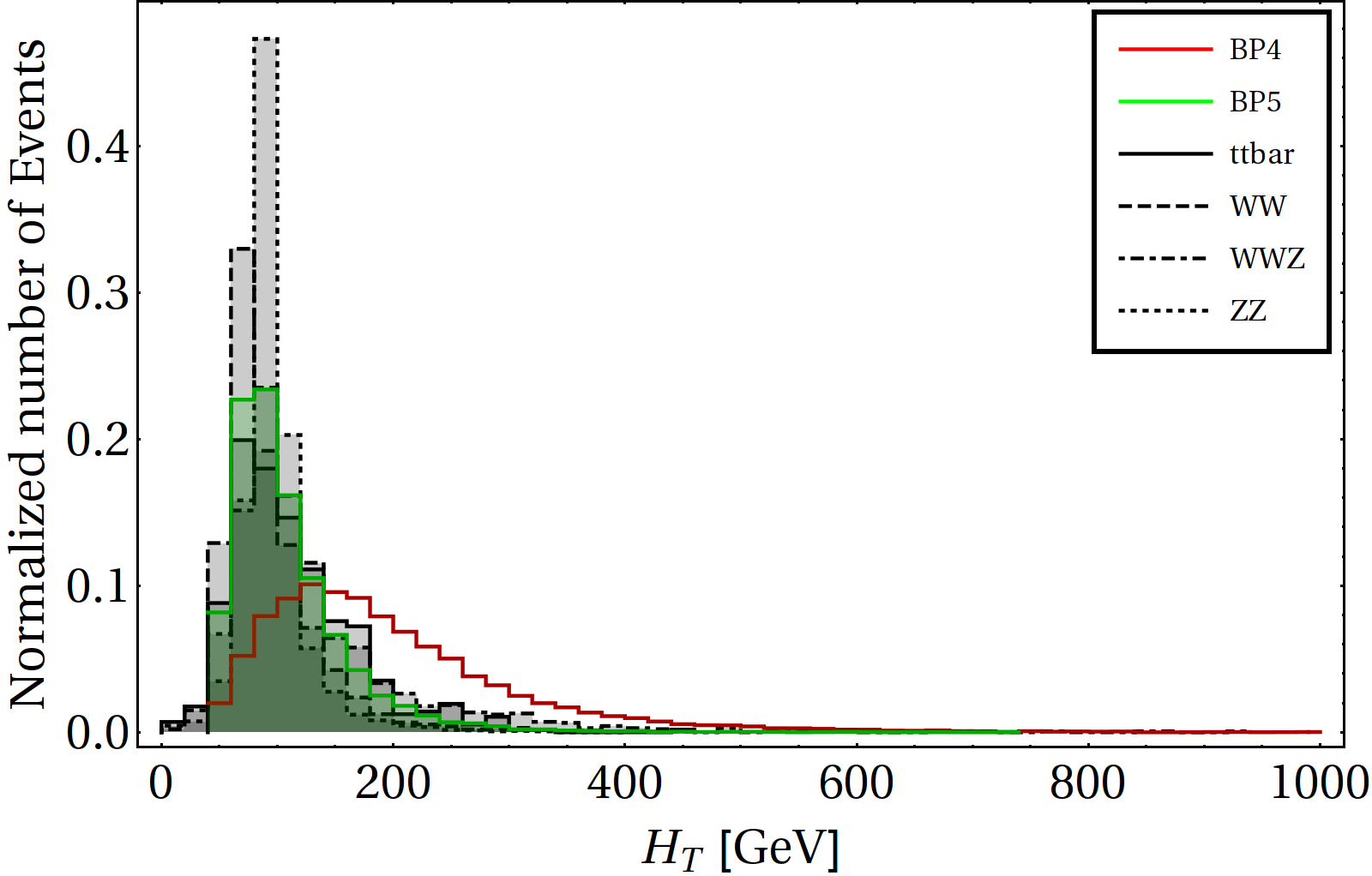} 
$$
\caption{Top left: Distribution of normalized number of signal and background events with MET for BP(1,2,3). Top right: Same as top left for BP(4,5). Bottom left: Distribution of normalized number of events with $H_T$ for BP(1,2,3). Bottom Right: Same as bottom left for BP(4,5). All simulations are done at $\sqrt{s}=14~\rm TeV$.}
\label{fig:metht}
\end{figure}

Here we would first like to show how the presence of new charge neutral gauge boson mediation can affect the pair production cross-section of the charged VLFs. This is explicitly tabulated in table~\ref{tab:prodcs} where we have listed the production cross-sections for our chosen BPs (table~\ref{tab:bp}) both in the presence and in the absence of $Z_{B-3L_{\tau}}$. As expected, in each case, the production via $Z$ and $Z_{B-3L_{\tau}}$ together is larger than that of only $Z$ mediation. The improvement, however, is not significant enough due to the reasons mentioned earlier.

\begin{table}[htb!]
\begin{center}
\begin{tabular}{|c|c|c|c|c|c|c|c|c|c|}
\hline
Benchmark & $\sigma_{pp\to\psi^+\psi^-}$ &  $\sigma_{pp\to\psi^+\psi^-}$  \\ [0.5ex] 
Point     &  ($Z+Z_{B-3L_{\tau}}$) &  (Only $Z$)   \\ [0.5ex] 
     &  (fb) &  (fb)   \\ [0.5ex]
\hline\hline

BP1 & 45.27 & 44.72 \\
\hline
BP2 & 3.82 & 3.71 \\
\hline
BP3 & 15.78 & 15.69 \\
\hline
BP4 & 46.76 & 46.62  \\
\hline
BP5 & 32.12 & 32.02  \\
\hline\hline
\end{tabular}
\end{center}
\caption {Production cross-section of charged VLF pairs for the chosen BPs in table~\ref{tab:bp} in presence of $Z_{B-3L_{\tau}}$ (left column) and in presence of only SM $Z$ (right column).} 
\label{tab:prodcs}
\end{table}

\begin{table}[htb!]
\begin{center}
\begin{tabular}{|c|c|c|c|c|c|c|c|c|c|}
\hline
Benchmark & $\slashed{E_T}$ &  & $\sigma^{\rm OSD}$  \\ [0.5ex] 
Points     & (GeV)  &  & (fb)    \\ [0.5ex] 
\hline\hline

BP1 & $>100$ && 0.82 \\
& $>200$ && 0.21 \\
& $>300$ && 0.06 \\
\cline{4-4}
\cline{1-2}
BP2 & $>100$ && 0.53 \\
& $>200$ &  & 0.24 \\
& $>300$ && 0.08 \\
\cline{4-4}
\cline{1-2}
BP3 & $>100$ &$H_T>$ 250 GeV& 0.04 \\
&$>200$ &  & 0.006 \\
&$>300$ && 0.001 \\
\cline{4-4}
\cline{1-2}
BP4 & $>100$ && 1.45 \\
& $>200$ && 0.38 \\
& $>300$ && 0.12 \\
\cline{4-4}
\cline{1-2}
BP5 & $>100$ && 0.05 \\
& $>200$ && 0.01 \\
& $>300$ && 0.002 \\
\hline\hline
\end{tabular}
\end{center}
\caption {Variation of final state signal cross-section with MET cut for a fixed cut on $H_T>250~\rm GeV$. All simulations are done at $\sqrt{s}=14~\rm TeV$.} 
\label{tab:sig1}
\end{table}

%

\begin{table}[htb!]
\begin{center}
\begin{tabular}{|c|c|c|c|c|c|c|c|c|c|}
\hline
Processes & $\sigma_{\rm production}$ & $\slashed{E_T}$ & & $\sigma^{\rm OSD}$  \\ [0.5ex] 
 & (pb) & (GeV)  &&  (fb)    \\ [0.5ex] 
\hline\hline

$t\bar t$ & & $>100$ && 0 \\
 & 814.64  & $>200$ && 0 \\
&& $>300$ && 0 \\
\cline{5-5}
\cline{1-3}
$W^+W^-$ && $>100$ &  & 5.99 \\
& 99.98 & $>200$ && 0.99 \\
&& $>300$ && 0 \\
\cline{5-5}
\cline{1-3}
$W^+W^-Z$ && $>100$ & $H_T>$ 250 GeV & 0.05 \\
& 0.15 & $>200$ && 0.02 \\
&& $>300$ && 0.009\\
\cline{5-5}
\cline{1-3}
$ZZ$ && $>100$ && $<1$ \\
& 14.01 &$>200$&& 0 \\
&&$>300$&& 0 \\
\hline
\hline
\end{tabular}
\end{center}
\caption {Variation of final state SM background cross-section with MET cut for a fixed cut on $H_T>250~\rm GeV$. All simulations are done at $\sqrt{s}=14~\rm TeV$.} 
\label{tab:bck}
\end{table}

In Fig.~\ref{fig:metht} we have shown the distribution of normalised number of events with respect to MET (upper panel) and $H_T$ (lower panel) for all the chosen BPs. In the same plot we have also shown the distribution from dominant SM backgrounds that can mimic our signal. For the SM the only source of MET are the SM neutrinos, which are almost massless with respect to centre of mass energy of the collider. As a result, the MET and $H_T$ distribution for SM peaks up at a lower value, while for the model MET arises from the DM $\psi_1$ (on top of the SM neutrinos) which is massive, and hence corresponding distribution for the signals are much flattened.  Noteworthy feature here is that, for larger $\Delta M$ the signal distributions are well separated from that of the background. This is due to the fact that the peak of the MET distribution is determined by how much of $p_T$ is being carried away by the missing particle ({\it i.e,} the DM), which in turn depends on the mass difference of charged and neutral component of the VLF {\it i.e,} $\Delta M$. Hence for larger $\Delta M$ the DM carries away most of the $p_T$ making the distribution much flatter, while for smaller $\Delta M$ the distribution peaks up at lower value as the produced DM particles are not boosted enough. As a consequence, in the LHS of top left panel of Fig.~\ref{fig:metht} we see BP3 (in blue) is completely submerged in the SM background, while BP1 (in red) can still be distinguished to some extent. BP2, because of large $\Delta M$ has a rather flattened distribution (in green) and therefore can be easily distinguished from the background with judicious choice of cuts. On the top right panel of Fig.~\ref{fig:metht} we have shown the MET distribution for BP4 (red) and BP5 (green). Here we also see the same consequence: with comparatively larger $\Delta M$ BP4 can be separated from the background, while BP5 shows no excess over the SM background. This trend is similar for $H_T$ distribution, which we have shown in the bottom panel of Fig.~\ref{fig:metht}.

\begin{figure}[htb!]
$$
\includegraphics[scale=0.5]{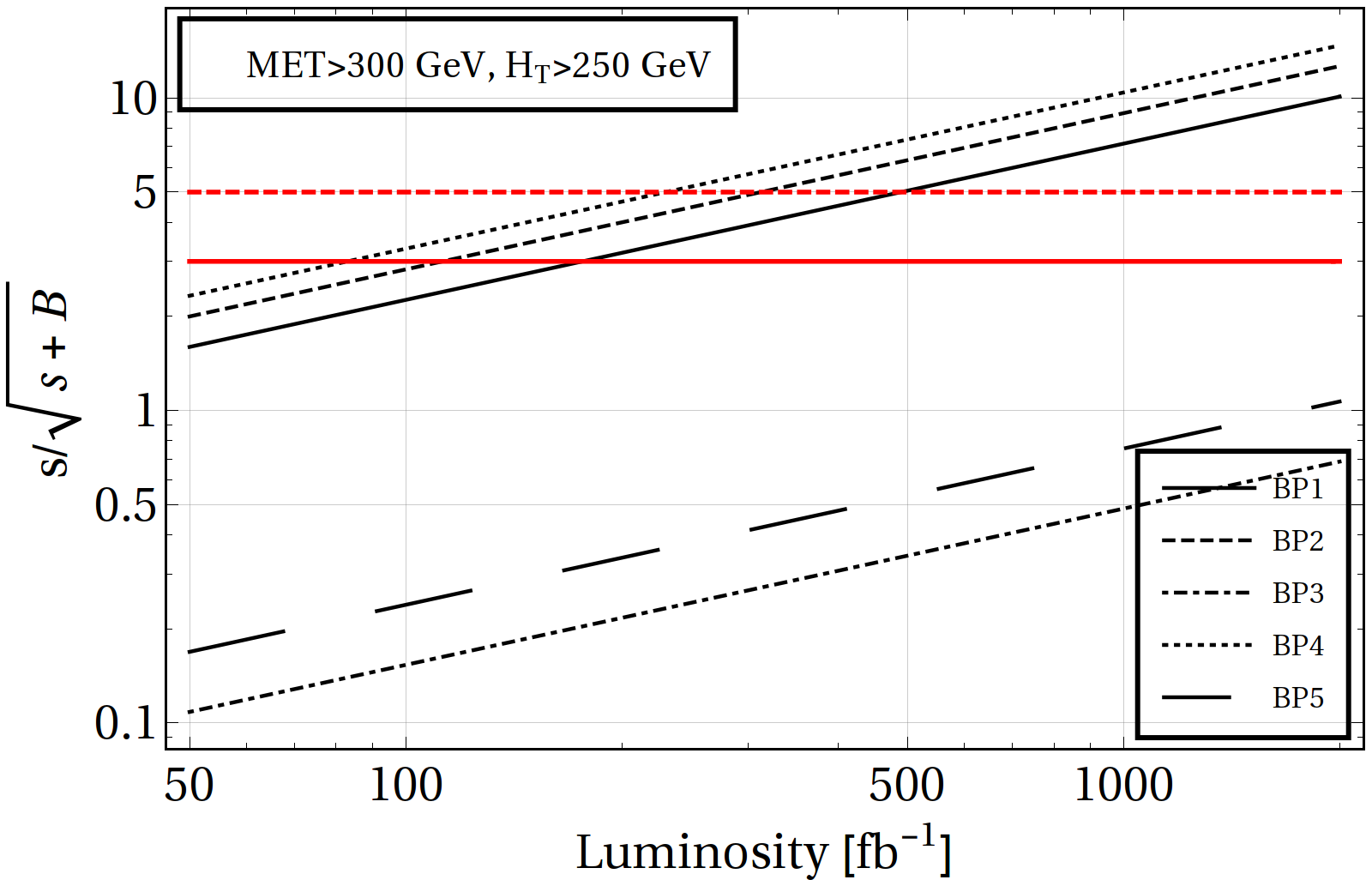} 
$$
\caption{Significance of benchmark points BP(1-5) at the LHC in terms of integrated luminosity. The solid red and dashed red lines correspond to $3\sigma$ and $5\sigma$ discovery limits respectively.}
\label{fig:signi}
\end{figure}

From the distributions one can easily see, with a MET cut of $\slashed{E_T}\gsim 200~\rm GeV$ and a $H_T$ cut of $\gsim 250~\rm GeV$ one can get rid off the SM backgrounds keeping most of the signal events intact. This is also shown in table~\ref{tab:sig1} where we have demonstrated the cut-flow {\it i.e,} how the number of events vary with MET cut, while the $H_T$ cut is kept fixed $H_T>250~\rm GeV$. On top of that we have also imposed the invariant mass cut over $Z$-window such that no events lie in the range: $|m_Z-15|\lsim m_{\ell^+\ell^-}\lsim |m_Z+15|$ in order to reduce the SM $Z$ background as explained earlier. Corresponding cut-flow for dominant SM backgrounds are also tabulated in table~\ref{tab:bck}. As one can see, the $t\bar t$ background is completely killed by imposing zero jet veto. Amongst other backgrounds, $ZZ$ also vanishes because of imposition of the $m_{\ell\ell}$ cut and $WW$ is also killed by putting a hard MET cut of 300 GeV. The only remaining background is due to $WWZ$ but that also becomes insignificant due to large MET cut.  


We finally plot the signal significance for the BPs in Fig.~\ref{fig:signi} by choosing the final state events with $\slashed{E_T}>300~\rm GeV$ and $H_T>250~\rm GeV$. BP2 and BP4 can reach $5\sigma$ discovery for an integrated luminosity $\sim 200~\rm fb^{-1}$ as they have the advantage of large $\Delta M$ which helps them to distinguish from the SM background as explained earlier. Due to comparatively smaller $\Delta M$ BP1 can reach a 5$\sigma$ discovery at a slightly higher luminosity $\sim 500~\rm fb^{-1}$. BP3 and BP5 can only be probed at the very high luminosity (HL-LHC). Here we would also like to emphasize the fact that this model may also be probed at the collider via stable charged track signature for $\Delta M\lsim 80~\rm GeV$. In that case the decay of the charged VLFs happen via off-shell $W^{\pm}$ and the decay width can be small enough for small $\sin\theta$ giving rise to charged tracks of length $\sim\mathcal{O}(\rm cm)$. This has been explored in details in~\cite{Bhattacharya:2017sml,Barman:2019tuo} and hence we refrain from discussing it again here.


\section{Summary and conclusion}
\label{sec:conclusion}

We have proposed a flavoured gauge extension of the singlet-doublet fermionic dark matter model by considering $B-3L_{\tau}$ as the additional gauge quantum number which naturally stabilises the DM without the need of additional discrete symmetries. The model also requires the existence of a singlet right handed neutrino (RHN) in order to be anomaly free. This RHN, along with another one or two singlet RHNs (having zero $B-3L_{\tau}$ charges) can take part in generating light neutrino masses via type I seesaw mechanism. The neutrino sector and the DM sector, however, are not very closely related as the bounds on the VEVs of the non-standard scalars from correct neutrino mass requirement is rather lose. This is attributed to the fact that light neutrino mass in the right ballpark can be generated keeping the scalar VEVs $\sim\mathcal{O}(\rm TeV)$ scale by tuning the new Yukawa couplings $(y_\ell,y_{\tau_3})$ accordingly. The family non-universal nature of this $B-3L_{\tau}$ gauge symmetry also helps us to avoid strong bounds from the LHC searches. The relatively lighter $Z^{\prime}$ boson  plays a role in generating dark matter relic abundance, leading to an enlarged parameter space satisfying DM related constraints compared to the purely singlet-doublet model. The pseudo-Dirac nature of the DM forbids the inelastic scattering via heavy neutral gauge boson allowing the DM to live over a huge parameter space satisfying both relic abundance and direct search constraints. Thus the parameter space remians valid upto DM mass of a few TeV for singlet-doublet VLF mixing of $\sin\theta\lsim 0.5$. A substantial portion of the parameter space however merges with the neutrino floor for smaller $\sin\theta$.

Apart from the motivations from dark matter and neutrino mass generation, the model also provides a solution to the electroweak vacuum metastability problem due to extended scalar sector. The model not only gives rise to a stable electroweak vacuum but also keeps it perturbative all the way upto the Planck scale. The requirements of vacuum stability and perturbativity however, significantly constrains the parameter space allowed purely from dark matter related constraints. As an effect of cumulative bound from relic abundance, direct search and stability and perturbativity of the scalar potential, the model substantially constraints the singlet-doublet Yukawa $Y\lsim 0.3$ and the gauge coupling $g_{B-3L_\tau}\lsim 0.25$. However, the mass difference between the heavier and lighter physical states of the VLF: $M_{\psi_2}-M_{\psi_1}=\Delta M$ can still be large enough $\sim 500~\rm GeV$ providing opportunity for the model to be probed at the LHC via hadronically quiet dilepton final states with missing energy excess. This is again attributed to the pseudo-Dirac nature of the VLFs due to which larger $\sin\theta$ is allowed from direct search, and hence large $\Delta M$ is possible to achieve.

We finally discussed possible signatures at colliders by analysing some of the benchmark points of the model which satisfy all theoretical and experimental bounds. We particularly focus on purely leptonic final states with missing energy and show that with judicious choice of cuts on different kinematical variables (eg. MET, $H_T$ etc) it is indeed possible to attain a $5\sigma$ discovery potential for the model at the high luminosity LHC. Apart from leptonic final states, the model may also be probed via displaced vertex signature due to the off-shell decay of the charged VLF to SM leptons and neutrino for $\Delta M\lsim m_W$.

%
\acknowledgements
DB acknowledges the support from IIT Guwahati start-up grant (reference number: xPHYSUGI-ITG01152xxDB001), Early Career Research Award from DST-SERB, Government of India (reference number: ECR/2017/001873) and Associateship Programme of Inter University Centre for Astronomy and Astrophysics (IUCAA), Pune. BB and PG would like to thank Triparno Bandyopadhyay and Subhaditya Bhattacharya for useful discussions during very early stage of this work. BB would also like to thank Krishnanjan Pramanik for computational helps.

\clearpage
\appendix
\section{Invisible Higgs and $Z$ decays}
\label{sec:invdecay}

The combination of SM channels yields an observed (expected) upper limit on the SM Higgs branching fraction of 0.24 at 95 \%\ CL~\cite{Khachatryan:2016whc} with a total decay width $\Gamma=4.07\times 10^{-3}~\rm GeV$. This gives rise to an allowed Higgs invisible decay branching fraction of 0.24(0.23). SM $Z$ boson, on the other hand, can also decay to invisible final states, and hence constrained from observation: $\Gamma_{inv}^{Z}=499\pm 1.5~\rm MeV$~\cite{PhysRevD.98.030001}. So, if $Z$ is allowed to decay invisibly, the decay width should not be more than 1.5 MeV. In our case, the decays of SM Higgs and SM $Z$ are only possible to $\psi_1\psi_1$ pairs as other invisible decay modes are kinematically forbidden because of large $\Delta M$. These decay widths are given by:

\bea
\Gamma_{h_1\to\psi_1\bar\psi_1}=\frac{Y c_{12}^2~c_{13}^2 \sin^2\theta \cos^2\theta}{8\pi} m_{h_1}\left(1-\frac{4 M_{\psi_1}^2}{m_{h_1}^2}\right)^{3/2} 
\eea

and

\bea
\Gamma_{Z\to\psi_1\bar\psi_1}=\frac{m_Z}{48 \pi} \frac{e^2 \sin^4\theta}{s_W^2 c_W^2} \left(1+\frac{2 M_{\psi_1}^2}{m_Z^2}\right) \sqrt{1-\frac{4 M_{\psi_1}^2}{m_Z^2}},
\eea

with $c_W=m_{W}/m_Z$ is the Weinberg angle where $m_{W(Z)}$ is the mass of SM $W(Z)$ boson.

\begin{table}[htb!]
\begin{center}
\begin{tabular}{|c|c|c|c|c|c|c|c|c|c|}
\hline
Benchmark & $Br_{inv}^{higgs}$ &  $\Gamma_{inv}^{Z}$ (MeV)  \\ [0.5ex] 
Point     &                    &             (MeV)  \\ [0.5ex] 
\hline\hline 

BP1 & - & - \\
\hline
BP2 & 0.019 & - \\
\hline
BP3 & - & -  \\
\hline
BP4 & 0.072 & 0.073 \\
\hline
BP5 & - & - \\
\hline\hline
\end{tabular}
\end{center}
\caption {Invisible Higgs branching ratio and invisible $Z$ decay width for different benchmark points tabulated in table~\ref{tab:bp}. ``-'' stands for cases where $M_{\psi_1}>m_Z/2$ and/or $>m_{h_1}/2$.} 
\label{tab:invdecay}
\end{table}

Now, for our chosen BPs we would like to see whether these bounds are applicable or not. First note that, Higgs invisible decay is possible for BP2 and BP4, while $Z$ can decay invisibly to DM pairs only for BP4. Rest of the benchmarks are safe from such bounds as DM mass is much above than SM Higgs or SM $Z$ mass. In table~\ref{tab:invdecay} we have tabulated the invisible branching fraction ($Br_{inv}^{higgs}$) for SM Higgs (left column) and invisible $Z$ decay width $\Gamma_{inv}^{Z}$ for SM $Z$ (right column). 

\section{RG equations at one loop}\label{ap2}
\begin{align}
&(4\pi)^2\beta_{g_{1}}=\frac{43}{6}g_1^3,\\
&(4\pi)^2\beta_{g_{2}}=-\frac{17}{6}g_2^3,\\
&(4\pi)^2\beta_{g_{3}}=-7g_1^3,\\
&(4\pi)^2\beta_{g_{B-3L_{\tau}}}=\frac{757}{24}g_{B-3L_{\tau}}^3,\\
&(4\pi)^2\beta_{\lambda_H}=\lambda_1^2+\lambda_2^2+24\lambda_H^2+\frac{3 g_1^4}{8}+\frac{9 g_2^4}{8}-3\lambda_Hg_1^2-9\lambda_H g_2^2-6 y_t^4+12\lambda_Hy_t^2\nonumber\\
&+\frac{3g_1^2g_2^2}{4}+4\lambda_HY^2+4\lambda_H(y_{\alpha_1}^2+y_{\alpha_2}^2+y_{\tau_3}^2)-2 Y^4-2(y_{\alpha_1}^4+y_{\alpha_4}^2+y_{\tau_3}^4),\\
&(4\pi)^2\beta_{\lambda_\Phi}=2\lambda_1^2+\lambda_{S\Phi}^2+20\lambda_\Phi^2+\frac{243}{8} g_{B-3L_{\tau}}^4-27\lambda_\Phi g_{B-3L_{\tau}}^2+8\lambda_\Phi y_\chi^2-16 y_\chi^4,\\
&(4\pi)^2\beta_{\lambda_S}=20\lambda_S^2-108 \lambda_Sg_{B-3L_{\tau}}^2+2\lambda_2^2+\lambda_{S\Phi}^2+486 g_{B-3L_{\tau}}^4+\lambda_S(y_{13}^2+y_{23}^2)-\frac{1}{8}(y_{13}^4+y_{23}^4)-\frac{y_{13}^2y_{23}^2}{4},\\
&(4\pi)^2\beta_{\lambda_1}=2\lambda_2\lambda_{S\Phi}+4\lambda_1^2+12\lambda_H\lambda_1+8\lambda_\Phi\lambda_1+6\lambda_1y_t^2-\frac{9\lambda_1 g_2^2}{2}-\frac{27}{2}g_{B-3L_{\tau}}^2\lambda_1-\frac{3g_1^2}{2}\lambda_1+2\lambda_1Y^2\nonumber\\
&+2\lambda_1(y_{\alpha_1}^2+y_{\alpha_2}^2+y_{\tau_3}^2+4\lambda_1y_\chi^2-16Y^2y_\chi^2),\\
&(4\pi)^2\beta_{\lambda_2}=4\lambda_2^2-54\lambda_2g_{B-3L_{\tau}}^2+2\lambda_1\lambda_{S\Phi}+6\lambda_2y_t^2+12\lambda_H\lambda_2+8\lambda_S\lambda_2-\frac{3\lambda_2g_1^2}{2}-\frac{9\lambda_2g_2^2}{2}\nonumber\\
&+\frac{\lambda_2}{2}(y_{13}^2+y_{23}^2)+2\lambda_2Y^2+2\lambda_2(y_{\alpha_1^2}+y_{\alpha_2}^2+y_{\tau_3}^2)-y_{\alpha_1}^2y_{\tau_3}^2-y_{\alpha_2}^2y_{\alpha_3^2}-y_{\alpha_1}^2y_{\alpha_2}^2,\\
&(4\pi)^2\beta_{\lambda_{S\Phi}}=8\lambda_S\lambda_{S\Phi}+4\lambda_2\lambda_1+4\lambda_{S\Phi}^2+243 g_{B-3L_{\tau}}^4+8\lambda_{S\Phi}\lambda_\Phi-\frac{135}{2}g_{B-L}^2\lambda_{S\Phi}\nonumber\\&+\frac{\lambda_{S\Phi}}{2}(y_{13}^2+y_{23}^2)+4\lambda_{S\Phi}y_\chi^2,\\
&(4\pi)^2\beta_{y_t}=\frac{9 y_t^3}{2}-8g_3^2y_t-\frac{17}{12}g_1^2y_t-\frac{9}{4}g_2^2y_t-\frac{2 g_{B-3L_{\tau}}^2}{3}y_t+Y^2y_t+y_t(y_{\alpha_1}^2+y_{\alpha_2}^2+y_{\tau_3}^2),\\
&(4\pi)^2\beta_{Y}=3Yy_t^2-\frac{3Yg_1^2}{4}-
\frac{9Yg_2^2}{4}-\frac{27Y}{8}g_{B-3L_{\tau}}^2+\frac{5 Y^3}{2}+Y((y_{13}^2+y_{23}^2)+2y_\chi^2),
\end{align}
\section{Unitarity}
\label{sec:apunit}

Quartic couplings of the scalar potential which are shown in Eq.~\eqref{co-positivity} are also constrained from tree level perturbative unitarity. The unitarity bounds are related with scattering amplitude as: $|\mathcal{M}| \leq 8 \pi$ .
 $\mathcal{M}$ be the scattering amplitude for any $2=2$ process which can be expressed in terms of partial waves as follows: 
\bea
\mathcal{M}=16\pi \sum_{l=0}^{\infty} a_{l}(2l+1) P_l(\cos\theta),
\eea
where $P_l(\cos\theta)$ is the Legendre polynomial of order $l$, $a_l$ be the partial wave amplitude and $\theta$ be the scattering angle. To implement unitarity bound in our case we form an amplitude matrix $M={\mathcal{M}}_{i =j}$ where $i$ and $j$ correspond to all possible two particle state. And each eigenvalue of this amplitude matrix, $M$ should lie within $8\pi$ (i.e. $|e_i| \leq 8 \pi$)in order to maintain unitarity. The amplitude matrix $M$ is decomposed of 22 neutral charged (NC) and 6 singly charged(SC) two particles state which is given by:
\begin{align}
M=\left(
\begin{array}{cc}
  (M^{NC})_{22\times 22}&  0  \\
 0 & (M^{SC})_{6\times 6}  \\
\end{array}
\right).
 \label{eq:PUnitarity}
\end{align}
The charged neutral two particles staTes which are formed the sub-matrix, $M^{NC}$ are given by:
\bea\nonumber
| G^+ G^- \rangle,~| \frac{h~h}{\sqrt2} \rangle,~| \frac{z_1~z_1}{\sqrt2} \rangle,~| \frac{\phi~\phi}{\sqrt2} \rangle,~| \frac{z_2~z_2}{\sqrt2} \rangle, ~| \frac{s~s}{\sqrt2} \rangle,~~| \frac{z_3~z_3}{\sqrt2} \rangle,~|\phi~z_2\rangle,~|\phi~s\rangle,~|\phi~z_3\rangle,~ \\
~|s~z_2\rangle,~|z_2~z_3\rangle,~|s~z_3\rangle,~|h~z_1\rangle,~|h~\phi\rangle,~|h~z_2\rangle,~|h~s\rangle,~|h~z_3\rangle,~|\phi~z_1\rangle,~|z_1~z_2\rangle,~|s~z_1\rangle,~|z_1~z_3\rangle ;\nonumber
\eea
\bea
M^{NC}={\tiny{
\left(
\begin{array}{cccccccccccccccccccccc}
 4 \lambda_H & \sqrt{2} \lambda_H & \sqrt{2} \lambda_H & \frac{\lambda_1}{\sqrt{2}} & \frac{\lambda_1}{\sqrt{2}} & \frac{\lambda_2}{\sqrt{2}} & \frac{\lambda_2}{\sqrt{2}} & 0 & 0 & 0 & 0 & 0 & 0 & 0 & 0 & 0 & 0 & 0 & 0 & 0 & 0 & 0 \\
 \sqrt{2} \lambda_H & 3 \lambda_H & \lambda_H & \frac{\lambda_1}{2} & \frac{\lambda_1}{2} & \frac{\lambda_2}{2} & \frac{\lambda_2}{2} & 0 & 0 & 0 & 0 & 0 & 0 & 0 & 0 & 0 & 0 & 0 & 0 & 0 & 0 & 0 \\
 \sqrt{2} \lambda_H & \lambda_H & 3 \lambda_H & \frac{\lambda_1}{2} & \frac{\lambda_1}{2} & \frac{\lambda_2}{2} & \frac{\lambda_2}{2} & 0 & 0 & 0 & 0 & 0 & 0 & 0 & 0 & 0 & 0 & 0 & 0 & 0 & 0 & 0 \\
 \frac{\lambda_1}{\sqrt{2}} & \frac{\lambda_1}{2} & \frac{\lambda_1}{2} & 3 \lambda_\Phi  & \lambda_\Phi  & \frac{\lambda_{\mathcal{S}\Phi}}{2} & \frac{\lambda_{\mathcal{S}\Phi}}{2} & 0 & 0 & 0 & 0 & 0 & 0 & 0 & 0 & 0 & 0 & 0 & 0 & 0 & 0 & 0 \\
 \frac{\lambda_1}{\sqrt{2}} & \frac{\lambda_1}{2} & \frac{\lambda_1}{2} & \lambda_\Phi  & 3 \lambda_\Phi  & \frac{\lambda_{\mathcal{S}\Phi}}{2} & \frac{\lambda_{\mathcal{S}\Phi}}{2} & 0 & 0 & 0 & 0 & 0 & 0 & 0 & 0 & 0 & 0 & 0 & 0 & 0 & 0 & 0 \\
 \frac{\lambda_2}{\sqrt{2}} & \frac{\lambda_2}{2} & \frac{\lambda_2}{2} & \frac{\lambda_{\mathcal{S}\Phi}}{2} & \frac{\lambda_{\mathcal{S}\Phi}}{2} & 3 \lambda_{\mathcal{S}} & \lambda_{\mathcal{S}} & 0 & 0 & 0 & 0 & 0 & 0 & 0 & 0 & 0 & 0 & 0 & 0 & 0 & 0 & 0 \\
 \frac{\lambda_2}{\sqrt{2}} & \frac{\lambda_2}{2} & \frac{\lambda_2}{2} & \frac{\lambda_{\mathcal{S}\Phi}}{2} & \frac{\lambda_{\mathcal{S}\Phi}}{2} & \lambda_{\mathcal{S}} & 3 \lambda_{\mathcal{S}} & 0 & 0 & 0 & 0 & 0 & 0 & 0 & 0 & 0 & 0 & 0 & 0 & 0 & 0 & 0 \\
 0 & 0 & 0 & 0 & 0 & 0 & 0 & 2 \lambda_\Phi  & 0 & 0 & 0 & 0 & 0 & 0 & 0 & 0 & 0 & 0 & 0 & 0 & 0 & 0 \\
 0 & 0 & 0 & 0 & 0 & 0 & 0 & 0 & \lambda_{\mathcal{S}\Phi} & 0 & 0 & 0 & 0 & 0 & 0 & 0 & 0 & 0 & 0 & 0 & 0 & 0 \\
 0 & 0 & 0 & 0 & 0 & 0 & 0 & 0 & 0 & \lambda_{\mathcal{S}\Phi} & 0 & 0 & 0 & 0 & 0 & 0 & 0 & 0 & 0 & 0 & 0 & 0 \\
 0 & 0 & 0 & 0 & 0 & 0 & 0 & 0 & 0 & 0 & \lambda_{\mathcal{S}\Phi} & 0 & 0 & 0 & 0 & 0 & 0 & 0 & 0 & 0 & 0 & 0 \\
 0 & 0 & 0 & 0 & 0 & 0 & 0 & 0 & 0 & 0 & 0 & \lambda_{\mathcal{S}\Phi} & 0 & 0 & 0 & 0 & 0 & 0 & 0 & 0 & 0 & 0 \\
 0 & 0 & 0 & 0 & 0 & 0 & 0 & 0 & 0 & 0 & 0 & 0 & 2 \lambda_{\mathcal{S}} & 0 & 0 & 0 & 0 & 0 & 0 & 0 & 0 & 0 \\
 0 & 0 & 0 & 0 & 0 & 0 & 0 & 0 & 0 & 0 & 0 & 0 & 0 & 2 \lambda_H & 0 & 0 & 0 & 0 & 0 & 0 & 0 & 0 \\
 0 & 0 & 0 & 0 & 0 & 0 & 0 & 0 & 0 & 0 & 0 & 0 & 0 & 0 & \lambda_1 & 0 & 0 & 0 & 0 & 0 & 0 & 0 \\
 0 & 0 & 0 & 0 & 0 & 0 & 0 & 0 & 0 & 0 & 0 & 0 & 0 & 0 & 0 & \lambda_1 & 0 & 0 & 0 & 0 & 0 & 0 \\
 0 & 0 & 0 & 0 & 0 & 0 & 0 & 0 & 0 & 0 & 0 & 0 & 0 & 0 & 0 & 0 & \lambda_2 & 0 & 0 & 0 & 0 & 0 \\
 0 & 0 & 0 & 0 & 0 & 0 & 0 & 0 & 0 & 0 & 0 & 0 & 0 & 0 & 0 & 0 & 0 & \lambda_2 & 0 & 0 & 0 & 0 \\
 0 & 0 & 0 & 0 & 0 & 0 & 0 & 0 & 0 & 0 & 0 & 0 & 0 & 0 & 0 & 0 & 0 & 0 & \lambda_1 & 0 & 0 & 0 \\
 0 & 0 & 0 & 0 & 0 & 0 & 0 & 0 & 0 & 0 & 0 & 0 & 0 & 0 & 0 & 0 & 0 & 0 & 0 & \lambda_1 & 0 & 0 \\
 0 & 0 & 0 & 0 & 0 & 0 & 0 & 0 & 0 & 0 & 0 & 0 & 0 & 0 & 0 & 0 & 0 & 0 & 0 & 0 & \lambda_2 & 0 \\
 0 & 0 & 0 & 0 & 0 & 0 & 0 & 0 & 0 & 0 & 0 & 0 & 0 & 0 & 0 & 0 & 0 & 0 & 0 & 0 & 0 & \lambda_2 \\
\end{array} 
\right) 
}} \nonumber \\
\eea 
And the singly charged two particle states for the sub-matrix $M^{SC}$  are as follows :
\bea\nonumber
|G^+~h\rangle,~|G^+~z_1\rangle,~|G^+~\phi\rangle,~|G^+~z_2\rangle,~|G^+~s\rangle,~|G^+~z_3\rangle ~~;
\eea
\bea
M^{SC}=\left(
\begin{array}{cccccc}
 2 \lambda_H & 0 & 0 & 0 & 0 & 0 \\
 0 & 2 \lambda_H & 0 & 0 & 0 & 0 \\
 0 & 0 & \lambda_1 & 0 & 0 & 0 \\
 0 & 0 & 0 & \lambda_1 & 0 & 0 \\
 0 & 0 & 0 & 0 & \lambda_2 & 0 \\
 0 & 0 & 0 & 0 & 0 & \lambda_2 \\
\end{array}
\right)
\eea
Each of distinct eigenvalues of the amplitude matrix, $M$ will be bounded from tree level unitarity as :
\bea
|\lambda_{H}| \leq 4 \pi , ~~|\lambda_{\mathcal{S}}| \leq 4\pi , \nonumber \\
|\lambda_1| \leq 8\pi,~~|\lambda_2| \leq 8 \pi,~~|\lambda_{\mathcal{S}\Phi}|\leq 8\pi, \nonumber \\
|x_{1,2,3}| \leq 16 \pi ,
\eea
where, $x_{1,2,3}$ are the cubic roots of the following polynomial equation:
\bea
x^3+x^2 (-12 \lambda_H-8 \lambda_{\mathcal{S}}-8 \lambda_\Phi )+x \left(-8 \lambda_1^2-8 \lambda_2^2+96 \lambda_H \lambda_{\mathcal{S}}+96 \lambda_H \lambda_\Phi +64 \lambda_{\mathcal{S}} \lambda_\Phi -4 \lambda_{\mathcal{S}\Phi}^2\right)
\nonumber \\
+64 \lambda_1^2 \lambda_{\mathcal{S}}-32 \lambda_1 \lambda_2 \lambda_{\mathcal{S}\Phi}  +64 \lambda_2^2 \lambda_\Phi -768 \lambda_H \lambda_{\mathcal{S}} \lambda_\Phi +48 \lambda_H \lambda_{\mathcal{S}\Phi}^2 =0 ~.~~~~~~~
\eea

\section{Relevant Feynmann Diagrams for DM (co-)annihilation}
\label{sec:feyn-diag}
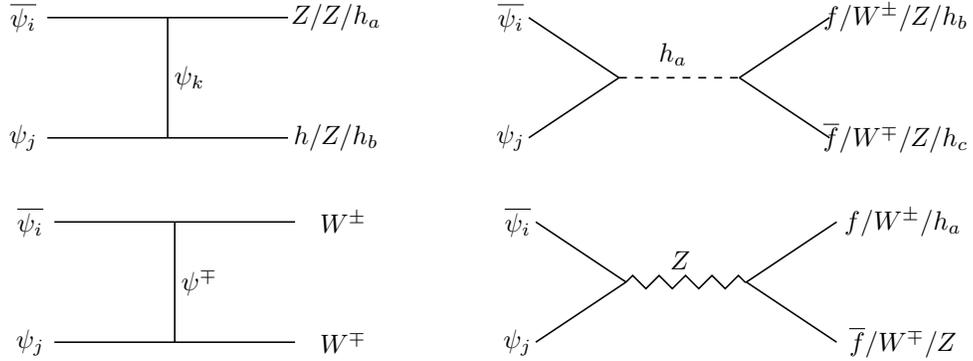
\begin{figure}[htb!]
\begin{center}
    \begin{tikzpicture}[line width=0.6 pt, scale=0.8]
        \draw[solid] (-3,1.0)--(-1.0,1.0);
        \draw[solid] (-3,-1.0)--(-1.0,-1.0);
        \draw[solid] (-1.0,1.0)--(-1.0,-1.0);
        \draw[solid] (-1.0,1.0)--(1.0,1.0);
        \draw[solid] (-1.0,-1.0)--(1.0,-1.0);
        \node at (-3.4,1.0) {$\overline{\psi_i}$};
        \node at (-3.4,-1.0) {$\psi_j$};
        \node [right] at (-1.05,0.0) {$\psi_k$};
        \node at (1.8,1.0) {$Z/Z/h_{a}$};
        \node at (1.8,-1.0) {$h/Z/h_{b}$};
        \draw[solid] (5.0,1.0)--(6.5,0.0);
        \draw[solid] (5.0,-1.0)--(6.5,0.0);
        \draw[dashed] (6.5,0.0)--(8.5,0.0);
        \draw[solid] (8.5,0.0)--(10.0,1.0);
        \draw[solid] (8.5,0.0)--(10.0,-1.0);
        \node at (4.7,1.0) {$\overline{\psi_i}$};
        \node at (4.7,-1.0) {$\psi_j$};
        \node [above] at (7.4,0.05) {$h_{a}$};
        \node at (11.1,1.0) {$f/W^\pm/Z/h_{b}$};
        \node at (11.1,-1.0) {$\overline{f}/W^\mp/Z/h_{c}$};
     \end{tikzpicture}
 \end{center}
\begin{center}
    \begin{tikzpicture}[line width=0.6 pt, scale=0.8]
        \draw[solid] (-3,1.0)--(-1.0,1.0);
        \draw[solid] (-3,-1.0)--(-1.0,-1.0);
        \draw[solid](-1.0,1.0)--(-1.0,-1.0);
        \draw[solid] (-1.0,1.0)--(1.0,1.0);
        \draw[solid] (-1.0,-1.0)--(1.0,-1.0);
        \node at (-3.4,1.0) {$\overline{\psi_i}$};
        \node at (-3.4,-1.0) {$\psi_j$};
        \node [right] at (-1.05,0.0) {$\psi^\mp$};
        \node at (1.8,1.0) {$W^\pm$};
        \node at (1.8,-1.0) {$W^\mp$};
        \draw[solid] (5.0,1.0)--(6.5,0.0);
        \draw[solid] (5.0,-1.0)--(6.5,0.0);
        \draw[snake] (6.5,0.0)--(8.5,0.0);
        \draw[solid] (8.5,0.0)--(10.0,1.0);
        \draw[solid] (8.5,0.0)--(10.0,-1.0);
        \node at (4.7,1.0) {$\overline{\psi_i}$};
        \node at (4.7,-1.0) {$\psi_j$};
        \node [above] at (7.4,0.05) {$Z$};
        \node at (11.1,1.0) {$f/W^\pm/h_a$};
        \node at (11.1,-1.0) {$\overline{f}/W^\mp/Z$};
     \end{tikzpicture}
 \end{center}
\caption{ Annihilation ($i=j$) and Co-annihilation ($i\neq j$) type number changing processes for Vector 
like fermionic DM in the model. Here $i,j,k=1,2$;~ $a,b,c=1,2,3$~and ~ $f$ stands for SM fermions. }
\label{fd:an-coan}
 \end{figure}
\begin{figure}[htb!]
\begin{center}
    \begin{tikzpicture}[line width=0.6 pt, scale=0.8]
        \draw[solid] (-3,1.0)--(-1.0,1.0);
        \draw[solid] (-3,-1.0)--(-1.0,-1.0);
        \draw[solid] (-1.0,1.0)--(-1.0,-1.0);
        \draw[solid] (-1.0,1.0)--(1.0,1.0);
        \draw[solid] (-1.0,-1.0)--(1.0,-1.0);
        \node at (-3.8,1.0) {$\overline{\psi_i}/\psi_i$};
        \node at (-3.8,-1.0) {$\psi^-/\psi^+$};
        \node [right] at (-1.05,0.0) {$\psi_j$};
        \node at (1.8,1.0) {$Z/h_a$};
        \node at (1.9,-1.0) {$W^\mp/W^\mp$};
        \draw[solid] (5.0,1.0)--(6.5,0.0);
        \draw[solid] (5.0,-1.0)--(6.5,0.0);
        \draw[snake] (6.5,0.0)--(8.5,0.0);
        \draw[solid] (8.5,0.0)--(10.0,1.0);
        \draw[solid] (8.5,0.0)--(10.0,-1.0);
        \node at (4.3,1.0) {$\overline{\psi_i}/\psi_i$};
        \node at (4.3,-1.0) {$\psi^-/\psi^+$};
        \node [above] at (7.4,0.05) {$W^\pm$};
        \node at (11.4,1.0) {$f/h_a/W^\mp/W^\mp$};
        \node at (11.1,-1.0) {$\overline{f^\prime}/W^\pm/\gamma/Z$};
     \end{tikzpicture}
 \end{center}
\begin{center}
    \begin{tikzpicture}[line width=0.5 pt, scale=0.9]
        \draw[solid] (-3,1.0)--(-1.0,1.0);
        \draw[solid] (-3,-1.0)--(-1.0,-1.0);
        \draw[solid](-1.0,1.0)--(-1.0,-1.0);
        \draw[solid] (-1.0,1.0)--(1.0,1.0);
        \draw[solid] (-1.0,-1.0)--(1.0,-1.0);
        \node at (-3.6,1.0) {$\overline{\psi_i}/\psi_i$};
        \node at (-3.6,-1.0) {$\psi^-/\psi^+$};
        \node [right] at (-1.05,0.0) {$\psi^\mp$};
        \node at (1.8,1.0) {$W^\mp/W^\mp$};
        \node at (1.8,-1.0) {$\gamma/Z$};
     \end{tikzpicture}
 \end{center}
\caption{ Feynmann diagrams for co-annihilation type number changing processes of $\psi_i ~(i=1,2)$ with the 
charged component $\psi^\pm$ to SM particles. Here $f$ and $f^\prime$ stand for SM fermions ($f \neq f^\prime$). }
\label{co-ann-2}
 \end{figure}
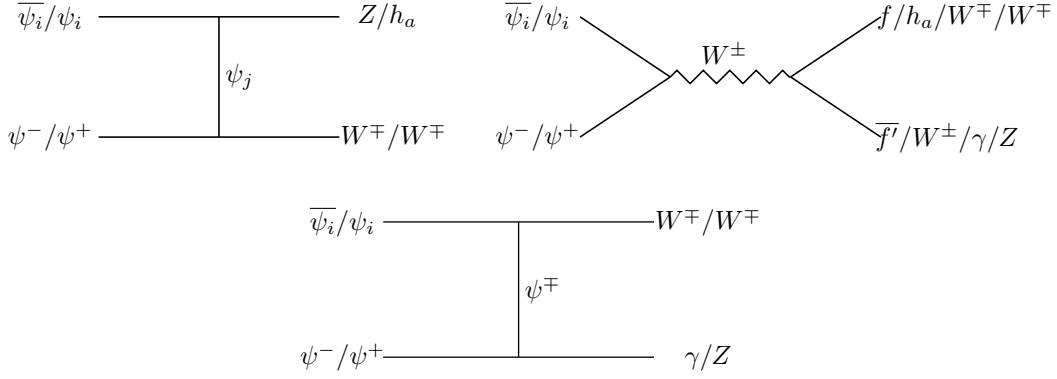
\begin{figure}[htb!]
\begin{center}
    \begin{tikzpicture}[line width=0.6 pt, scale=0.8]
        \draw[solid] (-3,1.0)--(-1.0,1.0);
        \draw[solid] (-3,-1.0)--(-1.0,-1.0);
        \draw[solid] (-1.0,1.0)--(-1.0,-1.0);
        \draw[solid] (-1.0,1.0)--(1.0,1.0);
        \draw[solid] (-1.0,-1.0)--(1.0,-1.0);
        \node at (-3.4,1.0) {$\psi^+$};
        \node at (-3.4,-1.0) {$\psi^-$};
        \node [right] at (-1.05,0.0) {$\psi_i$};
        \node at (1.8,1.0) {$W^+$};
        \node at (1.8,-1.0) {$W^-$};
        \draw[solid] (5.0,1.0)--(6.5,0.0);
        \draw[solid] (5.0,-1.0)--(6.5,0.0);
        \draw[snake] (6.5,0.0)--(8.5,0.0);
        \draw[solid] (8.5,0.0)--(10.0,1.0);
        \draw[solid] (8.5,0.0)--(10.0,-1.0);
        \node at (4.6,1.0) {$\psi^+$};
        \node at (4.6,-1.0) {$\psi^-$};
        \node [above] at (7.4,0.05) {$\gamma/Z$};
        \node at (11.1,1.0) {$f/W^+$};
        \node at (11.1,-1.0) {$\overline{f}/W^-$};
     \end{tikzpicture}
 \end{center}
\begin{center}
    \begin{tikzpicture}[line width=0.6 pt, scale=0.8]
        \draw[solid] (-3,1.0)--(-1.0,1.0);
        \draw[solid] (-3,-1.0)--(-1.0,-1.0);
        \draw[solid](-1.0,1.0)--(-1.0,-1.0);
        \draw[solid] (-1.0,1.0)--(1.0,1.0);
        \draw[solid] (-1.0,-1.0)--(1.0,-1.0);
        \node at (-3.4,1.0) {$\psi^+$};
        \node at (-3.4,-1.0) {$\psi^-$};
        \node [right] at (-1.05,0.0) {$\psi^-$};
        \node at (1.8,1.0) {$\gamma/Z$};
        \node at (1.8,-1.0) {$\gamma/Z$};
        \draw[solid] (5.0,1.0)--(6.5,0.0);
        \draw[solid] (5.0,-1.0)--(6.5,0.0);
        \draw[snake] (6.5,0.0)--(8.5,0.0);
        \draw[solid] (8.5,0.0)--(10.0,1.0);
        \draw[solid] (8.5,0.0)--(10.0,-1.0);
        \node at (4.6,1.0) {$\psi^+$};
        \node at (4.6,-1.0) {$\psi^-$};
        \node [above] at (7.4,0.05) {$Z$};
        \node at (10.6,1.0) {$h_a$};
        \node at (10.6,-1.0) {$Z$};
     \end{tikzpicture}
 \end{center}
\caption{Feynmann diagrams for charged fermionic DM, $\psi^\pm$ annihilation to SM particles in final states. Here $a=1,2,3$ . }
\label{co-ann-3}
 \end{figure}
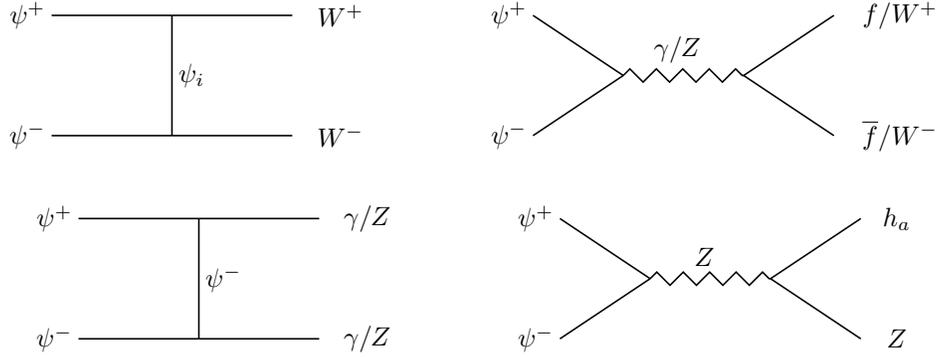
\begin{figure}[htb!]
\begin{center}
    \begin{tikzpicture}[line width=0.6 pt, scale=0.8]
        \draw[solid] (-3,1.0)--(-1.0,1.0);
        \draw[solid] (-3,-1.0)--(-1.0,-1.0);
        \draw[solid] (-1.0,1.0)--(-1.0,-1.0);
        \draw[solid] (-1.0,1.0)--(1.0,1.0);
        \draw[solid] (-1.0,-1.0)--(1.0,-1.0);
        \node at (-3.4,1.0) {$\overline{\psi_i}$};
        \node at (-3.4,-1.0) {$\psi_j$};
        \node [right] at (-1.05,0.0) {$\psi_k$};
        \node at (4.2,1.0) {$Z_{B-3L_\tau}/h_{a}/Z/Z_{B-3L_\tau}/Z_{B-3L_\tau}$};
        \node at (4.2,-1.0) {$h_{a}/Z_{B-3L_\tau}/Z_{B-3L_\tau}/Z/Z_{B-3L_\tau}$};
         \end{tikzpicture}
 \end{center}    
\begin{center}
    \begin{tikzpicture}[line width=0.6 pt, scale=0.8]        
        \draw[solid] (5.0,1.0)--(6.5,0.0);
        \draw[solid] (5.0,-1.0)--(6.5,0.0);
        \draw[snake] (6.5,0.0)--(8.5,0.0);
        \draw[solid] (8.5,0.0)--(10.0,1.0);
        \draw[solid] (8.5,0.0)--(10.0,-1.0);
        \node at (4.7,1.0) {$\overline{\psi_i}$};
        \node at (4.7,-1.0) {$\psi_j$};
        \node [above] at (7.4,0.05) {$Z_{B-3L_\tau}$};
        \node at (11.6,1.0) {$Z_{B-3L_\tau}/\tau/\nu_{\tau}/q$};
        \node at (11.1,-1.0) {$h_{a}/\overline{\tau}/\overline{\nu_{\tau}}/\overline{q}$};
        \draw[solid] (14.0,1.0)--(15.5,0.0);
        \draw[solid] (14.0,-1.0)--(15.5,0.0);
        \draw[dashed] (15.5,0.0)--(17.5,0.0);
        \draw[solid] (17.5,0.0)--(19.0,1.0);
        \draw[solid] (17.5,0.0)--(19.0,-1.0);
        \node at (13.7,1.0) {$\overline{\psi_i}$};
        \node at (13.7,-1.0) {$\psi_j$};
        \node [above] at (16.4,0.05) {$h_a$};
        \node at (20.1,1.0) {$Z_{B-3L_\tau}$};
        \node at (20.1,-1.0) {$Z_{B-3L_\tau}$};
     \end{tikzpicture}
 \end{center}
\caption{ Additional Feynmann diagrams for DM, $\psi_i$  due to presence of new gauged paricle $Z_{B-3L_{\tau}}$ in the model: Annihilation ($i=j$) and Co-annihilation ($i\neq j$). Here $i,j,k=1,2$;~ $a=1,2,3$ ~ and q stand for SM quarks. }
\label{fd:an-coan1}
 \end{figure}
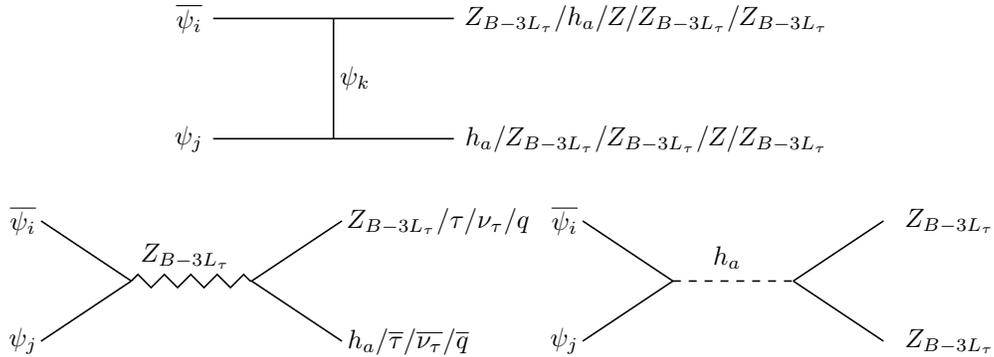
\begin{figure}[htb!]
\begin{center}
    \begin{tikzpicture}[line width=0.6 pt, scale=0.8]
        \draw[solid] (-3,1.0)--(-1.0,1.0);
        \draw[solid] (-3,-1.0)--(-1.0,-1.0);
        \draw[solid] (-1.0,1.0)--(-1.0,-1.0);
        \draw[solid] (-1.0,1.0)--(1.0,1.0);
        \draw[solid] (-1.0,-1.0)--(1.0,-1.0);
        \node at (-3.6,1.0) {$\overline{\psi_i}/\psi_i$};
        \node at (-3.6,-1.0) {$\psi^-/\psi^+$};
        \node [right] at (-1.05,0.0) {$\psi_k$};
        \node at (2.2,1.0) {$Z_{B-3L_\tau}$};
        \node at (2.2,-1.0) {$W^\mp$};
        \draw[solid] (5,1.0)--(7.0,1.0);
        \draw[solid] (5,-1.0)--(7.0,-1.0);
        \draw[solid] (7.0,1.0)--(7.0,-1.0);
        \draw[solid] (7.0,1.0)--(9.0,1.0);
        \draw[solid] (7.0,-1.0)--(9.0,-1.0);
        \node at (4.4,1.0) {$\overline{\psi_i}/\psi_i$};
        \node at (4.4,-1.0) {$\psi^-/\psi^+$};
        \node [right] at (7.0,0.0) {$\psi^\mp$};
        \node at (10.2,1.0) {$W^\mp$};
        \node at (10.2,-1.0) {$Z_{B-3L_\tau}$};
         \end{tikzpicture}
 \end{center}    
\caption{ Feynmann diagrams for co-annihilation processes of $\psi_i ~(i=1,2)$ with the 
charged component $\psi^\pm$ to SM $W^\pm$ and BSM $Z_{B-3L_\tau}$. }
\label{fd:an-coan}
 \end{figure}
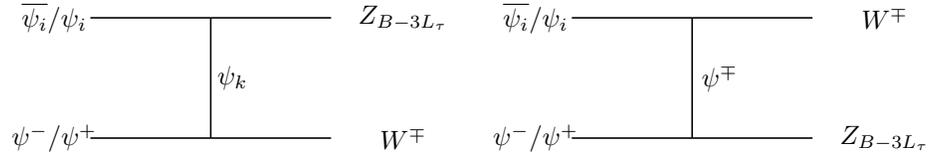
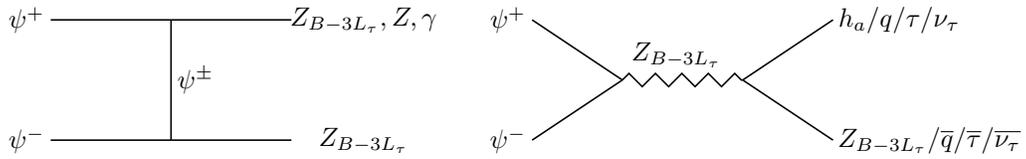
\begin{figure}[htb!]
\begin{center}
    \begin{tikzpicture}[line width=0.6 pt, scale=0.8]
        \draw[solid] (-3,1.0)--(-1.0,1.0);
        \draw[solid] (-3,-1.0)--(-1.0,-1.0);
        \draw[solid] (-1.0,1.0)--(-1.0,-1.0);
        \draw[solid] (-1.0,1.0)--(1.0,1.0);
        \draw[solid] (-1.0,-1.0)--(1.0,-1.0);
        \node at (-3.4,1.0) {$\psi^+$};
        \node at (-3.4,-1.0) {$\psi^-$};
        \node [right] at (-1.05,0.0) {$\psi^\pm$};
        \node at (2.2,1.0) {$Z_{B-3L_{\tau}},Z,\gamma$};
        \node at (2.2,-1.0) {$Z_{B-3L_{\tau}}$};
        \draw[solid] (5.0,1.0)--(6.5,0.0);
        \draw[solid] (5.0,-1.0)--(6.5,0.0);
        \draw[snake] (6.5,0.0)--(8.5,0.0);
        \draw[solid] (8.5,0.0)--(10.0,1.0);
        \draw[solid] (8.5,0.0)--(10.0,-1.0);
        \node at (4.6,1.0) {$\psi^+$};
        \node at (4.6,-1.0) {$\psi^-$};
        \node [above] at (7.4,0.05) {$Z_{B-3L_{\tau}}$};
        \node at (11.1,1.0) {$h_a/q/\tau/\nu_{\tau}$};
        \node at (11.6,-1.0) {$Z_{B-3L_{\tau}}/\overline{q}/\overline{\tau}/\overline{\nu_{\tau}}$};
     \end{tikzpicture}
 \end{center}
\caption{New Feynmann diagrams for charged fermionic DM, $\psi^\pm$ annihilation due to presence of new gauged paricle $Z_{B-3L_{\tau}}$. Here $a=1,2,3$. }
\label{co-ann-4}
 \end{figure}

\clearpage

\providecommand{\href}[2]{#2}\begingroup\raggedright\endgroup
\end{document}